\documentclass[10pt]{article}
\usepackage[french,english]{babel}
\usepackage{amsmath}
\usepackage{amsfonts}
\usepackage{amssymb}
\usepackage{graphicx}                              
\usepackage{times}
\usepackage[makeroom]{cancel}   
\begin{document}
%
October 21th, 2014   \hfill  

\vskip 7cm
{\baselineskip 12pt
\begin{center}
{\bf 
\hbox{\hskip -5mm REFRACTIVE PROPERTIES OF GRAPHENE
                IN A MEDIUM-STRONG EXTERNAL MAGNETIC FIELD}}
\end{center}
}
\baselineskip 16pt
\arraycolsep 3pt  
%
\vskip .2cm
\centerline{O.~Coquand
     \footnote[1]{Ecole Normale Sup\'erieure, 61 avenue du Pr\'esident
Wilson, F-94230 Cachan}
     \footnote[2]{ocoquand@ens-cachan.fr}
,\; B.~Machet
     \footnote[3]{Sorbonne Universit\'e, UPMC Univ Paris 06, UMR 7589,
LPTHE, F-75005, Paris, France}
     \footnote[4]{CNRS, UMR 7589, LPTHE, F-75005, Paris, France.}
     \footnote[5]{Postal address:
LPTHE tour 13-14, 4\raise 3pt \hbox{\tiny \`eme} \'etage,
          UPMC Univ Paris 06, BP 126, 4 place Jussieu,
          F-75252 Paris Cedex 05 (France)}
    \footnote[6]{machet@lpthe.jussieu.fr}
     }
\vskip 1cm

{\bf Abstract:} 1-loop quantum corrections are shown to induce large effects
on the refractive index $n$ inside a graphene strip
in the presence of a constant and uniform external magnetic field $B$ orthogonal to it.
To this purpose, we use the tools of Quantum Field Theory  to calculate
the photon propagator at 1-loop inside  graphene in position space,
which leads to an effective vacuum polarization in a brane-like theory
of  photons interacting with  massless 
electrons at locations  confined  inside the thin strip  (its
longitudinal spread is considered to be infinite).
The effects factorize into  quantum ones, controlled by the value of
$B$ and that of the electromagnetic coupling $\alpha$, 
and a transmittance function $U$ in which the geometry of the sample and 
the resulting confinement of the $\gamma\,e^+ e^-$ vertices  play  major roles.
They only concern the so-called ``transverse-magnetic'' polarization of
photons, which suggests (anisotropic) electronic spin resonance  of the  graphene-born
virtual electrons.  We consider photons  inside the visible spectrum
and  magnetic fields  in the range 1-20\; Tesla.
At $B=0$, quantum effects depend very weakly
on  $\alpha$ and $n$ is essentially controlled by  $U$; we recover, then,
 an opacity for visible light of the same order of magnitude $\pi \alpha_{vac}$
 as measured experimentally.



\newpage\tableofcontents

\newpage\listoffigures\newpage

\section{Introduction. Main features of the calculation}\label{section:intro}

Constant magnetic fields $B$ can induce, through the screening
of the Coulomb potential, dramatic effects on
the spectrum of hydrogen and on the critical number $Z_c$ of atoms
\cite{Vysotsky}\cite{MachetVysotsky} \cite{GodunovMachetVysotsky}.
However,  typical effects being ${\cal O}(\frac{\alpha \hbar e B}{m_e^2 c^2})$, 
 gigantic fields are needed, $B\geq 10^{12}\,T$, which are out of reach
on earth. It was also shown in \cite{Shabad} that such extreme
``supercritical'' magnetic fields could strongly modify the refraction of
light.
The property that the fine structure constant $\alpha$ in graphene largely exceeds
$1$ \cite{Goerbig} instead of its vacuum value
$\alpha_{vac}\simeq\frac{1}{137}$ was a sufficient motivation
to investigate whether  sizable effects could be obtained at lower cost in
there.

While graphene in a constant, uniform external magnetic field
 is usually associated with the
so-called ``abnormal quantum hall effect'' \cite{Goerbig} \cite{CastroNeto},
we have found that one can also expect optical effects for electromagnetic
waves in the visible spectrum and at ``reasonable''
values of the external $B$ not exceeding $20\,T$.

Since we are concerned with the refractive index, the main
object of our study is the propagator of the photon (with incoming momentum
$q$) inside graphene, and, more specially  its quantum corrections at 1-loop.
They originate from the creation, inside the medium,
of virtual $e^+ e^-$ pairs, which can then propagate everywhere
before annihilating, again inside graphene.  We therefore need to constrain the
two $\gamma\,e^+e^-$ vertices to lie  in the
interval $[-a, +a]$ along the direction $z$ of the magnetic field,
perpendicular to the surface of graphene.
To this purpose,
we evaluate the photon propagator in position space, and integrate
the ``$z$'' coordinates of the two vertices from $-a$ to $+a$ instead of the
usual infinite interval of customary Quantum Field Theory.
This strategy sets of course the intrinsic limitation of our calculations
that they are only valid inside graphene.

The next feature to be accounted for is that, in the close vicinity of the
Dirac points of graphene, electronic excitations are massless with a linear
dispersion relation of the type $p_0 = v_F |\vec p|$, where $p_0$ is the
energy of the particle and $v_F$ is the
Fermi velocity $v_F \ll c$ \cite{Goerbig} \cite{CastroNeto}.
 This is obtained in the tight-binding
approximation, which leads to a massless Dirac-like Hamiltonian for these
excitations, in which $c$ is replaced with $v_F$.
This raises the issue of which electronic propagator  we
have to insert in the quantum loop. At first sight, the natural candidate
corresponds to the massless Lagrangian
\begin{equation}
\bar\psi \left(\gamma_0 p_0 - v_F(\gamma_1 p_1
-\gamma_2 p_2 -\bcancel{\gamma_3 p_3}) -\bcancel{m c}^2\right)\psi
\label{eq:dirgraph}
\end{equation}
(we have restored the appropriate factors with dimension $[velocity]$),
which corresponds  to the effective Dirac-like
Hamiltonian of  graphene electrons. 
However, as explained  in subsection \ref{subsub:geneprop},
there are  strong motivations for putting inside the loop 
Dirac-like excitations propagating like in vacuum, that is with $c$
instead of $v_F$.
The first is that,  while electron/positron excitations are
created and annihilated inside graphene, they can then
propagate in the whole space. Actually, for the idealized graphene strip
with infinite horizontal spreading $L\to\infty$ that we are considering,
the Coulomb energy of an electron, expected to vary like $1/L$, can 
be neglected, and virtual electrons spend much
more time in the ``bulk'' (outer space) than inside graphene.
The second reason concerns energy-momentum conservation at the
vertices
\footnote{
We also checked that, if $v_F$ is used inside the electron
propagators, the value of the refractive index at $B\not=0$ grows
to unreasonably large values ($\geq 100$) and the opacity at $B=0$ gets also
spoiled by factors $\propto \frac{c}{v_F}$.}
.\newline
The last issues concern whether we may keep $m=0$ and $p_3=0$.
When doing a perturbative expansion, propagators of internal lines are
always the ones corresponding to the classical Lagrangian.
In the problem under scrutiny, internal electron lines must therefore correspond
 to the effective classical Hamiltonian of graphene at the Dirac
points.
Since doing perturbation amounts to  calculating  quantum fluctuations,
this choice amounts
to selecting a  ``classical'' starting point (vacuum) for perturbation
theory, which is ``graphene''.
Would quantum corrections  trigger, for example, large ``chiral symmetry
breaking", serious doubts should be cast on this choice and on the
reliability of the procedure (there are good reasons to think that we are
safe, look for example at subsection \ref{subsec:reson}). \newline
Therefore,  the propagator of virtual
electrons that we shall use is that of massless Dirac electrons with
$p_3=0$ which propagate like in vacuum, but in the presence of a constant,
uniform external $B$; its expression is given by the Schwinger formalism
\cite{Schwinger} \cite{DittrichReuter}. That no $v_F$ is
introduced in there makes finally that
the Fermi velocity appears nowhere in our formul{\ae},
except, implicitly, inside the electromagnetic coupling
$\alpha$ that we shall vary from its vacuum value
$\alpha_{vac}=\frac{e^2}{4\pi \epsilon_0 \hbar c}=
\frac{1}{137}$ up to $\alpha \simeq 2$, which roughly
corresponds to its effective value inside graphene
\footnote{In most of the paper, we shall nevertheless 
 keep the dependence on $\hbar$ and $c$, to make conspicuous the dimension
of the parameters entering the calculations. They are only skipped when no
confusion can arise.}.

The calculation of the photon propagator at 1-loop yields a
1-loop vacuum polarization tensor $\Pi^{\mu\nu}$
 that can be plugged in the light-cone
equations derived according to the pioneering work of Tsai and Erber
\cite{TsaiErber1}, and of Dittrich and Gies \cite{DittrichGies1}.
One of the salient features of  $\Pi^{\mu\nu}$ is that
it factorizes into a tensor $T^{\mu\nu}(\hat q, B)$,
which depends on $B$, $\hat q\equiv(q_0,q_1,q_2)$ and
on the electromagnetic coupling $\alpha$,
$\times$ a universal function $U$ which does not depend on the
magnetic field, nor of  $\alpha$.
While $U$  carries information concerning the geometry of
the sample and the confinement of the vertices, and shares
similarities with the so called ``transmittance'' function in optics or
``transfer function'' in electronics,
$T^{\mu\nu}(\hat q, B)$ gathers
quantum effects and those of the magnetic field. Its components 
$\mu,\nu=0,1,2$ are reminiscent of those of  vacuum polarization in $2+1$
dimensions in the presence of $B$; however, for the system under concern,
$T^{33}=-T^{00}\not=0$  plays the dominant role. At the limit
$B=\infty$ they are the only components
that subsist, as expected from the $D\to D-2$ dimensional reduction that takes
place in this case (see for example\cite{ShabadUsov})
(only the $0$ and $3$ components of the photon then couple).

From the dimensional point of view, we  consider the graphene strip 
as a truly 3+1 dimensional object,  the thickness $2a$ of which is
very  small as compared with its flat extension; nowhere have we made the
premise that the underlying physics is 2+1 dimensional. Classically, 
virtual electron-positron pairs created at the lowest Landau level on the
Dirac cone have  a vanishing momentum $p_3=0$ in the direction of
$B$; however, the large quantum fluctuations $\sim \hbar/a$ that
arise due to the confinement of $\gamma\,e^+ e^-$ vertices inside the
medium allow them to eventually evolve in the whole 3+1 space.
In this setup, the direction of  $B$ has a twofold
importance: first, due to the dimensional reduction $D \to D-2$ mentioned
above; secondly because  the  vertex confinement and the related quantum
fluctuations and momentum exchanges 
largely influence the behavior of the refractive index.
In particular, forgetting about $a\not=0$ erases
the transmittance function and the leading $1/\sin\theta$ behavior
of the refractive index.

The quantum fluctuations of electronic momentum  in the  direction of
$B$ get transferred to the photon.  That the resulting photonic
momentum non-conservation should not exceed $\hbar/a$ yields a
quantum upper bound $n \leq n_{quant}$ for the refractive index.

The effects of confinement that we exhibit should not be put hastily
in correspondence
with the ones that have, for example,  been investigated  in
 \cite{Perez} for a finite longitudinal size $L$ of graphene.
A major difference is indeed
that we are concerned here with the confinement in the
``short'' direction, the thickness $2a\approx 350\, pm$,
considering that its longitudinal spread $L$ is infinite
\footnote{The cyclotron radius $\ell_c = \sqrt{\frac{\hbar}{eB}}$ is
$\ell_c \approx 8.1\,10^{-9}\, (meter)$ at $B=10\,T$.}.
This makes the physical interpretation less intuitive 
 since no cyclotron radius can eventually, in our case, match the size of
graphene. However, our results exhibit the  remarkable property that, again
in relation with the dimensional reduction that takes place in the presence
of a strong external $B$,
only the propagation of  photons with ``parallel'' polarization gets
concerned (for the transverse
polarization,  the only solution that we found to the light-cone equation
is the trivial $n=1$). In this state of polarization, the oscillating  magnetic
field of the electromagnetic wave is orthogonal to the external constant $B$,
which is a typical situation  to induce the magnetic
resonance of the spins of the graphene-born electrons. 
Electron spin resonance may thus be
at the origin of the large sensitivity of the refractive
index to the external $B$. The large value of the electromagnetic coupling also
participates to producing macroscopic effects.

The refractive index $n=n_1+i\,n_2$
 is found to essentially depend on $\alpha$, on the angle of
incidence $\theta$, and on the ratio
$\Upsilon=\frac{c\sqrt{2\hbar eB}}{q_0}$.
In the absence of any external $B$, its dependence 
on the electromagnetic coupling fades away, and it is mainly constrained by
the sole property that electrons are created and annihilated inside
graphene.

A transition occurs at small angle of incidence $\theta_{min} \sim
\frac{1}{\Upsilon}$: no non-trivial solution with $|n_2| \ll n_1$ to the light-cone
equation exists anymore for $\theta < \theta_{min}$. This also corresponds
to $|n| \leq n_{max} \sim \Upsilon$. Since at
$\theta=0$ (normal incidence), the only solution to the light-cone equation
is  the trivial  $n=1$, getting reliable results in 
the zone of transition from $\theta_{min}$ down to $\theta=0$
requires more elaborate numerical techniques, which is left for a subsequent
work.

Our calculations and the corresponding expansions
 are made in the limit of a ``medium-strong'' $B$,
in the sense that $\sqrt{2\hbar eB} \gg q_0/c$,
 and are only valid at this limit such that, in particular, the limit $B\to 0$
cannot be taken. $B$ is however not considered to be ``infinite''
like in \cite{MachetVysotsky} \cite{GodunovMachetVysotsky} \cite{Godunov}.
In practice, in the case at hand, the leading terms in the expansion of the
electronic propagators in powers of $1/\tau eB \ll1$ ($\tau$ is the proper time)
do not contribute to the refractive index. The effects originate from
the subleading terms, and the final growing like $\sqrt{eB}$ of the
relevant components of the vacuum polarization tensor comes from the
integration over the transverse electronic degrees of freedom.

Expansions are also done at small values of the parameter
$\eta=\frac{aq_0}{\hbar c}$.
This condition is always satisfied for optical frequencies.
It also guarantees to stay in the linear part of the
electron spectrum close to the Dirac point, which is an essential
ingredient to use a ``Dirac-like'' effective Hamiltonian \cite{Goerbig}.

We are concerned with photons in the visible spectrum, which sets us very
far from geometrical optics, since 
the corresponding wavelengths are  roughly three orders of
magnitude larger that the thickness $2a$ of graphene.

We limit $B$, for the sake of experimental feasibility, to $20\,T$.
This upper bound also guarantees that the 4-fold degeneracy of the
Landau level at the Dirac point does not yet get lifted \cite{Zhang}.

Our results are summarized on the two plots of Figure\;\ref{fig:nBreal}.

The last section  deals with the case $B=0$, for which a dedicated
calculation is needed.  We show in this case that no $\theta_{min}$
exists and that, instead, as the angle of incidence gets smaller and
smaller, the refractive index $n$  goes continuously from ``quasi-real'' values to
complex values with larger $n_1$ and $n_2$. At very small values of
$\theta$, we recover an opacity of the same order
of magnitude  as the one measured experimentally \cite{Nair}. However, the same
problem as for $B\not=0$ exists concerning a smooth transition to
$\theta=0$. In addition, for $\alpha >1$, the index diverges at $s_\theta^2
\geq \frac{1}{\alpha}$, expressing problems of a fixed-order perturbative
expansion at strong coupling.

The paper is intended to be self-contained.
The amount of literature dedicated to graphene is very large and we cannot,
unfortunately, pay  a fair tribute to the whole of it. We only cite
the works that have been the most  used for writing the present one,
 but the reader can find, in particular inside the review
articles, references to most of the important papers.

\section{From the vacuum polarization to light-cone equations and to
the refractive index}\label{section:general}

\subsection{Conventions and settings. Notations}\label{subsec:settings}

\begin{figure}[h]
\begin{center}
\includegraphics[width=10 cm, height=7 cm]{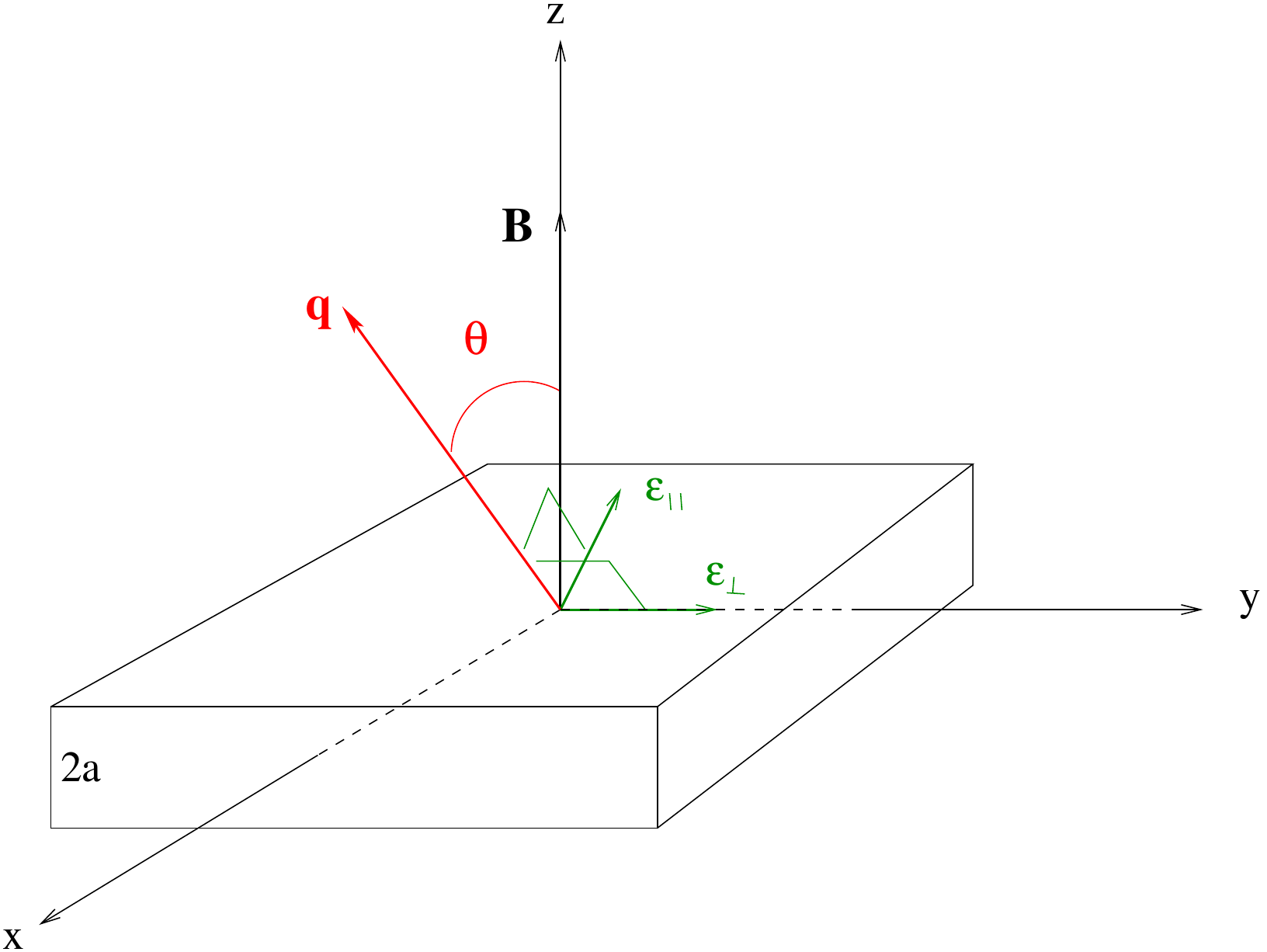}
\caption{$\vec B$ is perpendicular to the graphene strip of width $2a$.
The polarization vector $\vec\epsilon$, perpendicular to the momentum $\vec
q$ of the electromagnetic wave, is decomposed into $\vec\epsilon_\parallel $ in
the $(x,z)$ plane and $\vec\epsilon_\perp$ perpendicular to this
plane.} \label{fig:setup}
\end{center}
\end{figure}

Following Tsai-Erber \cite{TsaiErber1}, the constant and uniform
magnetic field  $\vec B$ is chosen to be parallel to the $z$ axis
and the wave vector $\vec q$ of the propagating photon 
to lie in the $(x,z)$ plane (see Figure~\ref{fig:setup})
\footnote{When no ambiguity can occur, we shall often omit the arrow on
3-dimensional vectors, writing for example $B$ instead of $\vec B$.}.

The $(\vec B, \vec q)$ angle $\theta$ is the ``angle of incidence''; 
since we are concerned with the propagation of light
{\em inside} graphene, $\theta$  is  the angle of incidence of light
{\em inside this medium}. The plane $(x,z)$ is the plane of incidence.

The polarization vector $\vec\epsilon$\; (which, by convention, refers to
the electric field) is decomposed
into $\vec\epsilon_\parallel= -\cos\theta\, \vec i + \sin\theta\, \vec k$,
in the $(x,z)$ plane,
 and $\vec\epsilon_\perp \parallel\vec j$, both orthogonal to
$\vec q$ ($\vec i, \vec j, \vec k$ are the unit vectors along the $x,y,z$
axes).
One has $\vec q= |\vec q|\,(\sin\theta\,\vec i + \cos\theta\,\vec
k)$.
$\vec\epsilon_\parallel$ is called ``parallel polarization'' and
$\vec\epsilon_\perp$ ``transverse polarization''. They are also called
respectively ``transverse magnetic'' and ``transverse electric'' by reference
to plane waves.
It must be noticed that, at normal incidence $\theta=0$, there is no longer
a plane $(\vec q, \vec B)$ such that these two polarizations can no longer be
distinguished.

We shall in the following use ``hatted'' letters for vectors living in the
Lorentz subspace $(0,1,2)$. For example
\begin{equation}
\hat q = (q^0,q^1,q^2),\quad q=(\hat q, q_3) = (q_0,q_1,q_2,q_3)=(q_0, \vec
q).
\end{equation}

Throughout this work we use the metric $(+,-,-,-)$ and mostly work in the
International Unit System (SI). It is however often convenient to express
energies in $eV$.

\subsection{The modified Maxwell Lagrangian and the light-cone equations}
\label{subsec:lightcone}

\begin{figure}[h]
\begin{center}
\includegraphics[width=6 cm, height=3 cm]{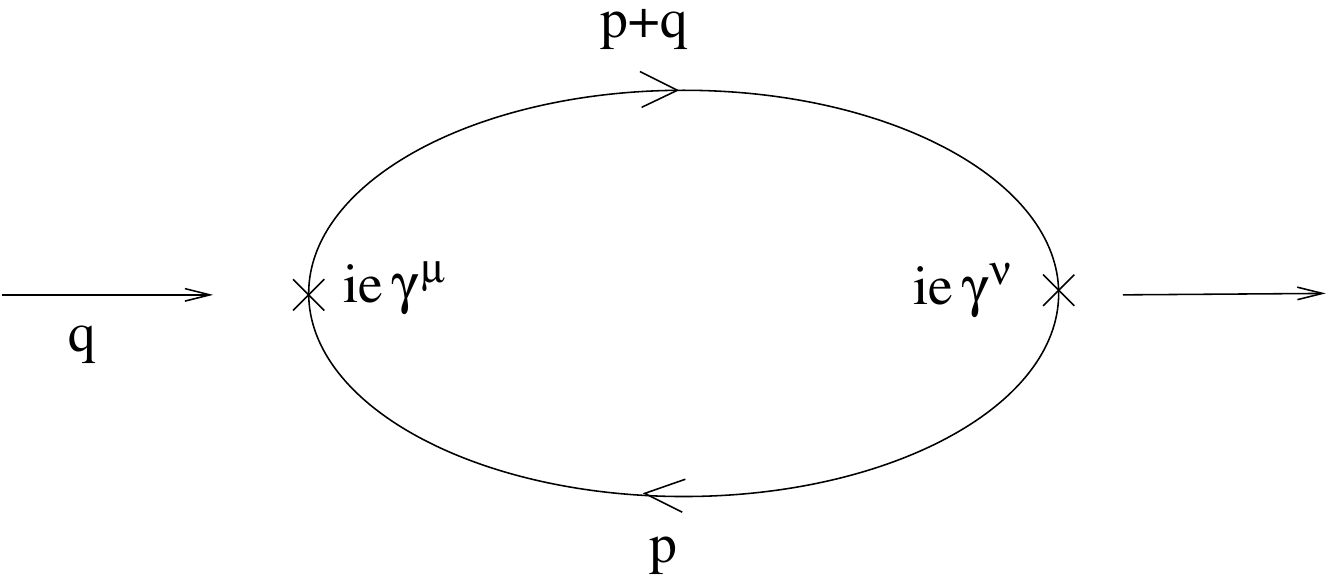}
\caption{The vacuum polarization $\Pi^{\mu\nu}(q)$.}
\label{fig:vacpol}
\end{center}
\end{figure}

Taking into account the contribution of
the vacuum polarization $\Pi_{\mu\nu}$ (see Figure \ref{fig:vacpol}) that
we shall calculate in section \ref{section:propagator},
 the Maxwell Lagrangian $-\frac14 F_{\mu\nu}(x)F^{\mu\nu}(x)$
 gets modified to
\cite{DittrichGies1}
\begin{equation}
{\cal L}(x,B)= -\frac14 F_{\mu\nu}(x)F^{\mu\nu}(x) -\frac12 \int d^4 y\;
A^\mu(x)\Pi_{\mu\nu}(x, y,B) A^\nu( y),
\label{eq:maxL}
\end{equation}
from which one gets the Euler-Lagrange equation
\begin{equation}
\Big( g_{\mu\nu}\,q^2 -q_\mu q_\nu + \Pi_{\mu\nu}(q, B)\Big)
A^\nu(q)=0.
\label{eq:maxeq}
\end{equation}
Left-multiplying (\ref{eq:maxeq}) with
\begin{equation}
A^\mu= \beta_1\,\epsilon^\mu_\perp + \beta_2\,\epsilon^\mu_\parallel,
\label{eq:Apol}
\end{equation}
yields the light-cone equation
\footnote{When $\Pi_{\mu\nu}$ is not present,  the only non-vanishing
elements are ``diagonal'',
$\epsilon^\mu_\perp \Big( g_{\mu\nu}\,q^2 -q_\mu
q_\nu\Big)\epsilon^\nu_\perp=q^2
=\epsilon^\mu_\parallel \Big( g_{\mu\nu}\,q^2 -q_\mu
q_\nu\Big)\epsilon^\nu_\parallel$, which yields
$A^\mu \Big( g_{\mu\nu}\,q^2 -q_\mu
q_\nu\Big) A^\nu = (\beta_1^2 + \beta_2^2) q^2$, and, accordingly,
the customary light-cone condition $q^2=0 \equiv q_0^2 -\vec q^2$.
If $\Pi_{\mu\nu}$ is transverse $\Pi_{\mu\nu}= (g_{\mu\nu}q^2
-q_\mu q_\nu)\Pi(q^2)$, the light-cone condition is $(\beta_1^2+\beta_2^2)q^2
(1+\Pi(q^2))=0$, that is, as usual, $q^2=0$. }

\begin{equation}
\begin{split}
& (\beta_1\epsilon^\mu_\perp + \beta_2\epsilon^\mu_\parallel)
\Big( g_{\mu\nu}\,q^2 -q_\mu q_\nu + \Pi_{\mu\nu}(q,  B)\Big)
(\beta_1\epsilon^\nu_\perp + \beta_2\epsilon^\nu_\parallel)=0,\cr
& \epsilon^\mu_\perp = (0,0,1,0),\quad
\epsilon^\mu_\parallel=(0,-c_\theta,0,s_\theta),\qquad
c_\theta\equiv\cos\theta,\ s_\theta\equiv\sin\theta.
\end{split}
\label{eq:lc00}
\end{equation}
As we shall see in sections \ref{section:propagator} and \ref{section:T},
$\Pi^{03}=0=\Pi^{13}=\Pi^{23}$, such 
the  light-cone equation (\ref{eq:lc00}) simplifies to
\begin{equation}
(\beta_1^2 + \beta_2^2)q^2 + \left(\beta_1^2 \Pi^{22}(q,B) + \beta_2^2 \left(
c_\theta^2\;
\Pi^{11}(q,B) +
s_\theta^2\; \Pi^{33}(q,B)\right) + 2\beta_1\beta_2\,c_\theta\,
\Pi^{12}(q,B)\right)
=0.
\label{eq:lcgen}
\end{equation}
$\vec q$ has been furthermore chosen to lie in the $(x,z)$ plane, so
$q_2=0$, which entails
(see (\ref{eq:summary})) $\Pi^{02}=0=\Pi^{20}, \Pi^{12}=0 =
\Pi^{21}$,
 and the light-cone equation finally shrinks to
\begin{equation}
(\beta_1^2 + \beta_2^2)\,q^2 + \left(\beta_1^2\, \Pi^{22}(q,B) + \beta_2^2 \left(
c_\theta^2\;
\Pi^{11}(q,B) + s_\theta^2\; \Pi^{33}(q,B)\right)\right)
=0.
\label{eq:lc0}
\end{equation}

Depending of the polarization of the photon, there are accordingly
two different light-cone  relations:\newline
$\bullet$\ for $A^\mu_\perp(q_0,q_1,0,q_3)$, $\beta_1=1,\beta_2=0$,

\begin{equation}
q^2 + \Pi^{22}(q,B) =0;
\label{eq:lcperp}
\end{equation}
$\bullet$\  for  $A^\mu_\parallel(q_0,q_1,0,q_3)$, $\beta_1=0,\beta_2=1$,
\begin{equation}
q^2 + \Big(c_\theta^2\; \Pi^{11}(q,B) + s_\theta^2\;
\Pi^{33}(q,B)\Big) =0.
\label{eq:lcpar}
\end{equation}
Notice   the occurrence of $\Pi^{33}$ in (\ref{eq:lcpar}),
which plays a major role and  would not be there in $QED_{2+1}$
\footnote{This is to be put in relation with
the property \cite{Miransky} that fermions from the lowest
Landau level only couple to the $(0,3)$ components of the photon at $B\to
\infty$ (see also subsection \ref{subsub:Fappro}).}
.

A remark is due concerning eq.~(\ref{eq:maxeq}). Its derivation from the
effective Lagrangian (\ref{eq:maxL}) relies on the property that, in
position space,
$\Pi^{\mu\nu}(x,y)$ is in reality a function of $(x-y)$ only. This is
however, as we shall see, not exactly the case here. $\Pi^{\mu\nu}$ depends
indeed on $(\hat x -\hat y)$ but individually on $x_3$ and $y_3$ (see the
first remark at the end of subsection \ref{subsub:graph}). Once the
dependence on $(x_3-y_3)$ has been extracted, there is a left-over dependence
on $y_3$, which finally yields for our results the dependence of the
refractive index on $u=\frac{y_3}{a} \in[-1,+1]$. We shall see however that
this dependence is always extremely weak, and we consider therefore 
the Euler-Lagrange equation (\ref{eq:maxeq})  to be valid to a very good approximation.

\subsection{The refractive index $\boldsymbol n$}\label{subsec:index}

We define it in a standard way by
\footnote{This is equivalent to $n=\frac{c}{v}$
 for a plane wave $e^{i(\vec q.\vec x -\omega t)}$ with phase velocity
$v=\frac{\omega}{|\vec q|}$.}
\begin{equation}
n= \frac{c|\vec q|}{q_0}.
\label{eq:ndef}
\end{equation}

In practice, $\Pi^{\mu\nu}$ is not only a function of $q$ and $B$,
but of the angle of incidence $\theta$ and of the relative depth $u$ inside
the graphene strip, $u\in [-1, +1] $.
The light-cone equations therefore translate into relations
$n=n(\theta,B,q_0,u)$ that we will write explicitly in section
\ref{section:lceqs}, after calculating the vacuum polarization.

\section{The photon propagator in $\boldsymbol x$-space and the
vacuum polarization $\boldsymbol{\Pi^{\mu\nu}}$}
\label{section:propagator}

The vacuum polarization $\Pi^{\mu\nu}$  to be introduced inside the light-cone
equations (\ref{eq:lcperp},\ref{eq:lcpar}) 
is  obtained by
calculating the photon propagator in position-space, while confining, at
the two vertices $\gamma\, e^+ e^-$, the corresponding $z$'s
to lie inside graphene, $z\in [-a,a]$.

It factorizes into $\Pi^{\mu\nu}(\hat q,q_3,\frac{y_3}{a},B)=
\frac{1}{\pi^2}\;T^{\mu\nu}(\hat q,B)\; U(\hat q, q_3, \frac{y_3}{a})$
in which  $U$ is a universal function that does not depend on the magnetic field,
nor on $\alpha$, that we also encounter when dealing
with the case of no external $B$. It is the Fourier transform of the
product of two functions:  the first, $\frac{\sin a k_3}{ak_3}$,  is the
Fourier transform of the ``gate function'' corresponding to
the graphene strip along $z$; the second  carries  the remaining
information attached to the confinement of the vertices.
Its analytical properties inside the complex plane control in particular
the ``leading'' $\frac{1}{\sin\theta}$ behavior of the refractive index
inside graphene.
The integration variable of this Fourier transform is $k_3$, the
difference between the momenta along $B$ of
the outgoing and incoming photons (see below).

This factorization can be traced back to $T^{\mu\nu}$  not depending on
$q_3$, for the simple reason that the propagators of electrons inside
graphene are evaluated at vanishing ``$p_3$'' momentum (in the direction of the
external $B$). 
An example of how factors combine is the following.
$\Pi^{\mu\nu}$ still includes an integration on $p_3$, which
factors out.
That the interactions of electrons are confined along $B$ triggers quantum
fluctuations of their momentum in this direction. Setting an ultraviolet
cutoff  $\pm\frac{\hbar}{a}$ on the $p_3$ integration (saturating the
Heisenberg uncertainty relation) makes  this integral  proportional to
$\frac{1}{a}$. This factor  completes, inside the integral
$\int dk_3$ defining $U$, the ``geometric'' $\frac{\sin ak_3}{ak_3}$
evoked above.

$k_3=s_3-q_3$ represents the amount of momentum non-conservation of photons
in the direction of $B$: it occurs by momentum exchange between photons
and (the quantum momentum fluctuations of) electrons. 
The integration $dk_3$  gets 
bounded by the rapid decrease of $\frac{\sin ak_3}{ak_3}$ for
$|k_3|$ larger than $\frac{\hbar}{a}$  and
this upper bound $|k_3|\leq \frac{\hbar}{a}$ is the same as the one
that we set for quantum fluctuations of
 the electron momentum $p_3$. So, the energy-momentum
non-conservation between  the outgoing and incoming photons cannot exceed
the uncertainty on the momentum of electrons 
due to the confinement of vertices. Momentum conservation for the photon
 is only recovered when $a\to\infty$ (limit of ``standard'' QFT).

\subsection{The 1-loop photon propagator in position space}
\label{subsec:gammaprop}

We calculate the 1-loop photon propagator
\begin{equation}
 \Delta^{\rho\sigma}(x,y)= \langle 0 \vert T A^\rho(x) A^\sigma(y)\vert
0\rangle
\end{equation}
and somewhat lighten  the notations, omitting symbols like T-product, 
\ldots, writing for example $G(\hat p)$ for $G(\hat p, B)$ etc.

Introducing the coordinates $u=(u_0,u_1,u_2,u_3)$ and $v=(v_0,v_1,v_2,v_3)$
of the two $\gamma\,e^+ e^-$ vertices one gets at 1-loop
\begin{equation}
 \Delta^{\rho\sigma}(x,y)=
 \int d^4u \int d^4v\; A^\rho(x) [(ie) A^\mu(u) \bar\psi(u)\gamma_\mu
\psi(u)]
[(ie)A^\nu(v) \bar\psi(v)\gamma_\nu \psi(v)] A^\sigma(y).
\end{equation}
Making the contractions for fermions etc \ldots yields
\begin{equation}
\begin{split}
 \Delta^{\rho\sigma}(x,y) &= e^2\int d^4u \int d^4v\; Tr
\int \frac{d^4q}{(2\pi)^4}\; e^{iq(u-x)}\Delta^{\rho\mu}(q)
\gamma_\mu \int \frac{d^4p}{(2\pi)^4}\; e^{ip(u-v)}G(p)
\gamma_\nu\cr
& \hskip 4cm \int \frac{d^4r}{(2\pi)^4}\; e^{ir(v-u)}G(r)
\int \frac{d^4s}{(2\pi)^4}\; e^{is(y-v)}\Delta^{\sigma\nu}(s).
\end{split}
\label{eq:start}
\end{equation}
In what follows we shall also omit the trace symbol ``$Tr$''.

\subsubsection{``Standard'' Quantum Field Theory}
\label{subsub:sQFT}

 One integrates $\int_{-\infty}^{+\infty}
d^4u$ and $\int_{-\infty}^{+\infty} d^4v$  for the four components of $u$ and
$v$. This gives:
\begin{equation}
 \Delta^{\rho\sigma}(x,y)=\int \frac{d^4q}{(2\pi)^4}\; e^{-iq(x-y)}
\Delta^{\rho\mu}(q)
\Delta^{\nu\sigma}(q)
\underbrace{e^2\int \frac{d^4p}{(2\pi)^4}\; \gamma_\mu G(p) \gamma_\nu
G(p+q)}_{i\Pi_{\mu\nu}(q)}.
\label{eq:stand}
\end{equation}
To obtain the sought for vacuum polarization, the two external photon propagators
$\Delta^{\rho\mu}(q)$ and $\Delta^{\nu\sigma}(q)$
have to be truncated, which gives the customary expression
\begin{equation}
 i\Pi_{\mu\nu}(q)=+e^2\int \frac{d^4p}{(2\pi)^4}\; \gamma_\mu G(p)
\gamma_\nu G(p+q).
\label{eq:pimunustand}
\end{equation}

\subsubsection{The case of graphene. $\boldsymbol{\gamma\, e^+ e^-}$
vertices   confined along $\boldsymbol z$ :
$\boldsymbol{\Pi^{\mu\nu}=\frac{1}{\pi^2}\; T^{\mu\nu}\times U}$}
\label{subsub:graph}

The coordinates $u_3$ and $v_3$ of the two vertices we do not integrate anymore
$\int_{-\infty}^{+\infty}$ but only $\int_{-a}^{+a}$ in which $2a$ is the
thickness of the graphene strip. This restriction {\em localizes the
interactions of electrons  with photons inside graphene}.

So doing, the results that we  get are only  valid inside graphene,
and we  therefore only focus on the ``optical properties'' of graphene.
Photons also interact with electrons outside graphene but this is
not of concern to us
\footnote{At least at 1-loop. At 2-loops and more,
virtual electrons propagating outside the medium due to their large
momentum fluctuations can interact, there, with virtual photons.}
 since we are studying how the propagation of photons is
influenced by their interactions with electrons inside the medium.

Decomposing in (\ref{eq:start})
 $du = d^3\hat u\, du_3,\; dv=d^3\hat v\, dv_3$, we get by
standard manipulations (see Appendix \ref{section:genform})
\begin{equation}
\begin{split}
\Delta^{\rho\sigma}(x,y)
&=
\int \frac{dp_3}{2\pi} \int \frac{dq_3}{2\pi}\int \frac{dr_3}{2\pi}
\int \frac{ds_3}{2\pi}\int_{-a}^{+a}du_3\; e^{iu_3(q_3+p_3-r_3)}
\int_{-a}^{+a}dv_3\; e^{iv_3(-p_3+r_3-s_3)}
\cr
& \int \frac{d^3\hat q}{(2\pi)^3}\;
e^{i\hat q(\hat y -\hat x)}
e^{iq_3(-x_3)} e^{is_3(y_3)}\Delta^{\rho\mu}(\hat q,q_3)
\Delta^{\sigma\nu}(\hat q, s_3)\
\underbrace{e^2 \int \frac{d^3\hat p}{(2\pi)^3}\;\gamma_\mu G(\hat p,B)
 \gamma_\nu G(\hat p+\hat q,B)}_{iT_{\mu\nu}(\hat q,B)},
\end{split}
\label{eq:genform}
\end{equation}
in which we introduced the tensor $T_{\mu\nu}(\hat
q,B)$ that is  calculated in section \ref{section:T}.

One of the main difference with standard QFT (subsection \ref{subsub:sQFT})
is that the tensor $T_{\mu\nu}$ that arises instead of the $\Pi_{\mu\nu}$
(\ref{eq:pimunustand})  does
not depend on $q_3$, but only on $\hat q$. The reason is that, as already
mentioned, the propagators of electrons in the loop are evaluated at
vanishing momentum in the direction of $B$. The calculation of $T_{\mu\nu}$
is performed in section \ref{section:T}. There, the explicit form of the
electron propagator in external $B$ will also be given.
Let us just notice here that, by its definition (see (\ref{eq:genform}) and
(\ref{eq:vapog})), the components $\mu,\nu=0,1,2$ of $T_{\mu\nu}$ are those
of a 2+1 dimensional vacuum polarization (in which the integration runs
over the variables $(p_0,p_1,p_2)$). However, the Lorentz indices
$\mu,\nu$ extend to 3 and, furthermore, it is precisely $T_{33}$ that will
play the leading role to determine the refractive index. The corresponding
physics cannot manifestly be reduced to 2+1 dimensions.

Notice that, despite the ``classical'' input $p_3=0$ for electrons created
inside graphene on the Dirac cone (see subsection \ref{subsec:eprop}),
 the photon propagator still involves the integration $\int dp_3$.

Now,
\begin{equation}
\int_{-a}^{+a} dx\; e^{itx} = 2\frac{\sin at}{t},
\end{equation}
such that
\begin{equation}
\begin{split}
& \Delta^{\rho\sigma}(x,y)= 4\int \frac{dq_3}{2\pi} \int \frac{ds_3}{2\pi}
e^{i(s_3y_3-q_3x_3)} L(a,s_3,q_3)
\int \frac{d^3\hat q}{(2\pi)^3}\; e^{i\hat q(\hat y -\hat x)}
\Delta^{\rho\mu}(\hat q,q_3) \Delta^{\sigma\nu}(\hat q, s_3)\;
iT_{\mu\nu}(\hat q,B),\cr
&\hskip 3cm \text{with}\quad L(a,s_3,q_3)=\int_{-\infty}^{+\infty} \frac{dp_3}{2\pi}  \frac{dr_3}{2\pi}
\;\frac{\sin a(q_3+p_3-r_3)}{q_3+p_3-r_3}
\; \frac{\sin a(r_3-p_3-s_3)}{r_3-p_3-s_3}.
\end{split}
\end{equation}

Going from the variables $r_3,p_3$ to the variables $p_3, h_3=r_3-p_3$
leads to
\begin{equation}
L(a,s_3,q_3)= \int_{-\infty}^{+\infty} \frac{dp_3}{2\pi}\; K(a,s_3,q_3),\quad
\text{with}\quad K(a,s_3,q_3)= 
\int_{-\infty}^{+\infty} \frac{dh_3}{2\pi}\;\frac{\sin
a(q_3-h_3)}{q_3-h_3}\; \frac{\sin a(h_3-s_3)}{h_3-s_3},
\end{equation}
and the photon propagator at 1-loop writes
\begin{equation}
\begin{split}
\hskip -1cm
\Delta^{\rho\sigma}(a,x,y) &= 4
\int_{-\infty}^{+\infty} \frac{d^3\hat q}{(2\pi)^3}\,e^{i\hat q(\hat y-\hat
x)}
\int_{-\infty}^{+\infty} \frac{ds_3}{2\pi}
\int_{-\infty}^{+\infty} \frac{dq_3}{2\pi} \;
e^{i(s_3y_3-q_3x_3)}\,
\Delta^{\rho\mu}(\hat q, q_3)\;K(a,s_3,q_3)\;
\Delta^{\nu\sigma}(\hat q, s_3) \;\mu\,
T_{\mu\nu}(\hat q,B),\cr
& \hskip 2cm \text{with}\quad \mu = \int_{-\infty}^{+\infty}
\frac{dp_3}{2\pi},\quad\text{which factors out}.
\end{split}
\label{eq:prop1}
\end{equation}
Last, going to the variable $k_3=s_3-q_3$
(difference of the momentum along $z$ of the incoming and outgoing photon),
one gets
\begin{equation}
K(a,s_3,q_3)\equiv \tilde K(a,k_3)
 = \frac12 \frac{\sin a(s_3-q_3)}{s_3-q_3}
= \frac12 \frac{\sin ak_3}{k_3}.
\label{eq:Kexp}
\end{equation}
To define the  vacuum polarization $\Pi_{\mu\nu}^{eff}$ from
(\ref{eq:prop1}) and (\ref{eq:Kexp}) we proceed like with (\ref{eq:stand}) in
standard QFT by truncating  two external photon propagators
$\Delta^{\rho\mu}(q)\equiv\Delta^{\rho\mu}(\hat q,q_3)$ and
$\Delta^{\nu\sigma}(q)\equiv\Delta^{\nu\sigma}(\hat q,q_3)$ off
$\Delta^{\rho\sigma}$. The mismatch
between $\Delta^{\nu\sigma}(\hat q,q_3)$ and $\Delta^{\nu\sigma}(\hat
q,s_3\equiv q_3+k_3)$ which occurs in (\ref{eq:prop1}) has to be accounted
for by writing symbolically (see subsection \ref{subsub:Feyn} for the explicit
interpretation)
$ \Delta^{\nu\sigma}(\hat q,q_3+k_3)= \Delta^{\nu\sigma}(\hat q,
q_3)[\Delta^{\nu\sigma}(\hat q, q_3)]^{-1}
\Delta^{\nu\sigma}(\hat q,q_3+k_3)$.
We therefore rewrite the photon propagator (\ref{eq:prop1}) as
\begin{equation}
\begin{split}
\Delta^{\rho\sigma}(a,x,y) &=4\mu
\int_{-\infty}^{+\infty} \frac{d^4 q}{(2\pi)^4}\,e^{i q(y- x)}
\Delta^{\rho\mu}(q)\; \Delta^{\nu\sigma}(q)\cr
& \left[\int_{-\infty}^{+\infty} \frac{dk_3}{2\pi}  \;
e^{ik_3 y_3}\, \tilde K(a,k_3)\;
[\Delta^{\nu\sigma}(\hat q, q_3)]^{-1}\Delta^{\nu\sigma}(\hat q, q_3+k_3)
\right] \; T_{\mu\nu}(\hat q,B)
\label{eq:prop2}
\end{split}
\end{equation}
Cutting off $\Delta^{\nu\sigma}(\hat q, q_3)$
leads then to the vacuum polarization $\Pi_{\mu\nu}$ 
\begin{equation}
\Pi_{\mu\nu}(\hat q, q_3, \frac{y_3}{a}, B)=
4\mu\int_{-\infty}^{+\infty} \frac{dk_3}{2\pi}  \;
e^{ik_3 y_3}\, \tilde K(a,k_3)\;
[\Delta^{\nu\sigma}(\hat q, q_3)]^{-1}\Delta^{\nu\sigma}(\hat q, q_3+k_3)
 \; T_{\mu\nu}(\hat q,B).
\label{eq:Pieff}
\end{equation}

The factor $\mu$, defined in (\ref{eq:prop1}),
 associated with the electron loop-momentum along $z$,
is potentially ultraviolet  divergent and needs to be regularized. 
In relation with the ``confinement'' along $z$ of the $\gamma\, e^+ e^-$
vertices, we shall consider that the electron momentum $p_3$ undergoes
quantum fluctuations
\begin{equation}
p_3\in[-\frac{\hbar}{a},+\frac{\hbar}{a}],
\end{equation}
with limits that saturate the Heisenberg uncertainty relation
\footnote{Since many photons and electrons are concerned, the system is
presumably gaussian, in which case one indeed expects the uncertainty
relation to be saturated.}
.
This amounts to taking
\begin{equation}
 p_3^m=\frac{\hbar}{a}
\label{eq:p3mdef}
\end{equation}
as an ultraviolet cutoff for the quantum electron momentum along $z$. Then
\begin{equation}
\mu \approx \frac{1}{2\pi}\;\frac{2\hbar}{a} = \frac{\hbar}{a\pi}.
\label{eq:rhoval}
\end{equation}
One gets accordingly, using also the explicit expression (\ref{eq:Kexp}) for
$\tilde K(a,k_3)$
\begin{equation}
\begin{split}
\Pi^{\mu\nu}(\hat q, q_3, \frac{y_3}{a}, B)
 &= \frac{1}{\pi^2}\;T^{\mu\nu}(\hat q,B)
\times  U(\hat q,q_3,\frac{y_3}{a}),\cr
\text{with}\quad U(\hat q, q_3, \frac{y_3}{a}) &= \int_{-\infty}^{+\infty} dk_3  \;
e^{ik_3 y_3}\, \frac{\sin ak_3}{ak_3}\;
[\Delta^{\nu\sigma}(\hat q, q_3)]^{-1}\Delta^{\nu\sigma}(\hat q,
q_3+k_3),
\end{split}
\label{eq:Pieff2}
\end{equation}
in which we have used the property  that
 $T_{\mu\nu}(\hat q,B)$ can be taken out of the integral because it does
not depend on $k_3$.
This demonstrates the result that has been announced and exhibits the
transmittance function $U(\hat q,q_3,\frac{y_3}{a})$
 which is independent of $B$ and of $\alpha$.

At the limit $a\to \infty$,  the position for  creation
and annihilation of electrons suffers an infinite uncertainty but its
momentum can be  defined with infinite precision: no quantum
fluctuation occurs for the momentum of electrons in the direction of $B$. 
Despite the apparent vanishing of $\mu$ at this limit,
our calculation remains meaningful. Indeed,
 the function $\frac{\sin ak_3}{ak_3}$ goes then to 
$\delta(k_3)$, which corresponds to the conservation of the photon momentum
along $z$ (the non-conservation of the photon momentum is thus seen to be
directly related to the quantum fluctuations of the electron momentum).
 This limit also corresponds to ``standard'' QFT, in which  $\hat
K(x)=\delta(x) \Rightarrow L(a,s_3,q_3)= \int_{-\infty}^{+\infty}
\frac{dp_3}{2\pi}\,
\frac{dr_3}{2\pi}\; \delta(q_3+p_3-r_3)\delta(r_3-p_3-s_3)
=\int \frac{dp_3}{2\pi}\; \delta(q_3-s_3)$.
Notice that, because our results are obtained for small values of the
parameter $\eta = aq_0$, their limit when $a\to \infty$ cannot be
obtained.

For $a<\infty$, momentum conservation along $z$  is only approximate:
then, the photon can exchange
momentum along $z$ with the quantum fluctuations of the electron momentum.
In general, the $\frac{\sin a k_3}{ak_3}$ occurring in $U$
provides for photons, by its fast decrease, the same cutoff $|k_3|
 \equiv |s_3-q_3| \leq \frac{\hbar}{a}=p_3^m$ along $z$ as for electrons.
As we shall see in subsection \ref{subsec:limtetanull},
this also provides an upper bound $n_{quant} \sim \frac{p_3^m}{q_0}$
for the refractive index, which can only be satisfied for $B \leq B^m \sim
11400\,T$.

The limit $a\to 0$ would correspond to infinitely thin graphene, infinitely
accurate positioning  of the creation and annihilation of
electrons, but to unbounded quantum fluctuations of their  momentum along
$B$.
Since $\frac{\sin x}{x} \to 1$ when $x\to 0$, no divergence can occur as
$a \to 0$, despite the apparent divergence of $p_3^m$ and $\mu$ (see also
subsections \ref{subsub:dimless} and \ref{subsub:limanul}).

By the choice (\ref{eq:p3mdef}), our model gets therefore suitably physically
 regularized both in the infrared and in the ultraviolet.

Notice that
the 1-loop photon propagator (\ref{eq:prop1}) still depends on the
difference $\hat y -\hat x$ but no longer depends on $y_3-x_3$ only,
it is now a function of both $y_3$ and $x_3$ (as already mentioned at the
end of subsection \ref{subsec:lightcone}, this ``extra'' dependence is in practice
very weak).

\subsection{The transmittance function $\boldsymbol{U(\hat q,q_3,\frac{y_3}{a})}$}

\subsubsection{The Feynman gauge}\label{subsub:Feyn}

We have seen that, when calculating the vacuum polarization
(\ref{eq:Pieff}), the mismatch between  $\Delta^{\nu\sigma}(\hat q,q_3)$,
chopped off to get $\Pi^{\mu\nu}$),
and $\Delta^{\nu\sigma}(\hat q,q_3+k_3)$ which effectively
 occurs in (\ref{eq:prop1}), has to be accounted for.
This is most easily done in the Feynman gauge for photons, in which their
propagators write
\begin{equation}
\Delta^{\mu\nu}(q) = -i\;\frac{g^{\mu\nu}}{q^2}.
\end{equation}
Thanks to  the absence of ``$q^\mu q^\nu/q^2$'' terms and as can be easily
checked for each component of $\Delta^{\rho\sigma}$,
 $[\Delta^{\nu\sigma}(\hat q, q_3)]^{-1}\Delta^{\nu\sigma}(\hat q,
q_3+k_3)$ can be simply written, then
$\frac{q_0^2-q_1^2-q_2^2 -q_3^2}{q_0^2-q_1^2-q_2^2 -(q_3+k_3)^2}$.
Accordingly, the expression for $U$ resulting from (\ref{eq:Pieff2})
 that we shall use from now onwards is
\begin{equation}
U(\hat q,q_3,\frac{y_3}{a})=\int_{-\infty}^{+\infty} dk_3  \;
e^{ik_3 y_3}\, \frac{\sin ak_3}{ak_3}\;
\frac{q_0^2-q_1^2-q_2^2 -q_3^2}{q_0^2-q_1^2-q_2^2 -(q_3+k_3)^2}.
\label{eq:Pieff4b}
\end{equation}
The analytical properties and pole
structure  of the integrand in the complex $k_3$ 
play, like for the transmittance in optics (or electronics), an essential
role.  Because they share many similarities, we have
given the same name to $U$.

\subsubsection{Going to dimensionless variables : $\boldsymbol{U(\hat
q,q_3,\frac{y_3}{a}) \to V(n,\theta,\eta,u)}$}\label{subsub:dimless}

Let us go to dimensionless variables. We define ($p_3^m$ is given in
(\ref{eq:p3mdef}))
\begin{equation}
\eta=\frac{q_0}{cp_3^m}=\frac{aq_0}{(\hbar c)},\quad
\zeta= \frac{\sqrt{2\hbar eB}}{p_3^m}=a\sqrt{\frac{2eB}{\hbar}},\quad
\Upsilon= \frac{\zeta}{\eta}=c\frac{\sqrt{2\hbar eB}}{q_0} \gg 1,\quad
u=\frac{y_3}{a}.
\end{equation}
It is also natural, in $U$, to go to the integration variable
$\sigma=\frac{k_3}{p_3^m}$, and to make appear the refractive index $n$
defined in (\ref{eq:ndef}) and the angle of incidence $\theta$ according to
\begin{equation}
q_2=0,\quad
q_1= |\vec q| s_\theta = n q_0 s_\theta,\quad
q_3= |\vec q| c_\theta = n q_0 c_\theta,\quad \theta \in ]0,\frac{\pi}{2}[,
\end{equation}
which, going to the integration variable $\sigma = ak_3=\frac{k_3}{p_3^m}$,
 leads to
\begin{equation}
U(\hat q,q_3,\frac{y_3}{a})=\frac{1-n^2}{a} V(n,\theta,\eta,u),\quad
V(n,\theta,\eta,u)  =\int_{-\infty}^{+\infty}
d\sigma\; e^{i\sigma u}\;
\frac{\sin\sigma}{\sigma}\frac{1}{1-n^2-\frac{\sigma}{\eta}
(2n\cos\theta+\frac{\sigma}{\eta})},
\label{eq:UVdef}
\end{equation}
and, therefore, to
\begin{equation}
\Pi^{\mu\nu}(\hat q,q_3,\frac{y_3}{a},B) = \frac{1}{\pi^2}\;T^{\mu\nu}(\hat q,B)
\frac{1-n^2}{a}\times  V(n,\theta,\eta,u).
\label{eq:PiV}
\end{equation}
We shall also call $V$ the {\em transmittance function}.

As  already deduced in subsection \ref{subsub:graph} from the smooth
behavior of the cardinal sine in the expression (\ref{eq:Pieff2}) of $U$,
the apparent divergence of (\ref{eq:PiV}) at $a\to 0$ is fake;
this can be checked by  expanding  $V$ at small $\eta \equiv
aq_0$, see (\ref{eq:Vexp}), (\ref{eq:Vexpcomp1}), (\ref{eq:Vcomp2}). The
expansions always start at ${\cal O}(\eta\equiv aq_0)^{\geq 1}$,
 which cancels the $\frac{1}{a}$ in (\ref{eq:PiV}).

\section{The  tensor
 $\boldsymbol{T^{\mu\nu}(\hat q,B)}$ at 1-loop
 in the presence of an external $\boldsymbol B$}
\label{section:T}

The tensor $T^{\mu\nu}$ that we compute in this section is
the one that arose in (\ref{eq:genform}) when calculating 
 the 1-loop photon propagator; it only depends on $(\hat q,B)$ (and
$\alpha$) and  writes
\begin{equation}
iT^{\mu\nu}(\hat q,B)= +e^2\int_{-\infty}^{+\infty} \frac{d^3\hat
p}{(2\pi)^3}\; Tr\; \left[\gamma^\mu G(\hat p,B)
\gamma^\nu G(\hat p+\hat q,B)\right],
\label{eq:vapog}
\end{equation}
in which $G(\hat p,B)$ is the  propagator of a massless Dirac electron at
$p_3=0$ (see section \ref{section:intro}) obtained
in the formalism of Schwinger \cite{Schwinger}\cite{Tsai1974-2}
 to account for the external magnetic field $B$.
$T^{\mu\nu}$
has dimension $[p]$, the appropriate dimension $[p]^2$ to fit
in the light-cone equations (\ref{eq:lc0}) being restored by the
transmittance $U$ which has also dimension $[p]$ (see eq.~(\ref{eq:Pieff2})).

\subsection{The electron propagator $\boldsymbol{G(\hat p, B)}$
in an external magnetic field}
\label{subsec:eprop}

\subsubsection{General expression. Why $\boldsymbol c$ and not
$\boldsymbol{v_F}$}\label{subsub:geneprop}

As mentioned in section \ref{section:intro}, we comment more here on
 the reasons why we choose the electron propagators inside the loop as Dirac-like
massless fermions with no reference to the Fermi velocity $v_F$ inside
graphene.

The first reason is that graphene-born (and annihilated)
electrons/positrons spend in practice much more time outside graphene than
inside.  Their average
life-time is $\tau_e \simeq \frac{\hbar}{\Delta{\cal E}}$ in which ${\Delta\cal E}$
is the average energy required to create a virtual particle,
 that we can consistently take to be $\simeq
\frac{q_0}{2}$, $q_0$ being the energy of the incoming photon.

On the other side, a characteristic time $t_g$ that they spend inside
graphene is the $z$ extension $\sim a$ divided by a velocity
$\frac{p_3}{m}$, that is $t_g \simeq \frac{a m}{p_3}$.
This argument is only valid when the Coulomb energy of the electron can be
neglected with respect to its kinetic energy.
This is expected at the limit  where the longitudinal spread $L$
of the graphene strip is ``infinite''. When the charge $+1$ is supposed to
be uniformly spread in the rest of the medium,  the average Coulomb
energy  of a graphene electron is then, indeed, expected to go like $1/L$ (see
Appendix \ref{section:coulomb}).
It is hereafter in such an ``idealized'' infinite
graphene strip that we shall propagate light.\newline
   $m$ is an effective mass for the electron and  we can take $p_3
\sim\frac{\hbar}{a}$, the quantum fluctuation linked to the confinement of
vertices (which is much larger that the photon momentum $|\vec
q|\sim\frac{q_0}{cn}$). If we assimilate $m$ with the effective cyclotron mass
\footnote{The Hamiltonian of graphene in a strong external $B$ exhibits
(see for example \cite{Goerbig}) a natural frequency
$\omega'= \sqrt{2}\frac{v_F}{\ell_c}$, in
which $v_F$ is the Fermi velocity and $\ell_c=\sqrt{\frac{\hbar}{eB}}$ is
 the cyclotron radius. This gives $\omega'=v_F \sqrt{2eB/\hbar}$.
 If one defines by analogy the cyclotron
mass by $\omega'=\frac{eB}{m_c}$, one gets $m_c=\frac{\sqrt{\hbar
eB}}{v_F\sqrt{2}}\approx .003\,\sqrt{B(T)}\,m_e$.
\label{foot:tg}}
 $m_c = \frac{\sqrt{\hbar eB}}{\sqrt{2}v_F}$, one gets $t_g \simeq
\frac{a^2\sqrt{eB}}{v_F\sqrt{2\hbar}}$. At $B =20\,T$,  $m_c \simeq
.014\,m_e$ and
  $\tau_e \simeq 6\,10^{-16}\,s \gg t_g \simeq 3.7\,10^{-18}\,s$.
As we shall see in
subsection \ref{subsec:reson}, the effective mass of the electron in this
process could even be much smaller.

The second argument concerns energy-momentum conservation at the $\gamma\,
e^+e^-$ vertices. A (massless) photon ($q_0 = c|\vec q|$)
 can never decay into two on-shell massless electrons with $p_0=v_F|\vec p|$ and
$r_0=v_F|\vec r|$ 
\footnote{Let $r=p+q$, in which $p$ and $r$ are associated with the
electron line and $q$ with the incoming photon. The photon being
on mass-shell, $q^2=0$, therefore energy-momentum conservation at the vertex
yields $(r_0-p_0)^2 - c^2(\vec r - \vec p)^2=0$.
On mass-shell ``graphene'' electrons  corresponding to $r_0=v_F |\vec r|$
and $p_0 = v_F|\vec p|$, the previous relation gives
$\frac{v_F^2}{c^2}=\frac{(\vec p - \vec r)^2}{(|\vec p| - |\vec r|)^2}$,
which cannot be fulfilled since the l.h.s is $<1$ while the r.h.s is $\geq
1$.},
but only into massless electrons with $p_0=c|\vec p|$ and $r_0=c|\vec r|$
\footnote{Then, the two electrons go in the same direction
(see for example \cite{Modanese}).}
. This argument could look dubious since, first,
 the electrons in the loop are not on-shell and, secondly, nature is full
of  particles which cannot decay into a  pair of heavy other particles.
However, in 2-body decays, increasing the energy $q_0$
 of the decaying particle enables to go
beyond the kinetic barrier due the large mass of the decay products.
This is not the case here, since the corresponding real decay can never
occur, and  it looks accordingly very hazardous to perform QFT calculations
with an  interaction Lagrangian derived from (\ref{eq:dirgraph}) by the
simple Peierls substitution $p^\mu \to p^\mu-\frac{e}{c} A^\mu$.

Following Schwinger ((\cite{Schwinger}, eqs.\, 2.7 to 2.10),
 we define the electron propagator as
\begin{equation}
G(x,y) = i\langle(\psi(x) \bar\psi(y))_+\rangle \Theta(x-y).
\end{equation}
\begin{equation}
G(x,y)=\Phi(x,y)\int \frac{d^4p}{(2\pi)^4}\; e^{ip.(x-y)} G(p),
\end{equation}
\begin{equation}
\Phi(x,y)=\exp\left[ie\int_y^x A(\xi)d\xi\right].
\label{eq:phase}
\end{equation}
In practice, the phase factors $\Phi$ (\ref{eq:phase})
disappear when we calculate the
vacuum polarization because the two of then combine into a closed path
integral which therefore vanishes. So, in what follows, we shall simply
 forget about $\Phi$. We shall also go to the notation $G(p,B)$ to recall
that we are working in the presence of an external $B$.

According to the remarks starting this subsection, and preserving, as stated in
section \ref{section:intro}, the properties that electrons, being
created inside graphene  correspond classically to massless
excitations with vanishing momentum $p_3$ along $z$
\footnote{When $p_3\not=0, m\not=0$, $-p_0^2$ should be replaced by $-p_0^2
+p_3^2+m^2$ in (\ref{eq:eprop1}),
and $\gamma^0p^0$ by $\gamma^0p^0-\gamma_3p_3+m$.\label{foot:eprops}}
, we shall  take their
propagator as
 \cite{Schwinger}\cite{Tsai1974-2}
\footnote{The expression (\ref{eq:eprop1}) is obtained after going from the
real proper-time $s$ of Schwinger to $\tau=is$ and switching to conventions
for the Dirac matrices and for the metric of space $(+,-,-,-)$
which are more usual today \cite{PeskinSchroeder}.}
\begin{equation}
\hskip -1cm  G(\hat p,B)=
\int_0^{\infty} d\tau\; \exp\left[
-\tau\left((-p_0^2)+ \frac{\tanh (e\tau B)}{e\tau B}(p_1^2+p_2^2)\right)
\right]
\left((\gamma^0 p^0)\big(1-i\gamma^1\gamma^2 \tanh(e\tau B)\big)-\frac{\gamma_1
p_1 + \gamma_2 p_2}{\cosh^2(e\tau B)}\right),
\label{eq:eprop1}
\end{equation}
which only depends on $\hat p$ and $B$.

\subsubsection{Expanding at ``large'' $\boldsymbol{B<\infty}$}
\label{subsub:bigB}

$\bullet$\ At the limit $B\to \infty$
\footnote{One considers then that $e\tau B$ also $\to \infty$, in which case,
in
(\ref{eq:eprop1})
$\tanh e\tau B \to 1, \cosh e\tau B \to \infty$.
This is only acceptable at $\tau \not=0$, but Schwinger's prescription is that
the integration over the proper time has to be made last.}
, (\ref{eq:eprop1}) becomes
\begin{equation}
G(\hat p,B)\stackrel{B=\infty}{\to} -e^{-\frac{p_\perp^2}{eB}}\;
\frac{\gamma^0 p^0}{p_0^2}(1- i\gamma^1\gamma^2),\quad
p_\perp^2 = p_1^2+p_2^2.
\label{eq:Ginf}
\end{equation}
The projector $(1-i\gamma^1\gamma^2)$  ensures that electrons in the lowest
Landau level only couple to the longitudinal $(0,3)$ components
of the photon \cite{Miransky}.

$\bullet$\ We shall in this work go one step further in the expansion of $G$ at large
$B$: we keep the first subleading terms in the expansions of $\tanh (\tau eB)$ and
$\cosh (\tau eB)$ of (\ref{eq:eprop1})
(this approximation does not allow  to take the limit $B\to
0$ since, for example, it yields $\tanh (\tau eB)
\to -1$ instead of $0$ and $\cosh^2(\tau eB) \to 3/4$ instead of $1$)
:
\begin{equation}
\tanh (\tau eB) \approx 1-2 e^{-2\tau eB},\quad \cosh^2(\tau eB) \approx
\frac{e^{2\tau eB}+2}{4} \Rightarrow \frac{1}{\cosh^2(\tau eB)} \approx
\frac{4\,e^{-2\tau eB}}{1+2\,e^{-2\tau eB}}.
\end{equation}

This gives (we note $ (\gamma p)_\perp=\gamma_1p_1+\gamma_2p_2$), still for
graphene,
\begin{equation}
\begin{split}
G(\hat p,B)\approx &\int_0^\infty d\tau \;
e^{-\tau (-p_0^2)}e^{-\frac{p_\perp^2}{eB}(1-2e^{-2\tau e B})}
(\gamma^0p^0)(1-i\gamma^1\gamma^2(1-2e^{-2\tau e B}))\cr
   &- 4 (\gamma p)_\perp \int_0^\infty d\tau \;
e^{-2\tau e B}\frac{1}{1+2e^{-2\tau e B}}e^{-\tau(-p_0^2)}e^{-\frac{p_\perp^2}{eB}
(1+2e^{-2\tau e B})}.
\end{split}
\end{equation}
We shall further approximate
 $e^{-\frac{p_\perp^2}{eB}(1-2e^{-2e\tau B})} \approx
e^{-\frac{p_\perp^2}{eB}}$, which can be seen to be legitimate  because the
exact integration yields subleading corrections $\propto 1/(eB)^2$,
while the ones that we keep are $\propto 1/eB$. This gives
\begin{equation}
\begin{split}
\hskip -5mm G(\hat p,B)\approx &
e^{-\frac{p_\perp^2}{eB}}\left(-\frac{\gamma^0p^0}{p_0^2}
(1-i\gamma^1\gamma^2)+2
\frac{\gamma^0p^0}{p_0^2-2eB}(-i\gamma^1\gamma^2)\right) 
-4 (\gamma p)_\perp e^{-\frac{p_\perp^2}{eB}}\int_0^\infty d\tau \;
\frac{1}{1+2e^{-2e\tau B}}e^{-\tau (-p_0^2+2eB)}.
\end{split}
\label{eq:gfinexact}
\end{equation}

One has
\begin{equation}
\int_0^\infty d\tau \;
\frac{1}{1+2e^{-2e\tau B}}e^{-\tau (-p_0^2+2eB)}=
(-2)^{-1+\frac{p_0^2}{2eB}}
\frac{\beta(-2,1-\frac{p_0^2}{2eB},0)}{2eB},
\label{eq:genint}
\end{equation}
such that  (\ref{eq:gfinexact}) rewrites
\begin{equation}
\hskip -.5cm
G(\hat p,B)
 = -e^{-\frac{p_\perp^2}{eB}} \left(
\frac{\gamma^0}{p^0}\left(1+i\gamma_1\gamma_2\;
\frac{p_0^2+2eB}{p_0^2-2eB}\right)
+4\frac{p_1\gamma_1+p_2\gamma_2}{2eB} F\Big(\frac{p_0^2}{2eB}\Big)
\right),\quad F(x)=(-2)^{(-1+x)}\beta(-2,1-x,0),
\label{eq:gfing2}
\end{equation}
in which $\beta$ is the incomplete beta function.

When $B < \infty$, corrections arise with respect to (\ref{eq:Ginf}),
 which exhibit in particular poles at
$p_0^2=2eB$ (first and 2nd term) and also $p_0^2 = 2n\,eB, n=1,2 \ldots$
 (second term)
\footnote{If we  work with massive electrons, one finds that their
mass squared $m_e^2$ gets replaced by $m_e^2 + 2n\,eB$ in the presence of
$B$. Massless electrons get accordingly replaced with excitations with mass
squared $2neB(\hbar/c^2)$.\label{foot:dynmass}}
.
They  are furthermore no longer proportional to the   projector
$(1-i\gamma^1\gamma^2)$. However, we shall see that the dependence of the
refractive index on $B$ and $\alpha$ stays mostly controlled by $\Pi^{33}$.

\subsubsection{Working approximation; low energy electrons}\label{subsub:Fappro}

\begin{figure}[h]
\begin{center}
\includegraphics[width=6 cm, height=4 cm]{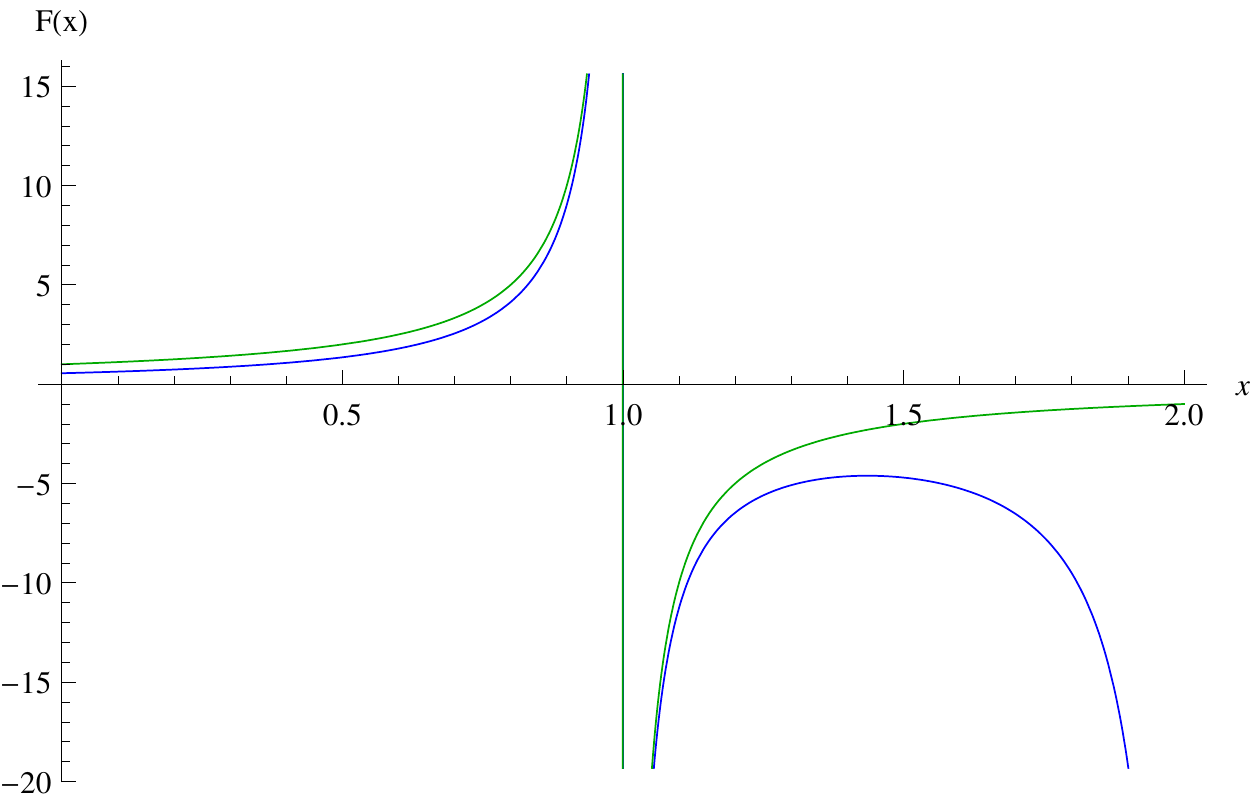}
\caption{The function $F(x)\;(blue)$ and its approximation
$\frac{1}{1-x}\;(green)$.}
\label{fig:F}
\end{center}
\end{figure}

The expression (\ref{eq:gfing2}) is still not very simple to use. This is
why we shall  further approximate $F(x)$ and take
\begin{equation}
F(x) \approx \frac{1}{1-x},
\label{eq:workapp}
\end{equation}
which amounts to only select, in there, the pole at $n=1$, $p_0^2 = 2eB$, and
neglect the other poles.
As can be seen on Figure \ref{fig:F}, the approximation
 (\ref{eq:workapp}) is 
 reasonable in the vicinity of this pole (as can be seen by  plotting)
for $0 \leq x \leq 1.5$, that is, setting back $\hbar$ and $c$,
$0 \leq p_0^2 \leq 1.5\times c^2(2\hbar eB)$.
This corresponds to electrons with energies  $\leq
c\sqrt{1.5}\sqrt{2\hbar eB}$.
 Since the spectrum of relativistic Landau levels in graphene is
$\epsilon_n = \pm v_F\sqrt{2n \hbar eB}$ \cite{Goerbig},
our approximation stays valid up to energies $\sim \sqrt{1.5} \frac{c}{v_F}
\epsilon_1 \sim 350\, \epsilon_1$, therefore in
a domain that largely exceeds the energy $\epsilon_1$ of the lowest Landau level
\footnote{At $B=20\,T$, the spacing of Landau levels in graphene is
$v_F \sqrt{2\hbar eB} \approx .16\,eV$. This energy scale goes up to
 $48\,eV$ when $v_F$ is replaced with $c$.\label{foot:landlev}}
.

In practice, this corresponds to electrons with energy $p_0 \leq 13
\sqrt{B(T)}\,eV$. This condition is always satisfied for optical
wavelengths; indeed the energy of photons range then between $1.5\,eV$ and
$3.5\,eV$, which is roughly twice the energy of the created virtual
electrons or positrons.

Notice that, at $x\equiv\frac{p_0^2}{2eB}=0$, which corresponds to $p_0=0$
(electrons with vanishing energy) or to $B\to \infty$,
$F(0)=\frac{\ln{3}}{2} \approx .55$ while our approximation goes to $1$.
A corresponding scaling down of $\alpha$ can  eventually be operated.

We shall therefore take in the following calculations
\footnote{see footnote \ref{foot:eprops}.}
\begin{equation}
\begin{split}
G(\hat p,B) &\approx
-e^{-\frac{p_1^2+p_2^2}{eB}} \left[
\frac{\gamma^0}{p^0}\left(1+i\gamma_1\gamma_2\;
\frac{p_0^2+2eB}{p_0^2-2eB}\right)
-4\frac{p_1\gamma_1+p_2\gamma_2}{p_0^2-2eB}
\right],\cr
&= -e^{-\frac{p_1^2+p_2^2}{eB}}\left[
\frac{p^0 \gamma^0}{p_0^2}(1-i\gamma_1\gamma_2)
+2i\frac{p^0\gamma^0}{p_0^2-2eB}\gamma_1\gamma_2 -4\frac{p_1\gamma_1+p_2\gamma_2}
{p_0^2-2eB}
\right],
\end{split}
\label{eq:ggap}
\end{equation}
which leads to  expressions  easy to handle, and enables to go a long way
analytically. 
In particular, setting the momentum along the direction of
 $B$ equal to $0$ for both electron propagators inside the loop
 makes their denominators only depend on $p_0$.
The integration of the transverse degrees of freedom $p_1,p_2$
 being elementary, the vacuum polarization can finally
be expressed only in terms of 1-dimensional convergent integrals  $\int dp_0$
(see subsection \ref{subsub:integrate}).
In the last line of (\ref{eq:ggap}) we have made the distinction between
three contributions: the one on the left corresponds to the only term which
is usually kept at $B=\infty$ (when $m\not=0$, $p_3\not=0$),
the middle one and the
one on the right are dropped at this same limit. However, in the following,
the right contribution will be seen to yield the leading components
of the vacuum
polarization tensor, due to the powers of $eB$ that arise when integrating
over the transverse degrees of freedom $p_1,p_2$ occurring in its numerator.

\subsection{Calculations and results}

There are two steps in the calculation: first  performing the traces
of the Dirac $\gamma$ matrices, then integrating over
the loop variables $\hat p=(p_0,p_1,p_2)$.

\subsubsection{Performing the traces of Dirac matrices}
\label{subsub:technique}

This  already yields
\begin{equation}
\Pi^{i3}=0=\Pi^{3i},\quad i=0,1,2.
\label{eq:Pii3}
\end{equation}

\subsubsection{Doing the integrations} \label{subsub:integrate}

Details of the calculation will be given somewhere else.
We just want here to present its main steps, taking the examples of
$\Pi^{00}$ and $\Pi^{33}$, which play the leading roles in the calculations
concerning the refractive index. After doing the traces, one gets
\begin{equation}
\begin{split}
iT^{00}(\hat q,B) &=
4\,e^2\int_{-\infty}^{+\infty} \frac{dp_0 dp_1 dp_2}{(2\pi)^3}
e^{-p_\perp^2/eB}e^{-(p+q)_\perp^2/eB}\cr
& \left(\frac{1}{p_0}\frac{1}{p_0+q_0}
+\frac{1}{p_0}\frac{p_0^2+2eB}{p_0^2-2eB}
\frac{1}{p_0+q_0}\frac{(p_0+q_0)^2 +2eB}{(p_0+q_0)^2-2eB}
+16 \frac{p_1(p_1+q_1)+p_2(p_2+q_2)}{(p_0^2-2eB)((p_0+q_0)^2-2eB)}
\right),
\end{split}
\label{eq:pi00}
\end{equation}
which decomposes into
\begin{equation}
\begin{split}
iT^{00}(\hat q,B) &= I(\hat q,B) + J(\hat q,B) + K(\hat q,B), \cr
I(\hat q,B) &= 4\,e^2\int_{-\infty}^{+\infty} \frac{dp_0 dp_1 dp_2}{(2\pi)^3}
e^{-p_\perp^2/eB}e^{-(p+q)_\perp^2/eB}
 \frac{1}{p_0}\frac{1}{p_0+q_0}, \cr
J(\hat q,B) &= 4\,e^2\int_{-\infty}^{+\infty} \frac{dp_0 dp_1 dp_2}{(2\pi)^3}
e^{-p_\perp^2/eB}e^{-(p+q)_\perp^2/eB}
\frac{1}{p_0}\frac{p_0^2+2eB}{p_0^2-2eB}
\frac{1}{p_0+q_0}\frac{(p_0+q_0)^2 +2eB}{(p_0+q_0)^2-2eB}, \cr
K(\hat q,B) &= 4\,e^2\int_{-\infty}^{+\infty} \frac{dp_0 dp_1 dp_2}{(2\pi)^3}
e^{-p_\perp^2/eB}e^{-(p+q)_\perp^2/eB}
16 \frac{p_1(p_1+q_1)+p_2(p_2+q_2)}{(p_0^2-2eB)((p_0+q_0)^2-2eB)}.
\end{split}
\label{eq:pi00c}
\end{equation}
Likewise, one gets
\begin{equation}
iT^{33}(\hat q,B) = I(\hat q,B) + J(\hat q,B) - K(\hat q,B).
\label{eq:pi33c}
\end{equation}
It is then convenient to integrate over the transverse degrees of freedom
$p_1,p_2$. This is done by going to
the variables of integration $u_1=p_1+\frac{q_1}{2}, u_2=p_2+\frac{q_2}{2}$
and canceling all terms which are odd in $u_1$ or $u_2$.
 This yields
\begin{equation}
\begin{split}
I(\hat q, B) &= \frac{\alpha}{\pi}eB\;e^{-q_\perp^2/2eB} B(q_0),\cr
J(\hat q, B) &= \frac{\alpha}{\pi}eB\;e^{-q_\perp^2/2eB} C(q_0,B),\cr
K(\hat q, B) &= \frac{8\alpha}{\pi}eB\;e^{-q_\perp^2/2eB}
(eB-q_\perp^2) D(q_0,B),
\end{split}
\end{equation}
 in which we have  introduced the (convergent) integrals
\begin{equation}
\begin{split}
B(q_0)&= \int_{-\infty}^{+\infty} dp_0\; \frac{1}{p_0}\frac{1}{p_0+q_0},\cr
C(q_0,B) &= \int_{-\infty}^{+\infty} dp_0\;
\frac{1}{p_0}\frac{p_0^2+2eB}{p_0^2-2eB}
\frac{1}{p_0+q_0}\frac{(p_0+q_0)^2 +2eB}{(p_0+q_0)^2-2eB},\cr
D(q_0,B)&= \int_{-\infty}^{+\infty} dp_0\;
 \frac{1}{(p_0^2-2eB)((p_0+q_0)^2-2eB)}.
\end{split}
\label{eq:Bdef}
\end{equation}
Note that two powers of $eB$  occur in $K$ due to the integration over
the transverse degrees of freedom.

\begin{figure}[h]
\begin{center}
\includegraphics[width=5 cm, height=3 cm]{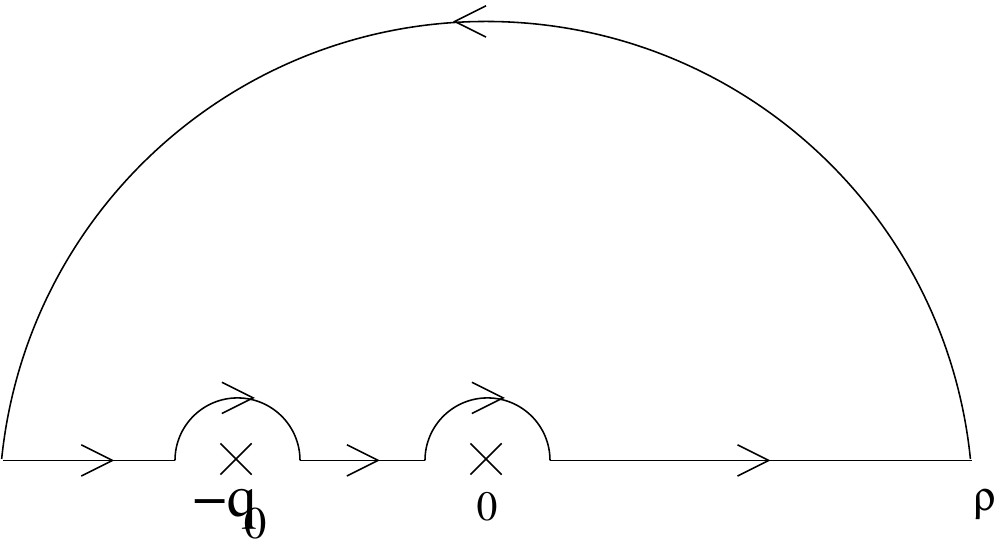}
\caption{The contour of integration for $B(q_0)$ and $C(q_0)$.}
\label{fig:contour}
\end{center}
\end{figure}

``Massless'' and ambiguous integrals of the type
 $\int_{-\infty}^{+\infty} d\sigma \;\frac{f(\sigma)}{\sigma}$ occurring in
$B(q_0), C(q_0,B), D(q_0,B)$ are
replaced, using the customary $+i\varepsilon$ prescription for the poles of
propagators in QFT dictated by causality, with
\begin{equation}
\lim _{\epsilon\to 0^+} \int_{-\infty}^{+\infty} d\sigma
\;\frac{f(\sigma)}{\sigma+i\epsilon} = -i\pi\,f(0)
+\lim_{\epsilon \to 0^+} \int_{|\sigma| >\epsilon}
\frac{f(\sigma)}{\sigma},
\end{equation}
which are just Cauchy integrals. This is nothing more than the
Sokhotski-Plemelj theorem \cite{SokhotskiPlemelj} :
\begin{equation}
\lim_{\varepsilon\to 0^+} \int_{-\infty}^\infty\frac{f(x)}{x\pm
i\varepsilon}\,dx = \mp i\pi f(0) + \lim_{\varepsilon\to 0^+}
\int_{|x|>\varepsilon}\frac{f(x)}{x}\,dx.
\end{equation}
It is easy to also check that the same result can be obtained, after
setting the $+i\varepsilon$ prescription, by integrating on the contour
described on Figure~\ref{fig:contour}. There, the two small 1/2 circles around the poles have
radii that $\to 0$. The large 1/2 circle has infinite radius.

This also amounts, for the poles ``on the real axis'',
 to evaluating $i\pi \sum residues$, that is $1/2$ of
what one would get if the poles were not on the real axis but inside the
contour of integration. The other poles that lie inside the contour of
integration are dealt with as usual by $2i\pi \times$ their residues.

So doing, one gets
\begin{equation}
\begin{split}
& B(q_0) =0 = C(q_0,B),\cr
& D(q_0,B)= 2i\pi
(-)\frac{1}{\sqrt{2eB}}\frac{1}{q_0^2-8eB},
\end{split}
\end{equation}
leading finally to 
\begin{equation}
I=0=J,\quad  K(\hat q,B)=
i \frac{2e^2}{\pi}\;e^{-\frac{q_\perp^2}{2eB}}
\sqrt{2eB}\;\frac{eB -  q_\perp^2}{q_0^2-8eB},
\label{eq:respi00}
\end{equation}
and, for $T^{00}(\hat q, B)$, to the first line of the set of equations
 (\ref{eq:summary}). From (\ref{eq:pi00c}), (\ref{eq:pi33c}) and
(\ref{eq:respi00}) one gets immediately $T^{33}(\hat q,B)=-T^{00}(\hat
q,B)$.

Notice that $T^{00}(\hat q, B)$ and $T^{33}(\hat q, B)$
 are controlled by $K(\hat q, B)$
which originates from the terms proportional to $4\frac{p_1\gamma_1 +
p_2\gamma_2}{p_0^2-2eB}$ in the electron propagator (\ref{eq:ggap}).
These terms are subleading with respect to the ones
proportional to $\frac{\gamma^0}{p^0}(1-i\gamma_1\gamma_2)$ 
and would have naively been dropped  in the 
limit $B\to\infty$.  However, in the calculation of $K(\hat q, B)$,
integrating over the transverse degrees of freedom brings two powers of
$eB$  which counter-balances the damping of $D(q_0,B)$ at large $B$
and finally makes $T^{00}$ and $T^{33}$ the  dominant components of the
vacuum polarization tensor.
Since the powers of
$p_1$ and $p_2$ stay the same, going to higher orders in the expansion at
``large $\tau eB$'' of the electron propagator would not change the result.

After all integrals have been calculated by similar techniques, one gets the
results displayed in subsection \ref{subsub:cavpol}.

\subsubsection{Explicit expression of $\boldsymbol{T^{\mu\nu}(\hat q, B)}$ at
1-loop}\label{subsub:cavpol}

\begin{equation}
\begin{split}
iT^{00}(\hat q,B) &= 4i\alpha\;\sqrt{2eB}\;e^{-\frac{q_1^2+q_2^2}{2eB}}\;
\frac{2eB -2(q_1^2+q_2^2)}{q_0^2 -4(2eB)} \stackrel{B\to\infty}{\simeq}
-i\alpha\;\sqrt{2eB}\;e^{-\frac{q_1^2+q_2^2}{2eB}},\cr
iT^{11}(\hat q,B) &= 4i\alpha\,
e^{-\frac{q_1^2+q_2^2}{2eB}}\sqrt{2eB}\,\frac{q_1^2-q_2^2}{q_0^2-4(2eB)}
\stackrel{B\to\infty}{\simeq}
i\alpha\,e^{-\frac{q_1^2+q_2^2}{2eB}}\,\frac{q_1^2-q_2^2}{\sqrt{2eB}}, \cr
iT^{22}(\hat q,B) &= -iT^{11}(\hat q,B),\quad
iT^{33}(\hat q,B) = -iT^{00}(\hat q,B)
,\cr
iT^{01}(\hat q,B) &= 2i\alpha\, e^{-\frac{q_1^2+q_2^2}{2eB}} q_1q_0\,
\frac{\sqrt{2eB}}{q_0^2-2eB} \stackrel{B\to\infty}{\simeq}
-i\alpha\,e^{-\frac{q_1^2+q_2^2}{2eB}}\,\frac{q_1q_0}{\sqrt{2eB}},\cr
iT^{02}(\hat q,B) &=   2i\alpha\, e^{-\frac{q_1^2+q_2^2}{2eB}} q_2q_0\,
\frac{\sqrt{2eB}}{q_0^2-2eB} \stackrel{B\to\infty}{\simeq}
-i\alpha\,e^{-\frac{q_1^2+q_2^2}{2eB}}\,\frac{q_2q_0}{\sqrt{2eB}},\cr
iT^{12}(\hat q,B) &= -16\alpha\, q_1q_2\, e^{-\frac{q_1^2+q_2^2}{2eB}}
\frac{\sqrt{2eB}}{q_0^2-4(2eB)} \stackrel{B\to\infty}{\simeq}
4\alpha\,e^{-\frac{q_1^2+q_2^2}{2eB}}\, \frac{q_1q_2}{\sqrt{2eB}},\cr
iT^{03}(\hat q,B)&=0,\quad iT^{13}(\hat q,B)=0,\quad iT^{23}(\hat q,B)=0.
\end{split}
\label{eq:summary}
\end{equation}

\subsubsection{Comments}\label{subsub:comments}

$\bullet$\ $T^{00}=-T^{33}$ are the only two components that
do not vanish when $B\to\infty$ (see also footnote \ref{foot:tetanull}
concerning $\theta\to 0$).

$\bullet$\  $T^{\mu\nu}$ is not transverse.
In our setup, which has in particular $q_2=0$, the four 
relations corresponding to $q_\mu T^{\mu\nu}(q)$ reduce to
\begin{equation}
q_\mu T^{\mu 0} \equiv q_0 T^{00}+ q_1 T^{10},\quad
q_\mu T^{\mu 1} \equiv q_0 T^{01}+ q_1 T^{11},\quad
q_\mu T^{\mu 2} \equiv 0,\quad
q_\mu T^{\mu 3} \equiv q_3 T^{33}.
\end{equation}
At the limit $B\to \infty$, they  shrink to
\begin{equation}
q_\mu T^{\mu 0} \equiv q_0  T^{00},\quad
q_\mu T^{\mu 1} \equiv 0,\quad
q_\mu T^{\mu 2} \equiv 0,\quad
q_\mu T^{\mu 3} \equiv q_3 T^{33}.
\label{eq:transinf}
\end{equation}
This non-transversality contrasts with the
formula (34) in Tsai-Erber \cite{TsaiErber1}
for the general $(3+1)$-dimensional
 vacuum polarization in an external $B$, which
they shown in their eq.~(36) to be transverse.
It can be traced back to classically setting respectively $p_3=0$ and
$p_3+q_3=0$ inside the two propagators of graphene-born electrons,
which cannot be achieved without $q_3=0$, which makes true the last
relation (\ref{eq:transinf}).
One should however not focus on  $T^{\mu\nu}$ because 
the transversality condition concerns the vacuum polarization
$\Pi^{\mu\nu}$,  $T^{\mu\nu}$  being only an intermediate step in
the calculation.  
We shall comment more about transversality in subsection \ref{subsec:MSM}.

\section{The light-cone equations and their solutions} \label{section:lceqs}

\subsection{Orders of magnitude}\label{subsec:magnitude}

In order to determine inside which domains we have to vary the dimensionless
parameters, it is useful to know the orders of magnitude of the
physical parameters involved in the study. 

$\bullet$\ The thickness of  graphene is $2a \approx 350\,pm$.

$\bullet$\  As stated in (\ref{eq:p3mdef}),
 $|p_3^{max}| \simeq \frac{\hbar}{a}$.
This gives $c \,p_3^{max} \simeq 1.8\,10^{-16} J$ or
$c\,p_3^{max} \simeq  1130\,eV \approx 2.2\,10^{-3}\,m_e$.

$\bullet$\  To $eB$ corresponds  $m^2 = \frac{\hbar e B}{c^2}$.
For example to $e B^m$ (see below) corresponds the mass $\frac{\sqrt{\hbar e
B^m}}{c} \approx
2~10^{-33}\,kg \approx 2~10^{-3}\,m_e \ll m_e$.

$\bullet$\ $[B] = \frac{[p]^2}{[e] \hbar}$ such that, to $(p_3^m)^2$
corresponds
$B^m \simeq \frac{\hbar}{e a^2} \approx 21400\,T$.

$\bullet$\ One has $\zeta\equiv\frac{\sqrt{2e\hbar B}}{p_3^m}=\sqrt{2\frac{B}{B^m}}$.
Since $B = \frac{\zeta^2}{2} B_m$, to $\zeta$ corresponds the
mass $\sqrt{2}\,\zeta \,10^{-3}\,m_e$.

$\bullet$\  We shall consider magnetic fields in the range $[1\,T,20\,T]$;
\begin{equation}
1\,T \leq B \leq 20\, T \Leftrightarrow 1/100 \leq \zeta \leq
\sqrt{20}/100.
\label{eq:Brange}
\end{equation}

$\bullet$\ The wavelength of visible light lies between $350\,nm$ and
$700\,nm$, which corresponds to frequencies $\nu$ between $4.3\,10^{14}\,Hz$ and
$7.9\,10^{14}\,Hz$, to  energies in the range $[3.5\,eV, 1.5\,eV]$
and to  $\eta=\frac{q_0}{cp_3^m}= \frac{2\pi a \nu}{c}$ such that
\begin{equation}
\text{visible light} \leftrightarrow 1.6\,10^{-3} \leq \eta \leq
2.9\,10^{-3} \ll 1.
\label{eq:etarange}
\end{equation}

\subsection{The light-cone equations}

It is now straightforward to write  the light-cone relations
(\ref{eq:lcperp}) and (\ref{eq:lcpar}) in the case of graphene.
We first express the relevant components 
$T^{11}, T^{22}, T^{33}$ in terms of dimensionless variables
\begin{equation}
\begin{split}
T^{11}(n,\theta,\eta,\zeta) &= 4\alpha\, e^{-(n_x^2+n_y^2)
\frac{\eta^2}{\zeta^2}}\; \zeta\,
\eta^2\,p_3^m \;\frac{n_x^2-n_y^2}{\eta^2-4\zeta^2},\cr
T^{22}(n,\theta,\eta,\zeta) &=
-T^{11}(\alpha,n,\theta,\eta,\zeta),\cr
T^{33}(n,\theta,\eta,\zeta) &= -4\alpha\, e^{-(n_x^2+n_y^2)
\frac{\eta^2}{\zeta^2}}\;
\zeta\, p_3^m\; \frac{\zeta^2 -2(n_x^2+n_y^2)\eta^2}{\eta^2-4\zeta^2},
\end{split}
\label{eq:summary2}
\end{equation}
in which $n_x = n s_\theta$ and, since $q_2=0$, $n_y=0$
\footnote{It is easy to see on (\ref{eq:summary}) that $T^{00}(\hat q, B)$ and
$T^{33}(\hat q,B)$ are also the only two components of $T^{\mu\nu}(\hat
q)$ that do not vanish at $\theta \to 0$.\label{foot:tetanull}}.
Then, (\ref{eq:lcperp}), (\ref{eq:lcpar}) and (\ref{eq:PiV}) lead  to
\begin{equation}
\begin{split}
& \star\ for\ A^\mu_\perp : (1-n^2)\left[1 +\frac{p_3^m}{\pi^2}\;\frac{1}{q_0^2}\;
T^{22}(n,\theta,\eta,\zeta)\; V(n,\theta,\eta,u)
\right]=0,\cr\;
& \star\ for\ A^\mu_\parallel : (1-n^2)\left[
1+\frac{p_3^m}{\pi^2}\;\frac{1}{q_0^2}
\Big( c^2_\theta\, T^{11}(n,\theta,\eta,\zeta) + s^2_\theta\,
T^{33}(n,\theta,\eta,\zeta)\Big)\;
V(n,\theta,\eta,u)\right]=0,
\end{split}
\label{eq:lcgeneral}
\end{equation}
and, using (\ref{eq:summary2}), to
\begin{equation}
\begin{split}
& \star\ for\ A^\mu_\perp : (1-n^2)\left[
1-\frac{4\alpha}{\pi^2}\,s_\theta^2 n^2\,
\frac{\zeta}{\eta^2-4\zeta^2}\;
e^{-(n s_\theta\,\frac{\eta}{\zeta})^2}\;
V(n,\theta,\eta,u)\right]
=0,\cr
& \star\ for\ A^\mu_\parallel :
(1-n^2)\left[
1 + \frac{4\alpha}{\pi^2}s_\theta^2
\left(c_\theta^2 n^2 \, \frac{\zeta}{\eta^2-4\zeta^2}
+\frac{\zeta}{\eta^2}
\frac{2\eta^2 n^2 s_\theta^2-\zeta^2}{\eta^2-4\zeta^2}
\right)
\,e^{-(n s_\theta\,\frac{\eta}{\zeta})^2}\,V(n,\theta,\eta,u)\right]
=0.
\end{split}
\label{eq:lc4}
\end{equation}
For each polarization, this defines an index $n=n(\alpha, u,\theta,\eta,\zeta)$.

At large values of $\Upsilon\equiv \frac{\zeta}{\eta}$, the second
contribution to the light-cone equation for $A^\mu_\parallel$ inside the
$(\ )$, which is that of $T^{33}$, largely dominates.

\subsection{Analytical expression for the transmittance $\boldsymbol
{V(n,\theta,\eta,u)}$}

In order to solve the light-cone equations (\ref{eq:lc4}),
 the first step is to compute
$V$, so as to get an algebraic equation for $n$.
$V$ as given by (\ref{eq:UVdef})
 is the Fourier transform of the function $x \mapsto
-\eta ^2 \frac{\sin x}{x(x-\sigma_1)(x-\sigma_2)}$ where 
\begin{equation}
\sigma_1 = -\eta \left(n c_\theta - \sqrt{1-n^2 s_\theta^2}\right),\quad
\sigma_2 = -\eta \left(n c_\theta + \sqrt{1-n^2 s_\theta^2}\right)
\label{eq:poles}
\end{equation}
are the poles of the integrand.
The Fourier transform of such a product of a cardinal sine with a
rational function is well known.
The result involves Heavyside functions of the imaginary parts of the
poles $\sigma_1, \sigma_2$, noted $\Theta_i^+$ for $\Theta_i
(\Im(\sigma_i))$ and $\Theta_i^-$ for $\Theta_i (-\Im(\sigma_i))$.
\begin{equation}
\hskip -1cm V(u, n, \theta, \eta)= \frac{-\pi \eta^2}{\sigma_1 \sigma_2 (\sigma_1 -
\sigma_2)}
\left[ (\sigma_1-\sigma_2)
+ \sigma_2 \left(\Theta_1^- e^{-i \sigma_1 (1-u)}+\Theta_1^+
e^{+i \sigma_1 (1+u)}\right)
-\sigma_1 \left(\Theta_2^- e^{-i \sigma_2
(1-u)}+\Theta_2^+ e^{+i \sigma_2 (1+u)}\right) \right].
\label{eq:transmit}
\end{equation}
The poles $\sigma_1, \sigma_2$
 are seen to control the behavior of $V$, thus  of $n$,
which depends  on the signs of their imaginary parts.
That the Fourier transform  is well defined
needs in particular that the poles have a non-vanishing imaginary
part. This requires either  $n \not\in {\mathbb R}$ or $n s_\theta >1$.

The case when the poles are real needs a special treatment.
A first possibility is to define  the integral
as a Cauchy integral, like we did when calculating $T^{\mu\nu}$,
arguing in particular of the $+i\varepsilon$ which is understood in the
denominator of the outgoing photon propagator. Then, $V$ is
calculated through contour integration in the complex plane.
This alternate method can also be used when the poles are complex. It is
comforting that the two methods give, at leading order in an expansion at
small $\eta$ and $n_2$ ($n_2$ is the imaginary part of the refractive
index) the same results. In particular, the
cutoff that is then needed to stabilize the integration on the large upper
1/2 circle turns out to be the same as the one that naturally arises in the
Fourier transform because of the $\frac{\sin \sigma}{\sigma}$
function.
The second, and  simplest, possibility, is to define everywhere in
(\ref{eq:transmit}) $\Theta(0)=\frac12$. It is equivalent to the previous
one, again at leading order in an expansion at small $\eta$ and $n_2$.
Then, one gets, at $u=0$ (which is always a very good approximation)
\begin{equation}
V(0,n,\theta,\eta)\stackrel{poles\in{\mathbb R}}{=}\frac{\pi}{1-n^2}
\left(1+\frac{\sigma_2\cos\sigma_1-\sigma_1\cos\sigma_2}{2\eta\sqrt{1-n^2s^2_\theta}}
\right).
\label{eq:Vrp}
\end{equation}

Last, if one shrinks $V$ to $\int_{-\infty}^{+\infty} e^{i\sigma
u}\frac{\sin\sigma}{\sigma}$, which means only accounting for the gate
function in the transmittance, it becomes $V=\pi$ inside graphene
(see also subsection \ref{subsub:Vsimp}).

\subsection{Solving the light-cone equations for
$\boldsymbol{A^\mu_\parallel}$ and  $\boldsymbol{n\in {\mathbb
R}> \frac{1}{\sin\theta}}$}

That $n \in {\mathbb R}$ largely simplifies the equations.
No non-trivial solution has been found for $n<\frac{1}{s_\theta}$ (see
subsection \ref{subsec:nosolinf}).

\subsubsection{Calculation of $\boldsymbol{V(n,\theta,\eta,u)}$}

Expanding $V$ at leading orders in $\eta$, one gets
\begin{equation}
\begin{split}
\star\ \Re(V) &= -\frac{\pi}{\sqrt{ n^2 s^2_\theta-1}}\;\eta +
 \frac12 \pi(1 + u^2) \eta^2 + {\cal O}(\eta^3),\cr
\star\ \Im(V) &= u\, n\,c_\theta\;\frac{\pi}{\sqrt{ n^2 s^2_\theta-1}}\;\eta^2
+ {\cal O}(\eta^3).
\end{split}
\label{eq:Vexp}
\end{equation}
The expansion for $\Im(V)$ in (\ref{eq:Vexp}) starts at ${\cal
O}(\eta^2)$ while that of  $\Re(V) = {\cal O}(\eta)$.

For $n\in {\mathbb R} > \frac{1}{s_\theta}$,  we replace in
(\ref{eq:poles})
$\sqrt{1-n^2s^2_\theta}$ with $i\sqrt{n^2s^2_\theta-1}$ and
 the two poles $\sigma_1$ and $\sigma_2$ of $V$ become
\begin{equation}
\sigma_1=-\eta\left(n \cos\theta -i  \sqrt{n^2 s_\theta^2 -1}\right), \qquad
\sigma_2=-\eta\left(n \cos\theta +i  \sqrt{n^2 s_\theta^2 -1}\right);
\label{eq:realpoles}
\end{equation}
the first term in $\Re(V)$ coincides with $\pm 2i\pi\times$ the
residue at the pole $\sigma_1$ or $\sigma_2$  that lies inside the
contour of integration when one calculates $V$ as a contour integral (see
(\ref{eq:resapp})).

\subsubsection{$\boldsymbol V$ at $\boldsymbol{\theta=0}$}
\label{subsub:Vtetanul}

At $\theta=0$ the poles $\sigma_1 = -\eta(n-1), \sigma_2=-\eta(n+1)$ are
real such that we set $\Theta(0)=\frac12$ in the general formula
 (\ref{eq:transmit}).
The expression (\ref{eq:transmit}) for $V$ becomes
\begin{equation}
V_{\theta=0} = -\frac{\pi \eta^2}{\sigma_1 \sigma_2}\left(
1+\frac{\sigma_2 \cos\sigma_1\, e^{iu\sigma_1} -\sigma_1
\cos\sigma_2\, e^{iu\sigma_2}}{\sigma_1-\sigma_2}\right).
\end{equation}
Using the explicit expressions of the poles just written and expanding the
$\cos$ and $\exp$ functions at small values of $\sigma_1,\sigma_2$
(we suppose that $n$ is much
smaller that its quantum upper limit $n_{quant}\sim\frac{1}{\eta}$ (see
subsection \ref{subsec:nquant}), and that, accordingly, $|\sigma_1|,
|\sigma_2|\ll 1$) one gets finally
\begin{equation}
V_{\theta=0} \approx \frac{\pi\eta^2}{2}(1+u^2).
\label{eq:Vtetanull}
\end{equation}
It will be used later to show, for $B\not=0$ as well as for $B=0$,
 that the only solution of the light-cone equations at $\theta=0$ is the trivial $n=1$.

\subsubsection{The imaginary parts of the light-cone equations}

The imaginary parts of both light-cone equations (\ref{eq:lc4}) shrink, for
$n$ real, to
\begin{equation}
\Im (V) =0.
\end{equation}
It is only rigorously satisfied at $u=0$, but, 
(\ref{eq:Vexp}) and  numerical
calculations show that, for values of $\eta$ in the visible spectrum
$\eta \in [1.6/1000, 2.9/1000]$,
$\Im(V) \ll \Re (V) <1$ and that $\Im(V)\approx 0$ is always an excellent
approximation.

\subsubsection{There is no non-trivial solution
for $\boldsymbol{A^\mu_\perp}$}

Detailed numerical investigations show that no solution exists for the
transverse polarization but the trivial solution $n=1$. We shall therefore
from now onwards only be concerned with photons $A^\mu_\parallel$ with a
parallel polarization (see Figure~\ref{fig:setup}).

\subsubsection{The light-cone equation for
$\boldsymbol{A_{\parallel}^{\mu}}$ and its solution }

Expanding  $V$ in powers of $\eta$ and neglecting $\Im(V)$ enables to
get, through standard manipulations,
 a simple analytical equation for the refractive index $n$. For $\Upsilon
\equiv \frac{\sqrt{2eB}}{q_0}\gg 1$ and $\eta < 3/1000$,
 the following accurate expression is
obtained by expanding (\ref{eq:lc4}) in powers of $\frac{1}{\Upsilon}$
\begin{equation}
(1-n^2)\left[
1-\frac{\alpha}{\pi} \Upsilon\;\frac{s^2_\theta}{\sqrt{n^2s^2_\theta -1}}
\left(1+\frac{-3n^2s^2_\theta -c^2_\theta +1/4}{\Upsilon^2}
\right)\right] =0,
\label{eq:nparfull}
\end{equation}
which leads consistently  to the non-trivial solution of the light-cone
equation (\ref{eq:lc4})
\begin{equation}
 n^2 \simeq \frac{1}{s_\theta^2}\ \frac{ 1+ \left(\frac{\alpha
 \Upsilon s_\theta^2}{\pi}\right)^2 \left( 1+\frac{1}{2\Upsilon^2}
\right)
 }{1+2\left(\frac{\alpha s_\theta}{\pi}\right)^2 (3s_\theta^2
 +c_\theta^2)}.
 \label{eq:solpar}
 \end{equation}

\subsubsection{Graphical results and comments}

\begin{figure}[h]
\begin{center}
\includegraphics[width=6 cm, height=5 cm]{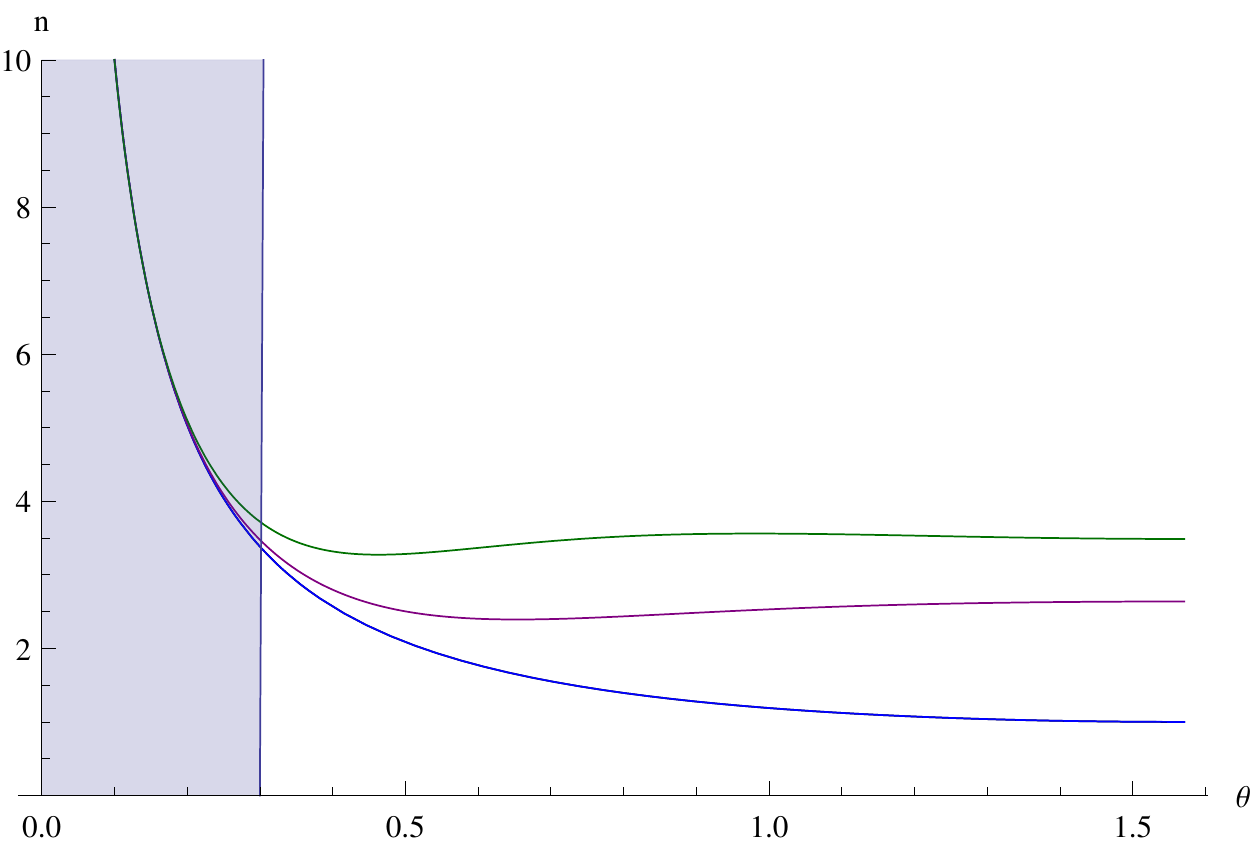}
\hskip 2cm
\includegraphics[width=6 cm, height=5 cm]{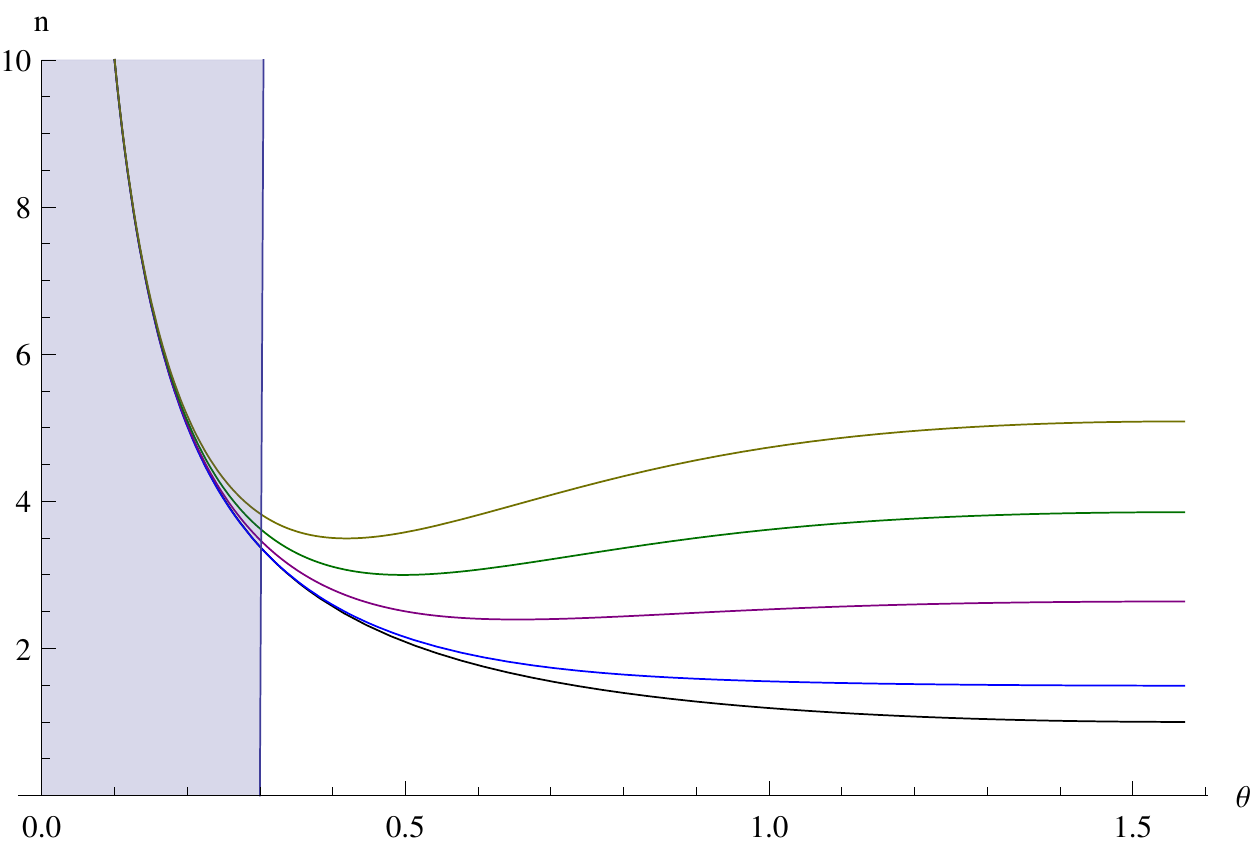}
\end{center}
\caption{The index $n\in{\mathbb R}$ for $A^\mu_\parallel$
 as a function of $\theta$. On the left we vary $\alpha =
1/137\,(blue), 1\,(purple), 2\,(green)$ at $\Upsilon=10$; on the right we vary
$\Upsilon = 5\,(blue), 10\,(purple), 15\,(green), 20\, (yellow)$ at $\alpha=1$. The
lower (black) curves are $1/\sin\theta$.}
\label{fig:nBreal}
\end{figure}

The results given in eq.~(\ref{eq:solpar}) are plotted on
Figure~\ref{fig:nBreal}.
On the left we vary $\alpha$ from $\frac{1}{137}$ to $2$
 at $\Upsilon =10$ and on the right we keep
$\alpha=1$ and vary $\Upsilon$ between $5$ and $20$. On both plots, the
black lower curve is $n=\frac{1}{\sin\theta}$ (on the left plot it cannot
be distinguished from the blue curve). We have shaded the domain of
low $\theta$ in which  $n$ must make a transition 
to another regime (see subsection \ref{subsec:limtetanull}).

$\star$\ The curves go asymptotically to $\frac{1}{s_\theta}$ when
$\theta \to 0$. However, we shall see that they should be truncated before
$\theta=0$).

$\star$\ At large angles, the effects are mainly of quantum nature, 
strongly influenced by the presence  of $B$ and largely depending on the
value of $\alpha$;
when $\theta$ gets smaller, one goes to
another regime in which the effects of confinement are the dominant ones.

1-loop effects are therefore potentially large at $\alpha
\geq 1$.  Furthermore, at reasonable values of $B$ and for photons
in the visible spectrum, the dependence on $B$ is strong. 

$\star$\ They increase with $\Upsilon=\frac{c\sqrt{2\hbar eB}}{q_0}$,
therefore inversely  to the energy of
the photon : low frequencies are favored for testing, and this limit is
fortunate since our expansions are done at $\eta = aq_0(/\hbar c) \ll 1$. 
As for the proportionality to $\sqrt{eB}$ for very large values of $eB$,
 it should be compared with the
corresponding factor $eB$ pointed at in \cite{Shabad} in the ``vacuum''.
The difference in powers can be easily traced back to the different
integrations in the course of the calculations. In our case, integrating
over the transverse electronic degrees of freedom yields a factor $(eB)^2$
while the remaining integral $D$ (\ref{eq:Bdef}) over $p_0$ yields a factor
$1/(eB)^{3/2}$. This apparently infinitely growing refraction with $eB$
should however stop at $B=B^m$,  above which new quantum effects are
expected (see subsection \ref{subsec:nquant}).

$\star$\ For $\eta \ll 1$ and $n>\frac{1}{s_\theta}$,
 the residues of $V$ at the poles $\sigma_1$ and $\sigma_2$ are 
\begin{equation}
res(\sigma_1)=-\frac{\eta}{2i\sqrt{n^2 s^2_\theta-1}} + {\cal O}(\eta^2)
 = -res(\sigma_2).
\label{eq:resapp}
\end{equation}
The agreement between $\Re(V)$ in the first line of (\ref{eq:Vexp}) and $\pm
2i\pi\ res(\sigma_1)$ is conspicuous. Indeed, it is easy to prove that  for $n\in
{\mathbb R}$, only one of the two poles lies inside the contour of
integration in the upper 1/2 complex $\sigma$-plane,
 which is the alternate method to calculate $V$.

$\star$\ In the approximation that we made, the refractive index does not
depends on $u$, the position inside the strip. This dependence, very weak,
only starts to appear through higher orders in the expansion of the
transmittance $U$ (or $V$).

$\star$\ $n$ does not depend explicitly on the
thickness $a$ (it depends only on $\Upsilon$, independent of $a$).
The limit $a\to 0$
(which is compatible with $\eta = aq_0 \ll 1$) is therefore ``smooth'' (see
also subsection \ref{subsub:limanul}).
At the opposite, the limit $a\to\infty$, which corresponds to forgetting
about the confinement of vertices and about the transmittance, to exact photon
momentum conservation $(k_3=0)$ 
cannot be taken reliably because it is in contradiction with $\eta \ll 1$.

\subsubsection{The ``leading''
$\boldsymbol{n \sim \frac{1}{\sin\theta}}$ behavior}
\label{subsub:leading}

It is easy to track the origin of the leading $\frac{1}{s_\theta}$
behavior of the index (we shall see below that the related divergence at
$\theta \to 0$ is fake).
It comes in the regime when the two poles of $V$ lie in different 1/2 planes,
such that $V$ can be safely approximated by $V \approx 2i\pi\times
residue(\sigma_1\ or\ \sigma_2)$.

Keeping only the leading terms $\propto \Pi^{33}$
 in the light-cone equation (\ref{eq:lcpar}) and using (\ref{eq:PiV})
yields (we factor out $(1-n^2)$ and forget about the trivial solution
$n=1$)
\begin{equation}
1+\frac{s_\theta^2}{\pi^2}\frac{p_3^m}{q_0^2}\;T^{33}\, V=0.
\label{eq:leading0}
\end{equation}
Using (\ref{eq:resapp}) gives then
\begin{equation}
1 - \frac{\alpha s^2_\theta}{\pi^2} \frac{\Upsilon}{\eta} \left(2i\pi 
\frac{\eta}{2i\sqrt{n^2s^2_\theta -1}}\right)=0.
\label{eq:leading1}
\end{equation}
The factor $\frac{\alpha s^2_\theta}{\pi^2} \frac{\Upsilon}{\eta}$, which
depends in particular of $\alpha$ and $B$,
originates from $\frac{s_\theta^2}{\pi^2} \frac{p_3^m}{q_0^2}T^{33}$
in (\ref{eq:leading0}), while the term inside $(\quad)$ comes from the
(residue of the) pole of $V$.
Eq.~(\ref{eq:leading1}) yields
\begin{equation}
n^2s^2_\theta - 1 = \left(\frac{\alpha s^2_\theta \Upsilon}{\pi} \right)^2,
\label{eq:leading2}
\end{equation}
in which we recognize the leading terms of the solution (\ref{eq:solpar}).

\subsubsection{The limit $\boldsymbol{a\to 0}$}
\label{subsub:limanul}

At this stage, we can understand why the limit of infinitely thin graphene
$a\to 0$ is delicate and should not, {\em a priori}, be taken from the
start.

Since $\Upsilon$ is independent of $a$, so is eq.~(\ref{eq:leading2})
\footnote{as long as $\eta = aq_0$ stays small since we made
expansions at small values of this parameter and our results are only valid
at this limit.}. However, this property arises after the cancellation
of two $\eta$ factors  in (\ref{eq:leading1}), one coming from
$\frac{p_3^m}{q_0^2} T^{33}$ and the second from the residue
(\ref{eq:resapp}) of $V$. Taking $a=0$ cancels the transmittance $V$ and
its poles, such that the $\sqrt{n^2 s^2_\theta-1}$ in (\ref{eq:leading1}),
 which yields the l.h.s. of (\ref{eq:leading2}) and the leading
$1/s_\theta$ behavior of $n$, fades away.
Notice however that, in the domain of (fairly large) values of $\theta$ 
 in which our results are reliable, this leading behavior
is not very constraining, specially at large values of $\alpha$ and $B$.

\subsubsection{The trivial solution $\boldsymbol{n=1}$}
 
To better understand the fate of the solution $n=1$, let us rewrite
(\ref{eq:transmit}) as
\begin{equation}
V(u,n,\theta,\eta)=\frac{\pi}{1-n^2}\left[
1+\frac{\sigma_2 \left(\Theta_1^- e^{-i \sigma_1 (1-u)}+\Theta_1^+
e^{+i \sigma_1 (1+u)}\right)
-\sigma_1 \left(\Theta_2^- e^{-i \sigma_2
(1-u)}+\Theta_2^+ e^{+i \sigma_2
(1+u)}\right)}{2\eta\sqrt{1-n^2s^2_\theta}}
\right],
\label{eq:transmit2}
\end{equation}
in which we have also used the expressions (\ref{eq:poles}) of $\sigma_1$
and $\sigma_2$.
At $n=+1$, $\sigma_1=0,\,\sigma_2=-2\eta c_\theta$ and, at $n=-1$,
$\sigma_1=2\eta c_\theta,\, \sigma_2=0$ such that, in both cases
(\ref{eq:transmit2}) writes (setting to $\frac12$ the two appropriate
$\Theta$ functions since the poles are real)
\begin{equation}
V(u,\pm 1,\theta,\eta)= \frac{\pi}{1-n^2}\times(1-1).
\end{equation}
Accordingly, the product $(1-n^2)\times V$ occurring in the light-cone equations
(\ref{eq:lc4}) vanishes for $n=\pm1$ such that, in particular, the trivial
solution $n=1$ always stays valid.

\subsubsection{The limit $\boldsymbol{\alpha = 0}$}

At $\alpha=0$ the contribution of the vacuum polarization vanishes and, as
is seen on (\ref{eq:nparfull}), the only solution is the trivial $n=1$.

To be complete, this limit should also operate smoothly on the nontrivial
solution (\ref{eq:solpar}). However, since (\ref{eq:solpar})
 was obtained by the expansion (see subsection \ref{subsub:bigB})
of $tanh(\tau eB)$ and $cosh(\tau eB)$ at large values of their argument
(large $B < \infty$), like the limit $B\to 0$, the limit $e \to 0$ cannot be safely
obtained in this framework. In particular the apparent limit $n
\stackrel{\alpha\to 0}{\to}
\frac{1}{s_\theta}$ that occurs in (\ref{eq:solpar}) should be considered
as fake.

\subsubsection{Shrinking the transmittance to the sole gate function}
\label{subsub:Vsimp}

To test the importance of the poles in the integrand of the transmittance
(\ref{eq:Pieff4b}) (\ref{eq:UVdef}), it is instructive
to arbitrarily shrink $V$  to the pure
geometric (Fourier transform of the) gate function. 
This drastic approximation forgets about the ratio of external photon
propagators at $k_3=0$ and $k_3 \not=0$.
One gets then $V=\pi$ inside graphene and
the light-cone equation (\ref{eq:lc4}) for $A^\mu_\parallel$ shrinks to
(we forget about the global factor $(n^2-1)$ and the trivial solution
$n=1$)
\begin{equation}
1+\frac{\alpha}{\pi} \frac{s_\theta^2}{\eta}e^{-\frac{n^2
s_\theta^2}{\Upsilon^2}}\left(\Upsilon-\frac{n^2(1+s_\theta^2)}{\Upsilon}
\right)=0.
\label{eq:ngatepar}
\end{equation}
Eq.~(\ref{eq:ngatepar}) has only real solutions and, accordingly, no
absorption.

\begin{figure}[h]
\begin{center}
\includegraphics[width=6 cm, height=5 cm]{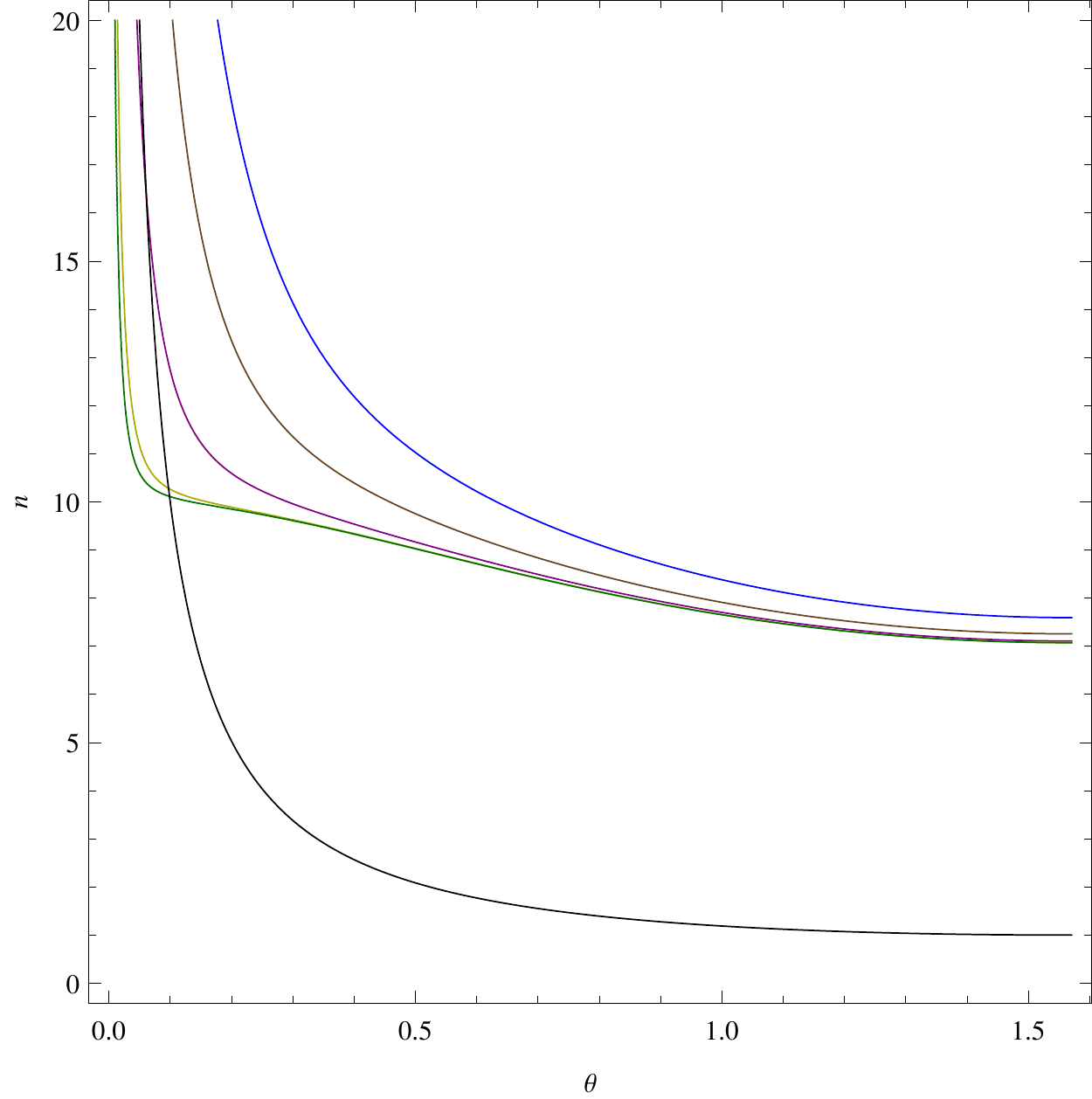}
\hskip 2cm
\includegraphics[width=6 cm, height=5 cm]{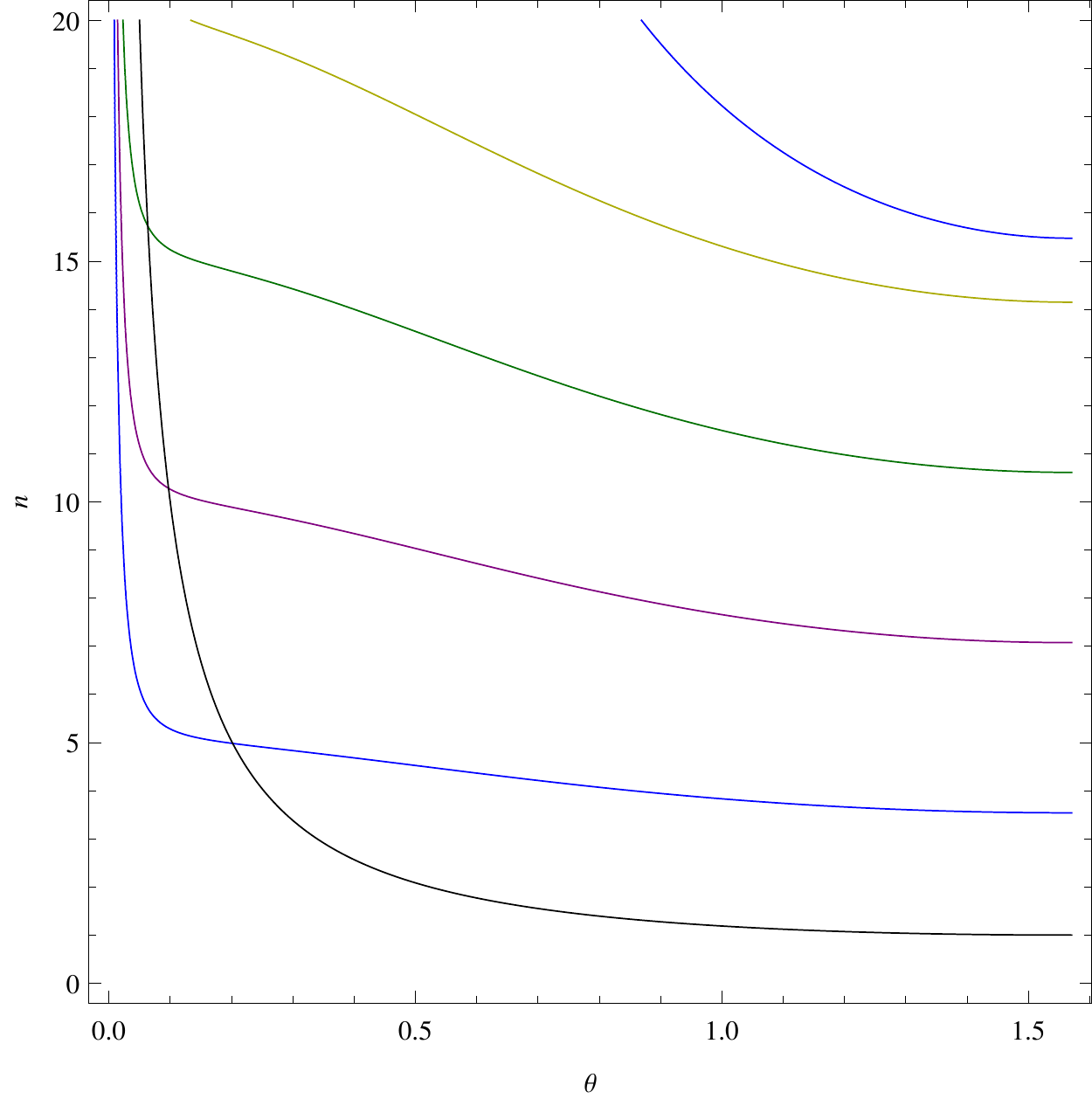}
\end{center}
\caption{The index $n$ for $A^\mu_\parallel$
 as a function of $\theta$ in the approximation $V=\pi$ inside graphene
(gate function).
On the left we vary $\alpha =
1/137\,(blue), 1/50\,(brown), 1/10\,(purple), 1\,(yellow), 2\,(green)$
at $\Upsilon=10, \eta=\frac{2}{1000}$; on the right we
vary
$\Upsilon = 5\,(blue), 10\,(purple), 15\,(green), 20\, (yellow)$ at
$\alpha=1$. The
lower (black) curves are $1/\sin\theta$.}
\label{fig:ngate}
\end{figure}

Results are summarized on Figure~\ref{fig:ngate}. On the left plots, we
keep $\Upsilon=10, \eta=\frac{2}{1000}$ and vary\newline $\alpha=1/137 \,(blue),
1/50\,(brown), 1/10\,(purple), 1\,(yellow), 2\,(green)$. On the right, we
keep $\alpha=1, \eta=\frac{2}{1000}$ and vary
 $\Upsilon = 5\,(blue),
10\,(purple), 15\,(green), 20\, (yellow)$. The black curves on both figures
are $n=\frac{1}{s_\theta}$. 

Like when using the full expression for $V$, the limit of small
$\theta$ is not reliable (in particular, a smooth transition to $n=1$ at
$\theta=0$ looks more unreachable than ever).
 In general, $n$ roughly grows like $\Upsilon$ and the role of
$\alpha$ has decreased, in particular at large values of $\theta$.

There exist other families of solutions  at larger values of $n$. A trace
of them can be seen for $\Upsilon=5$ (blue)
 in the upper 1/2 of the right plot in
Figure~\ref{fig:ngate}. They are due to the presence of the exponential
$e^{n^2s_\theta^2/\Upsilon^2}$ in (\ref{eq:ngatepar}).

The large differences that we get with respect to the full calculation
shows the importance of treating  the transmittance as a complex function
of a complex $n$ and of paying special
attention to its poles (in close relation with the fluctuations
of electron momentum due to the confinement of vertices).

\subsection{The transition $\boldsymbol{\theta \to 0}$}\label{subsub:transit}
\label{subsec:limtetanull}

\subsubsection{At $\boldsymbol{\theta=0}$}
\label{subsub:tetanul}

The simplest is to come back to the light-cone equation (\ref{eq:lc0}). At
$\theta=0$ $V$ is given by (\ref{eq:Vtetanull}) and, at this same limit,
$\Pi^{11}=-\Pi^{22}=0$ since not only $q_2=0$ by the choice of frame, but
also, now, $q_1 \equiv |\vec q|s_\theta=0$. The light-cone equation
shrinks then to
\begin{equation}
(\beta_1^2 + \beta_2^2) q^2 =0,
\end{equation}
which has for unique solution the trivial $n=1$.

\subsubsection{A cumbersome transition}

It is fairly easy to determine the value of $\theta$ below which our
calculations and the resulting approximate formula (\ref{eq:solpar}) may
not be trusted anymore.
There presumably starts a transition to another regime.

Our calculations stay valid as long as the two poles $\sigma_1$ and
$\sigma_2$ of the transmittance function $V$ lie in  different 1/2 planes.
This requires that their imaginary parts have opposite signs. Their
explicit expressions are given in (\ref{eq:impoles}) below. It is then
straightforward to get the following condition (we slightly anticipate and
consider $n=n_1+in_2 \in {\mathbb C}$)
\begin{equation}
\sigma_1\ \text{and}\ \sigma_2\ \text{in  different 1/2 planes} \Leftrightarrow
n_1^2 > \frac{1+n_2^2}{\tan^2\theta}.
\label{eq:condval}
\end{equation}
(\ref{eq:condval}) is always satisfied at $\theta=\frac{\pi}{2}$ and never
at $\theta=0$. Since $n_2 \approx 0$, the transition occurs at 
\begin{equation}
n_1(\theta) \approx n(\theta) \approx \frac{1}{\tan\theta},
\end{equation}
in which we can use (\ref{eq:solpar}) for $n$. Since at small $\theta$,
$\sin\theta \simeq \theta \simeq \tan\theta$, this condition writes
approximately
\begin{equation}
1\leq  \frac{1+\left(\frac{\alpha\Upsilon s_\theta^2
}{\pi}\right)^2\left(1+\frac{1}{2\Upsilon^2}\right)}
{1+2\left( \frac{\alpha s_\theta}{\pi}\right)^2 (3s^2_\theta+c^2_\theta)}
\Leftrightarrow \theta \geq \theta_{min}= \sqrt{\frac{2}{\Upsilon^2-\frac72}}.
\label{eq:thetalim}
\end{equation}
For example, at $\Upsilon =5$ it yields $\theta \geq .3$.
Notice that the
condition (\ref{eq:thetalim}) also sets a lower limit $\Upsilon >
\sqrt{\frac72}$. 

It is easy to get the value  $n_{max}$ of $n$ at $\theta = \theta_{min}
\simeq \frac{\sqrt{2}}{\Upsilon}$ given by (\ref{eq:thetalim}). Plugging
this value in (\ref{eq:solpar}) one gets
\begin{equation}
n_{max} \equiv n(\theta =\theta_{min})\approx \frac{\Upsilon}{\sqrt{2}}.
\label{eq:nmax}
\end{equation}

Seemingly, the solution (\ref{eq:solpar})
 that we have exhibited gets closer and closer to
the ``leading'' $\frac{1}{s_\theta}$ when $\theta$ becomes smaller and
smaller. It is however easy to show that this divergence is fake, by using
our result $n=1$ at $\theta=0$ deduced in subsection \ref{subsub:tetanul}.

The diverging solution
(\ref{eq:solpar}) cannot  be trusted down to $\theta =0$ at which $n=1$;
so, the true solution of the light-cone equation
 must cross the curve $n=\frac{1}{s_\theta}$ somewhere at
small $\theta$.
However, such a transition cannot exist.
This is most easily proved by showing that, at no value of
$\theta$,
$n=\frac{1}{s_\theta}$ can be a solution to the light-cone equation
(\ref{eq:lc4}).
Let us write $\sigma_1 = -\eta
\frac{c_\theta}{s_\theta}+\epsilon, \sigma_2 =
-\eta\frac{c_\theta}{s_\theta}-\epsilon$. The poles being real, $V$ can be
calculated by setting  $\Theta(0)=\frac12$ in (\ref{eq:transmit}), which
yields
\begin{equation}
\begin{split}
V & \stackrel{real\ poles}{\rightarrow}
-\frac{\pi \eta^2}{\sigma_1\sigma_2(\sigma_1-\sigma_2)}
\left(\sigma_1-\sigma_2 +\sigma_2 \cos\sigma_1 e^{i\sigma_1 u}-\sigma_1
\cos\sigma_2 e^{i\sigma_2 u}\right)\cr
&\hskip -1cm =-\frac{\pi \eta^2}{\sigma_1\sigma_2(\sigma_1-\sigma_2)}
\Big( \sigma_1-\sigma_2
+ \sigma_2 \cos \sigma_1 \cos u\sigma_1 -\sigma_1 \cos\sigma_2 \cos
u\sigma_2
+i\big(\sigma_2 \cos\sigma_1 \sin u\sigma_1 -\sigma_1 \cos\sigma_2 \sin
u\sigma_2\big)
\Big).
\end{split}
\label{eq:Vlim}
\end{equation}
and, in our case, at $u=0$,
\begin{equation}
V(u=0) \approx -\pi \frac{s^2_\theta}{c^2_\theta}\left( 1-\cos\Big(\eta^2
\frac{c^2_\theta}{s^2_\theta}\Big)-\eta\frac{c_\theta}{s_\theta}
\sin\Big(\eta\frac{c_\theta}{s_\theta}\Big)\right).
\end{equation}
The light-cone equation (\ref{eq:lc4}) for $A^\mu_\parallel$ writes then
\begin{equation}
\left(1-\frac{1}{s^2_\theta}\right)\left[
1+\frac{\alpha}{\pi}\frac{s^2_\theta}{c^2_\theta}\frac{1}{\zeta} \left(
c^2_\theta -\Upsilon^2 s^2_\theta \Big(1-\frac{2}{\Upsilon^2}\Big)\right)
\left(1-\cos\Big(\eta^2\frac{c^2_\theta}{s^2_\theta}\Big)
-\eta\frac{c_\theta}{s_\theta}\sin\Big(\eta\frac{c_\theta}{s_\theta}\Big)\right)
\right]=0,
\label{eq:lclim}
\end{equation}
in which we have incorporated the ``trivial'' term $(1-n^2)$.

 Eq.~(\ref{eq:lclim}) has no solution: therefore
the crossing that would make the connection between our diverging solution
and $n=1$ at $\theta=0$ cannot be realized
\footnote{We have even investigated the existence of such solutions using
the exact expression for $V$, with the same conclusion. One has to be
careful that, in this case, the two poles are identical, and the expression of
$V$ must therefore be adapted.}. This proves that the domain in
which we can trust our solution (\ref{eq:solpar}) cannot be extended down
to $\theta=0$
\footnote{Actually, we have extended our numerical calculations to values
of $\theta$ for which the two poles of $V$ lie in
the same 1/2 plane. They show that, in practice, the solution
(\ref{eq:solpar}) stays valid even in a small domain below
$\theta_{min}$.}.

This investigation is continued in subsection \ref{subsub:wallcomp} for
$n\in{\mathbb C}$ (see also  subsection \ref{subsec:bigsol}).

\subsection{The quantum upper bound $\boldsymbol{n<n_{quant}}$.
The threshold at $\boldsymbol{B=B^m}$}
\label{subsec:nquant}

Quantum Mechanics sets an upper bound $n_{quant}$ for the index.
It comes from a constraint that exists on the poles of the outgoing photon
propagator, which are also those of the transmittance $U$ : $|k_3|$, the
momentum exchanged with electrons along $B$ must be
smaller or equal to $\frac{\hbar}{a} = p_3^m$, the  cutoff of the (quantum)
momentum of the graphene-born electrons along $z$.
This translates for the poles (\ref{eq:poles})
of $V$ into
\begin{equation}
|\sigma_1| \leq 1,\quad |\sigma_2| \leq 1.
\label{eq:phiscond}
\end{equation}
For $n \in {\mathbb R} > \frac{1}{s_\theta}$, both conditions yield
\footnote{
for $n < \frac{1}{s_\theta}$, the condition
$nc_\theta \leq \sqrt{1-n^2s^2_\theta}$ must also hold, and then one must have
$n^2 \leq 1$ (the case $nc_\theta \geq\sqrt{1-n^2s^2_\theta}$ or,
equivalently $n^2 \geq 1$ has no solution).} 
\begin{equation}
n^2 \leq n^2_{quant} =\frac{1}{\eta^2} +1.
\label{eq:nquant}
\end{equation}
The existence of this bound is another clue showing that the index cannot diverge 
at small values of $\theta$, which shrinks  the domain of reliability of the
solution (\ref{eq:solpar}).

At the values of $\eta$ and $\Upsilon$ that we are operating
at (see subsection \ref{subsec:magnitude}),
 $n_{max}$ given in (\ref{eq:nmax}) is much smaller than the quantum limit
(\ref{eq:nquant}). However, when the energy of photons $q_0= \frac{\eta}{a}$
 increases, $n_{quant}$ decreases, its asymptotic
value being $1$ for infinitely energetic photons.

The case $\theta=0$ is special and is investigated directly. One has then
$\sigma_1 = -\eta(n-1), \sigma_2 = -\eta(n+1)$, such that $|\sigma_1|,
|\sigma_2| \leq 1$, that is
\begin{equation}
|n(\theta=0)| \stackrel{quantum}{\leq} \frac{1}{\eta}-1.
\label{eq:boundtetanul}
\end{equation}
To be compatible with $n=1$ at $\theta=0$ that we deduced in
subsection \ref{subsub:tetanul}, the bound (\ref{eq:boundtetanul}) requires $\eta
\leq \frac12$.

When $B$ and $\Upsilon\equiv c\frac{\sqrt{2\hbar eB}}{q_0}$ increase,
$\theta_{min}$ given in (\ref{eq:thetalim})
 decreases, while $n_{max}\equiv n(\theta_{min})$ given by
(\ref{eq:nmax}) increases. A point can be reached at which $n_{max}$
becomes equal to $n_{quant}$; it occurs
at $\eta \simeq \frac{\sqrt{2}}{\Upsilon}
\Leftrightarrow \zeta \simeq \sqrt{2}$, independently of $\eta$,
which corresponds (see subsection \ref{subsec:magnitude}) to
$B\simeq B^m \approx 21400\,T$. This gives a physical meaning to
$B^m$, which appears  as the (very large)
 magnetic field at which the two upper bounds
$n_{max}$ and $n_{quant}$ coincide. Still increasing $B$ would result in
$n_{max}$ exceeding the quantum limit. Beyond this limit, new
phenomena are  expected  which lie beyond the scope of this work.

\subsection{Going to $\boldsymbol{n\in {\mathbb C}}$}
\label{subsec:complex}

\subsubsection{The case of $\boldsymbol{A^\mu_\parallel }$}

\begin{figure}[h]
\begin{center}
\includegraphics[width=6 cm, height=4 cm]{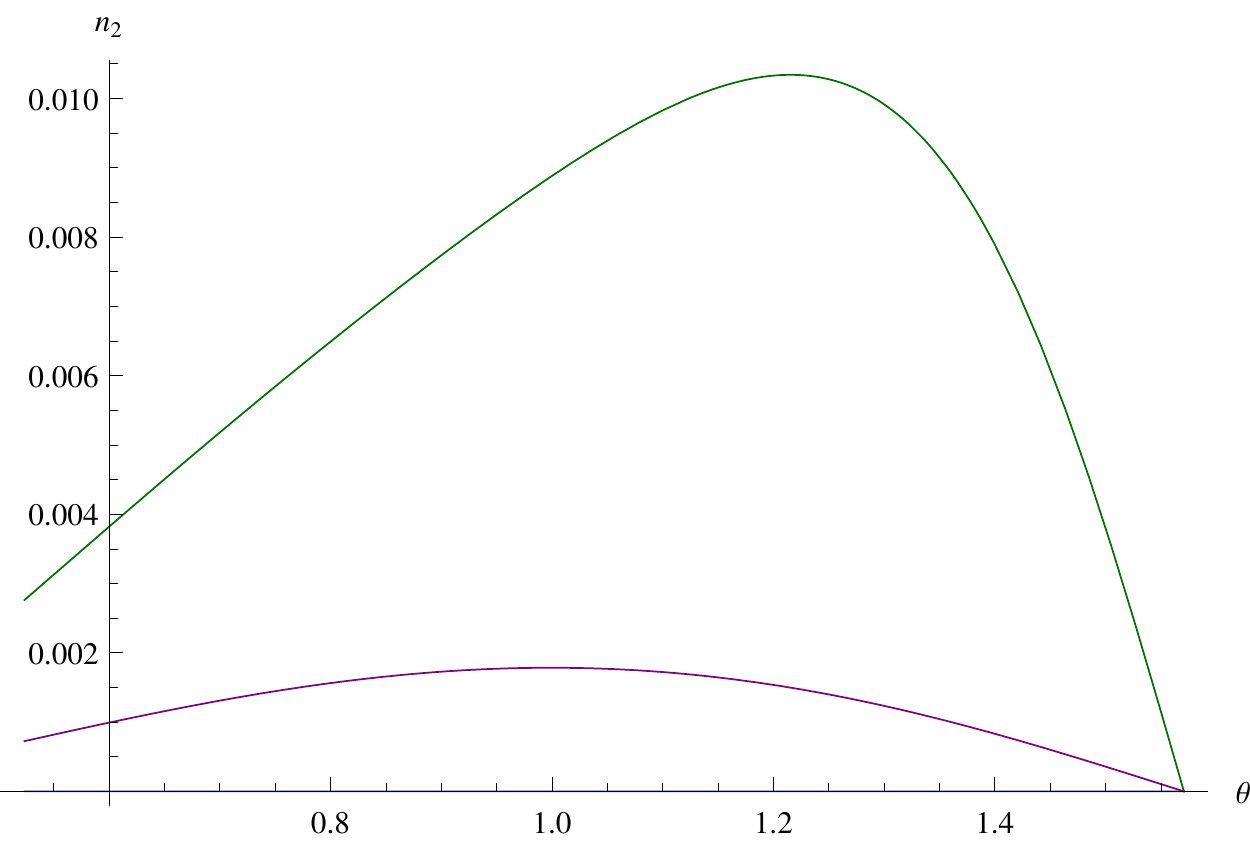}
\hskip 2cm
\includegraphics[width=6 cm, height=4 cm]{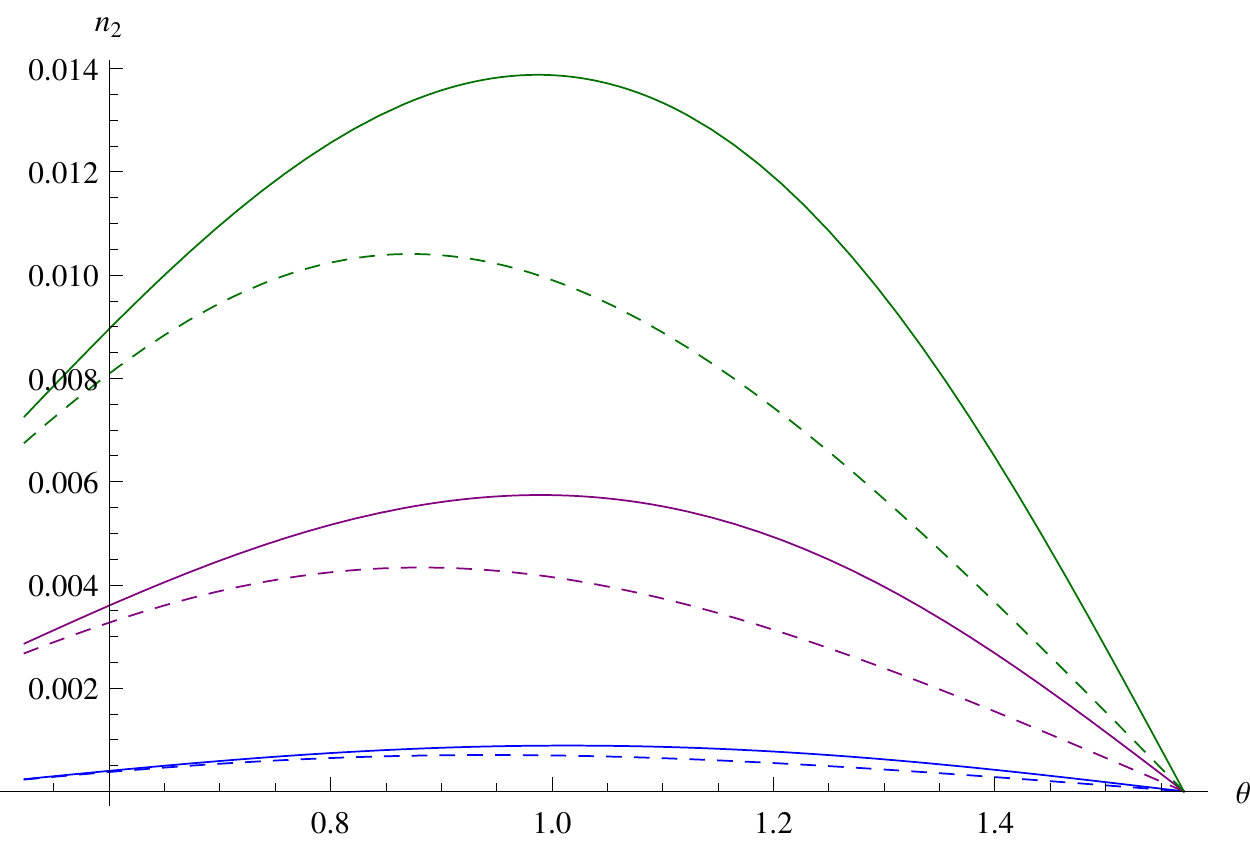}
\end{center}
\caption{The imaginary part $n_2$  of the  index $n$ for $A^\mu_\parallel$
 as a function of $\theta$. On the left
we vary $\alpha =
1/137\,(blue)$, $1\,(purple)$, $2\,(green)$ at $\Upsilon=5$; on the right we vary
$\Upsilon = 4\,(blue)$, $8\,(purple)$, $12\,(green)$ at $\alpha=1$. The dashed curves
on the right correspond to the rough approximation (\ref{eq:n2app}).}
\label{fig:nBcomptheta}
\end{figure}

\begin{figure}[h]
\begin{center}
\includegraphics[width=6 cm, height=4 cm]{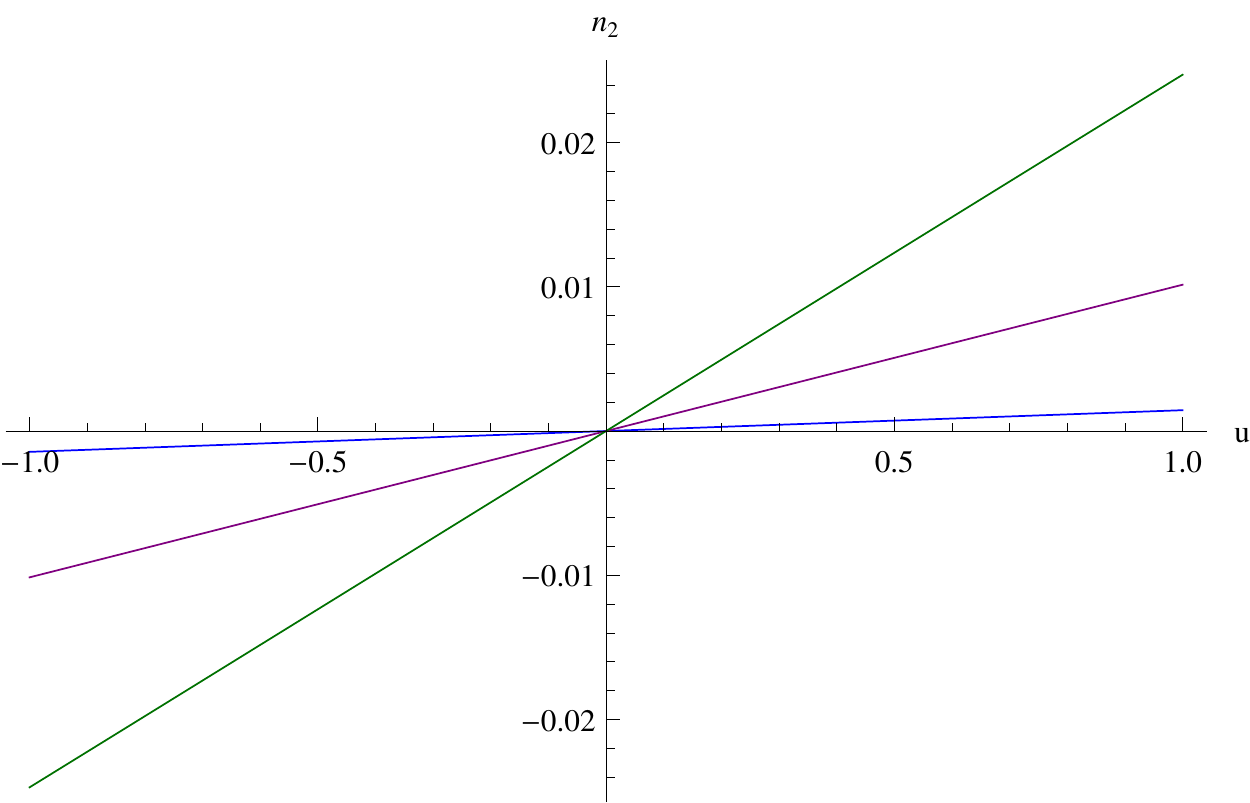}
\end{center}
\caption{The imaginary part $n_2$  of  index $n$ for $A^\mu_\parallel$
 as a function of $u$. We take $\alpha =1$, $\eta=5/1000$, and vary
$\Upsilon = 4\, (blue), 8\, (purple), 12\, (green)$.}
\label{fig:nBcompu}
\end{figure}

Numerical calculations can be performed in the general case of a complex
index $n=n_1 + i n_2$. They show in particular that $|n_2| \ll n_1$,
confirming the reliability of the
 approximation that we made in the main stream of this study
(we have limited them to values of $\theta$ large enough for our equations
to be valid). The results are displayed on Figure~\ref{fig:nBcomptheta}, in
which we plot $n_2$ as a function of $\theta$, varying $\alpha$ (left) and
$\Upsilon$ (right), and on Figure~\ref{fig:nBcompu} in which we plot $n_2$ as
a function of $u$, varying $\Upsilon$.

To this purpose, and because the real part of the light-cone equation only
gets very slightly modified, it is enough to consider the imaginary part of the
light-cone equation (\ref{eq:lc4}) for $A^\mu_\parallel$
 in which we plug,  for $n_1^2$, the
analytic expression (\ref{eq:solpar}). In practice, the expansion of this
equation at ${\cal O}(\eta^2)$ and ${\cal O}(n_2)$, which is a polynomial
of first order in $n_2$ is enough for our purposes
An important ingredient of the calculation is the expansion of
the transmittance $V$ at order ${\cal O}(\eta^2)$ and ${\cal O}(n_2)$, in
the case when its two poles lie in different 1/2 planes,
which writes
\begin{equation}
\begin{split}
\star\ \frac{1}{\pi}\Re(V) &= -\frac{\eta}{\sqrt{n_1^2s^2_\theta
-1}}+\frac{1}{2}(1+u^2)\eta^2 + \frac{u c_\theta (2n_1^2s^2_\theta -1)
}{(n_1^2 s^2_\theta -1)^{\frac32}}\eta^2 n_2 + \ldots,\cr
\star\ \frac{1}{\pi}\Im(V) &= \frac{u n_1 c_\theta}{\sqrt{n_1^2 s^2_\theta -1}}\eta^2
-\frac{n_1 s^2_\theta}{(n_1^2 s^2_\theta -1)^{\frac32}}\eta n_2 + \ldots
\end{split}
\label{eq:Vexpcomp1}
\end{equation}
The corresponding analytical expression for $n_2$, an odd function of $u$,
is long and  unaesthetic  and we only give it in  footnote \ref{foot:eqn2}
\footnote{The imaginary part of the light-cone equation for
$A^\mu_\parallel$ writes
\begin{equation}
\begin{split}
& M + N n_2 =0,\cr
& M =
u \zeta c_\theta s_\theta^2 (-1 + n_1^2 s_\theta^2) + 
 \frac{1}{4 \zeta}\eta^2 u c_\theta s_\theta^2
(1 - 4 n_1^2 c_\theta^2 - 12 n_1^2 s_\theta^2)
(-1 + n_1^2 s_\theta^2),\cr
& N =
-\frac{\zeta s_\theta^4}{\eta} - \frac{1}{\zeta}
  \eta^2 (1 + u^2) s_\theta^2 (c_\theta^2 + 3 s_\theta^2)
(-1 + n_1^2 s_\theta^2)^{\frac32} + \frac{1}{4 \zeta}
(-8 \eta c_\theta^2 s_\theta^2 - 25 \eta s_\theta^4 + 
    12 \eta n_1^2 c_\theta^2 s_\theta^4 + 36 \eta n_1^2 s_\theta^6).
\end{split}
\label{eq:eqn2}
\end{equation}
\label{foot:eqn2}}.
However a rough order of magnitude can be obtained with
very drastic approximations which lead to the equation
\begin{equation}
n_2 s^2_\theta \sim u \eta
c_\theta(n_1^2 s_\theta^2-1),
\label{eq:n2app}
\end{equation}
in which, like before, we can plug in the analytical formula
(\ref{eq:solpar}) for $n_1^2$. The corresponding curves  are the dashed ones in
Figure~\ref{fig:nBcomptheta}.
The agreement with the exact curves worsens as $\alpha$ increases.

As $B$ increases, it is no longer a reliable approximation to consider
$n\in{\mathbb R}$ : absorption becomes non-negligible.
The window of medium-strong $B$'s from 1 to 20\,T together with photons
in the visible range appears therefore  quite simple and special.
Outside this window, the physics is most
probably more involved and equations much harder to solve.

\subsubsection{The ``wall'' for $\boldsymbol{A^\mu_\parallel}$}
\label{subsub:wallcomp}

The situation is best described in the complex $(n_1,n_2)$ plane of the
solutions $n=n_1+in_2$ of the light-cone equation (\ref{eq:lc4}) for
$A^\mu_\parallel$. In the limit $\eta \ll \zeta \Leftrightarrow \Upsilon
\gg 1$, and neglecting the exponential $e^{-\frac{n^2 s^2_\theta}{\Upsilon^2}}$
which plays a negligible role, it decomposes into its real and imaginary parts
according to
\begin{equation}
\begin{split}
& \ast\ 1+\frac{\alpha}{\pi}
\frac{s^2_\theta}{\zeta}\left(1+\frac{1}{4\Upsilon^2}\right)\left[\left(\Upsilon^2
-(n_1^2-n_2^2)(1+s^2_\theta)\right) \Re(V) +2n_1n_2(1+s^2_\theta)
\Im(V)\right]=0,\cr
& \ast\ -2n_1n_2 (1+s^2_\theta) \Re(V) +
\left(\Upsilon^2-(n_1^2-n_2^2)(1+s^2_\theta\right)\Im(V)=0.
\label{eq:lccomp}
\end{split}
\end{equation}

All previous calculations favoring solutions with
low absorption $|n_2| \ll n_1$, it is  in this regime that we shall
investigate the presence of a ``wall'' at small $\theta$. To this purpose,
we shall plug into the light-cone equation (\ref{eq:lc4}) for
$A^\mu_\parallel$ the expansion of the transmittance $V$ that is written in
(\ref{eq:Vexpcomp1}).

\begin{figure}[h]
\begin{center}
\includegraphics[width=6 cm, height=5 cm]{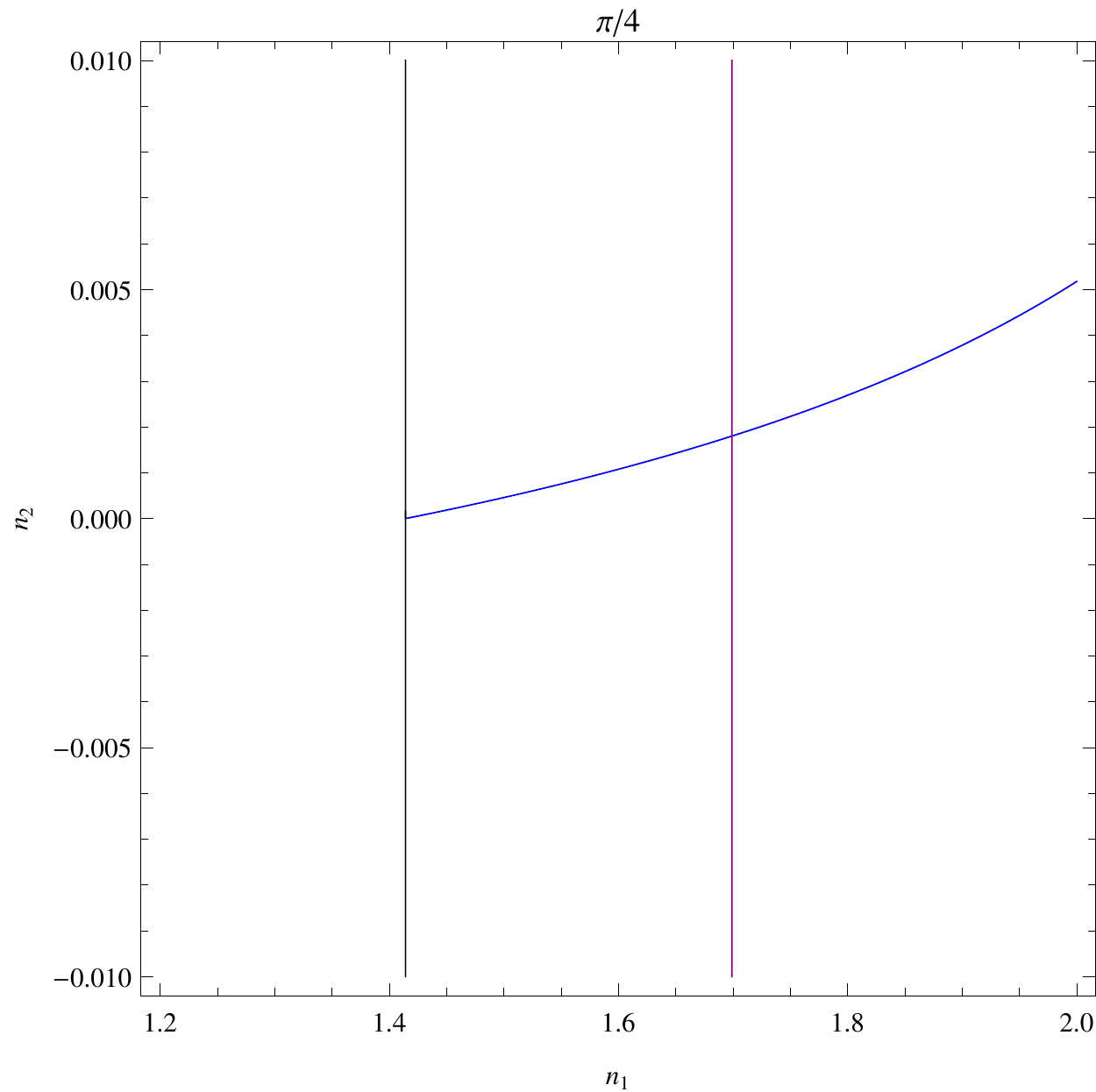}
\hskip 2cm
\includegraphics[width=6 cm, height=5 cm]{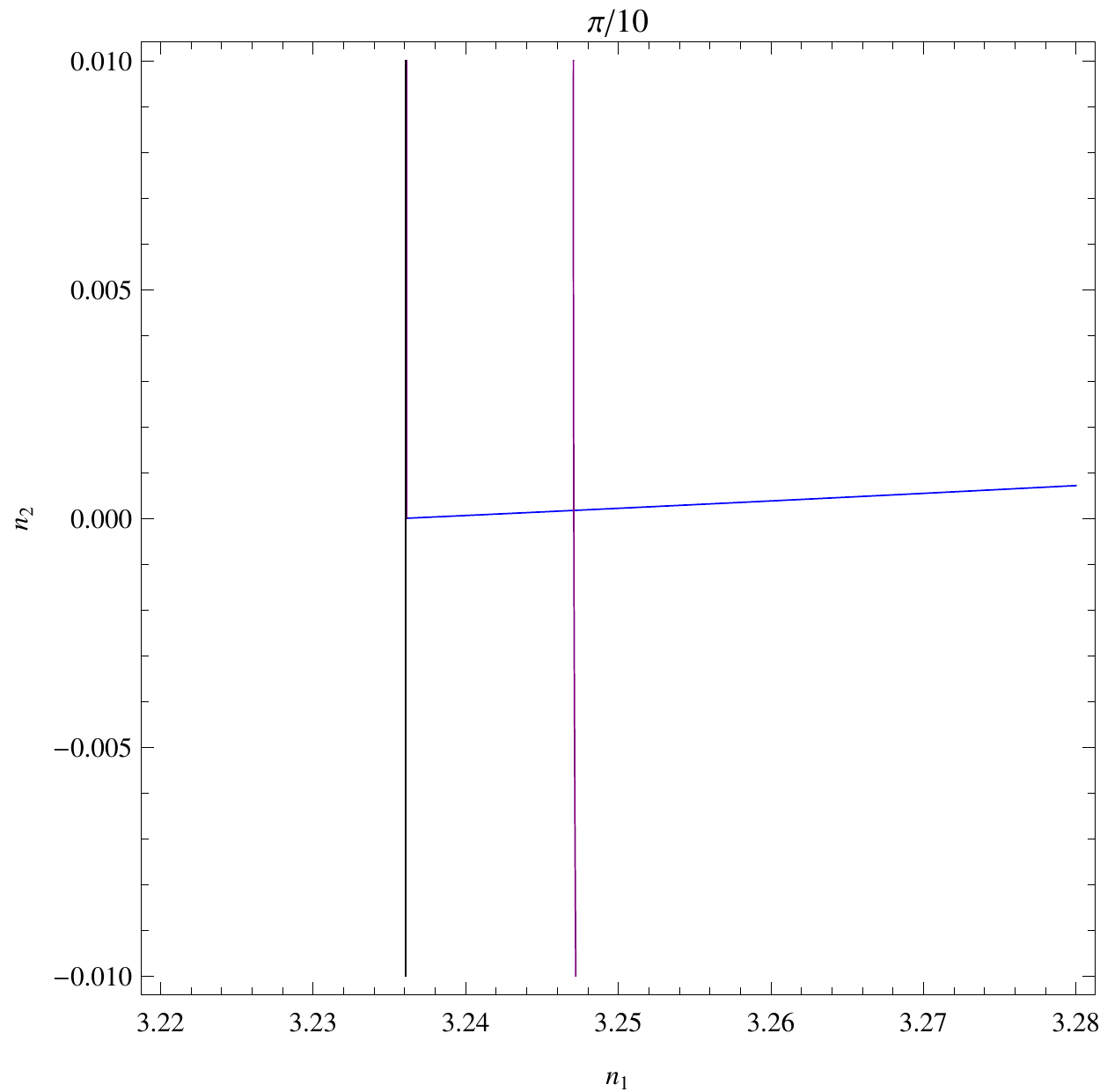}
\end{center}
\caption{The index $(n_1,n_2)$ for $A^\mu_\parallel$ at
$\theta=\frac{\pi}{4}$ (left) and $\theta=\frac{\pi}{10}$ (right).}
\label{fig:wall1}
\end{figure}

The situation at  $\theta=\frac{\pi}{4}$ (left) and $\theta=\frac{\pi}{10}$
 are depicted in Figure~\ref{fig:wall1}. The values
of the parameters are $\alpha=1, u=.5, \eta=\frac{5}{1000}, \Upsilon=5$.

The purple curve corresponds to the solutions of the real part of the
light-cone equation and the blue quasi-vertical line to the solution of its
real part. The intersection of the two curves yields the
solution $n=n_1+in_2$. We recover  $|n_2| \ll n_1$.
 The black vertical line on the left corresponds to $n_1=\frac{1}{s_\theta}$.

\begin{figure}[h]
\begin{center}
\includegraphics[width=6 cm, height=5 cm]{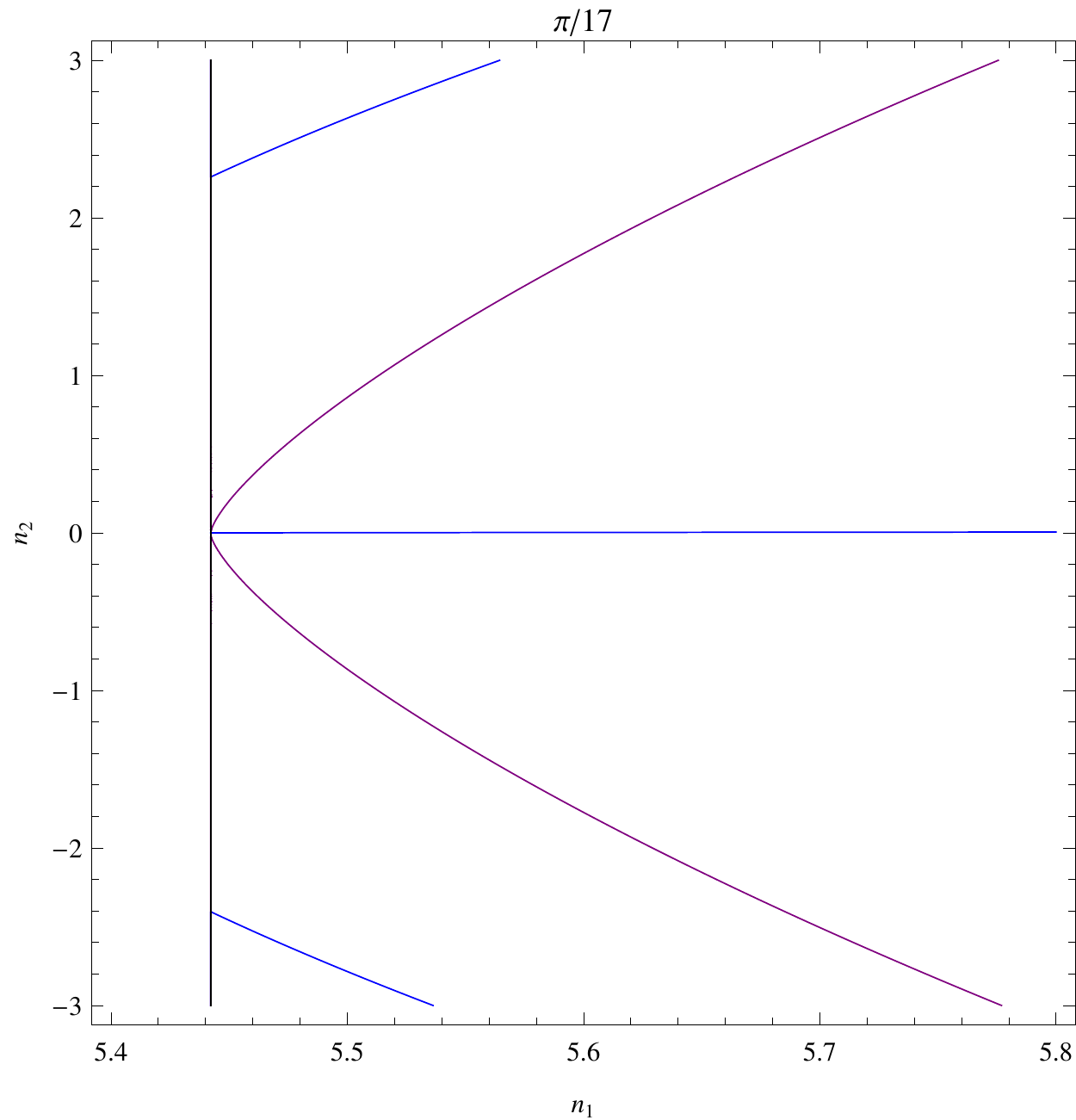}
\hskip 2cm
\includegraphics[width=6 cm, height=5 cm]{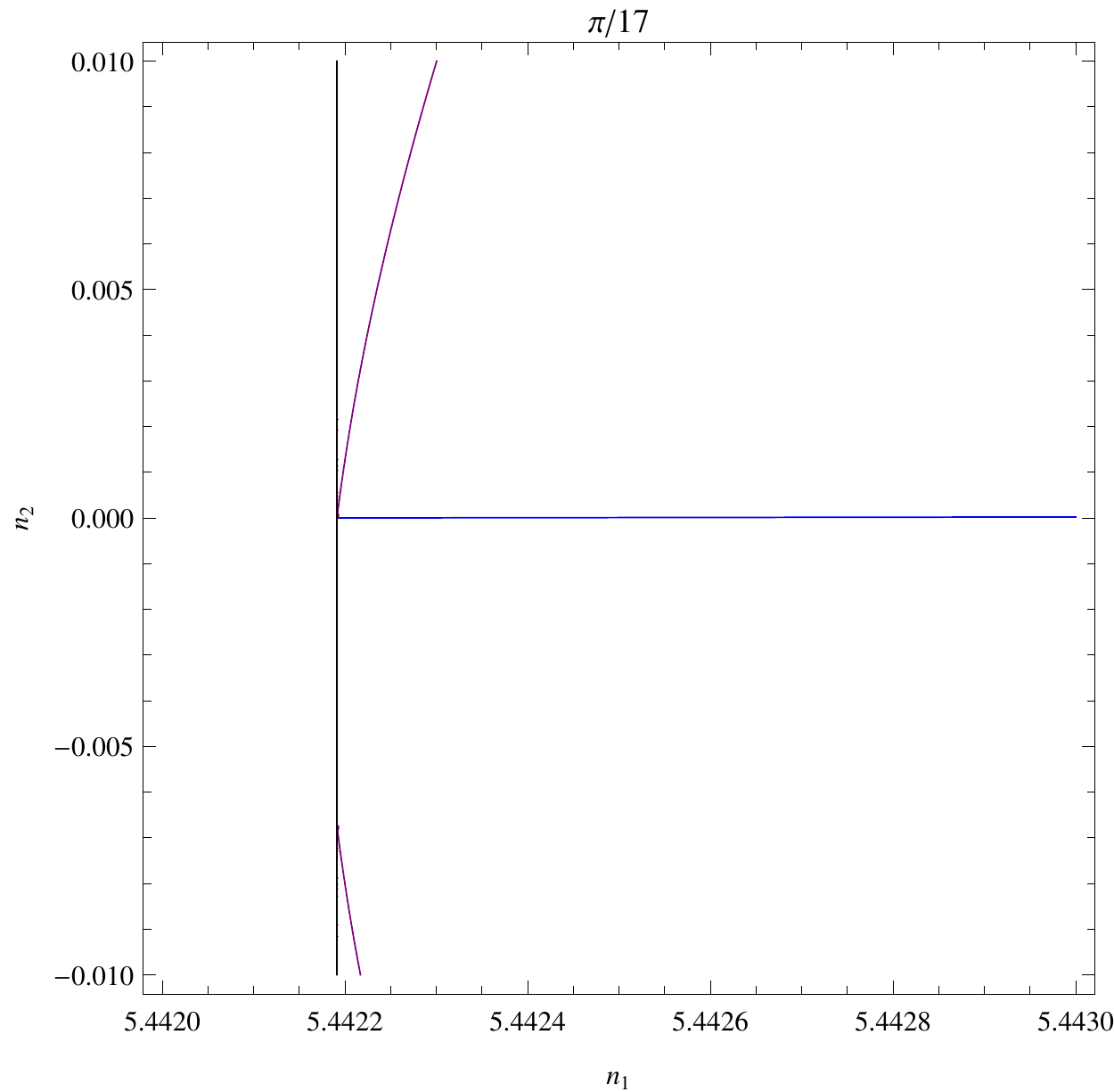}
\end{center}
\caption{The index $(n_1,n_2)$ for $A^\mu_\parallel$ at
$\theta=\frac{\pi}{17}$. The figure on the right is an enlargement of that
on the left.}
\label{fig:wall2}
\end{figure}

A transition brutally occurs close to $\theta=\frac{\pi}{14}$. Then the
solution at $|n_2| \ll n_1 ={\cal O}(1)$ disappears. It is clearly visible on
Figure~\ref{fig:wall2} below in which we plot the situation after the transition, for
$\theta=\frac{\pi}{17}$.

There is no more intersection between the solutions of the real (purple)
and imaginary (blue) parts of the light-cone equations, except at $n_2=0,
n_1=\frac{1}{s_\theta}$, which is a fake solution since we know that $n_1$
can never reach its ``asymptotic'' value $\frac{1}{s_\theta}$.

\subsubsection{An estimate of the angle of transition
$\boldsymbol{\theta_{min}}$}

This change of regime  is characterized by a brutal jump in the value of $n_2$,
which should be manifest on the imaginary part of the light-cone equation
(\ref{eq:lccomp}). A very reliable approximation can be obtained by
truncating $\Re(V)$ to its first term, in which case one gets
\begin{equation}
n_2 \approx (n_1^2 s^2_\theta-1)\frac{u\eta c_\theta\left(\Upsilon^2
-n_1^2(1+s^2_\theta)\right)}{s^2_\theta\left(\Upsilon^2
-n_1^2(1+s^2_\theta)\right)-2(1+s^2_\theta)(n_1^2 s^2_\theta-1)}
\end{equation}
which has a pole at (we use $s^2_\theta \ll 1$)
\begin{equation}
n_1^2 \approx \frac{2 + \Upsilon^2 s^2_\theta}{3s^2_\theta}.
\label{eq:n1max}
\end{equation}
This value for $n_1$ determines the maximum that can be reached when
$\theta$ decreases. Indeed, then, $n_2$ becomes out of control in the
framework of our approximations.
We also know that that $n_1$ should stay below $\frac{1}{s_\theta}$.
The intersection of (\ref{eq:n1max}) and
$\frac{1}{s_\theta}$ yields the lower limit for $\theta$
\begin{equation}
\theta_{min} \sim \frac{1}{\Upsilon}.
\label{eq:thetamin}
\end{equation}
(\ref{eq:thetamin}) is smaller than our previous estimate
(\ref{eq:thetalim}) obtained in the approximation $n\in{\mathbb R}$.

At $\Upsilon=5$ one gets $\theta_{min} \approx \frac{\pi}{15}$, which shows
the reliability of our estimate (the true transition numerically
 occurs between $\frac{\pi}{14}$ and $\frac{\pi}{15}$).

\subsubsection{The case of $\boldsymbol{A^\mu_\perp}$}

We only summarize below the steps that lead to the conclusion that no solution
to the refractive index except the trivial $n=1$ exists for the transverse
polarization.

Starting from the corresponding light-cone equation in (\ref{eq:lc4}), the
main task is to get the appropriate expression for the transmittance
function $V$.
To this purpose the starting point is the general expression
(\ref{eq:transmit}). We expand it in powers of $\eta$ in the sense that the
exponentials are expanded at ${\cal O}(\eta)$ or, eventually ${\cal
O}(\eta^2)$. No expansion in powers of $n_2$ is done because, if solutions
exist, they may occur for fairly larges values of $n_2$ (and $n_1$).

Since the sign of the imaginary parts of the poles $\sigma_1$ and
$\sigma_2$ obviously play a central role, it is also useful to extract
($c$ should not be confused here with the speed of light)
\begin{equation}
\begin{split}
\Im(\sigma_1) &= \eta \left(
-n_2 c_\theta + \frac{1}{\sqrt{2}}\sqrt{-c + \sqrt{c^2+d^2}}
\right),\cr
\Im(\sigma_2) &= \eta \left(
-n_2 c_\theta - \frac{1}{\sqrt{2}}\sqrt{-c + \sqrt{c^2+d^2}}
\right),\cr
 c &= 1-(n_1^2-n_2^2) s^2_\theta,\quad d= 2n_1n_2 s^2_\theta.
\end{split}
\label{eq:impoles}
\end{equation}

Straightforward manipulations on (\ref{eq:transmit}) show that:

* when $n_2 > 0$ ($\Rightarrow \Im(\sigma_2) <0$):
 if $\Im(\sigma_1)>0$, $V= \frac{-i\pi\eta}{\sqrt{1-n^2
s^2_\theta}}+\ldots$;
if $\Im(\sigma_1)<0$, $V= \frac{\pi\eta^2}{2}(1-u)^2+\ldots$\newline
* when $n_2 <0$  ($\Rightarrow \Im(\sigma_1) > 0$):
if $\Im(\sigma_2)>0$, $V=\frac{\pi \eta^2}{2} (1+u)^2+\ldots$;
if $\Im(\sigma_2) < 0$, $V =
\frac{-i\pi\eta}{\sqrt{1-n^2s^2_\theta}}+\ldots$

The cases when $V={\cal O}(\eta^2)$ correspond to $\sigma_1$ and $\sigma_2$
being in the same 1/2 complex $\sigma$-plane.

When $V=  \frac{-i\pi\eta}{\sqrt{1-n^2 s^2_\theta}}$,
its real and imaginary parts are given by
\begin{equation}
\Re(V) = \frac{\pi\,\eta\, n_1n_2\, s^2_\theta}{\sqrt{2}}
\frac{\sqrt{c + \sqrt{c^2+d^2}}}{\sqrt{c^2 + d^2}},\qquad
\Im(V) = \frac{-\pi\,\eta}{\sqrt{2}} \frac{\sqrt{-c+\sqrt{c^2+d^2}}}
{\sqrt{c^2+d^2}}.
\end{equation}
Numerical solutions of the light-cone equation show that no solution exists
that fulfill the appropriate criteria on the signs of $\Im(\sigma_1),
\Im(\sigma_2)$. For example, for $n_2<0$, one gets  solutions shared by both
the real and imaginary parts of the light-cone equations, but they satisfy
$\Im(\sigma_2)>0$ and must therefore be rejected.

The next step is to use the exact expression (\ref{eq:transmit})  of $V$,
but no acceptable solution exists (solutions with very large values of
$n_1$ and $n_2$, larger than $20$, are a priori rejected).

\subsection{There is no non-trivial solution $\boldsymbol{n\in{\mathbb R}
<\frac{1}{s_\theta}}$
 or $\boldsymbol{n=n_1+in_2,n_1 <\frac{1}{s_\theta}$ for $A^\mu_\parallel}$}
\label{subsec:nosolinf}

For $n\in{\mathbb R} <\frac{1}{s_\theta}$ the two poles $\sigma_1,\sigma_2$
of $V$ given in (\ref{eq:poles}) become real. One then defines $V$ as a
Cauchy integral, tantamount to setting $\Theta(0)=\frac12$ in
(\ref{eq:transmit}). One gets then
\begin{equation}
V \stackrel{\sigma_1,\sigma_2 \in{\mathbb R}}{=}
 -\frac{\pi\,\eta^2}{\sigma_1\sigma_2(\sigma_1-\sigma_2)}\left[
(\sigma_1-\sigma_2)+\sigma_2\,e^{i\sigma_1 u}\cos\sigma_1
-\sigma_1\, e^{i\sigma_2 u}\cos\sigma_2 \right].
\label{eq:Vrealpoles}
\end{equation}
No solution is then found to the light-cone equation (\ref{eq:lc4}).

Likewise, careful numerical investigations show that no complex solution
$n=n_1+in_2$ to this equation exists for $n_1 <\frac{1}{s_\theta}$.

\subsection{Conjectural interpretation in terms of electron spin resonance}
\label{subsec:reson}

The modified Maxwell Lagrangian that we used in subsection \ref{subsec:lightcone}
describes the interaction inside graphene between  electrons and an
electromagnetic wave in the presence of a constant uniform external magnetic field. 
We have shown that the effects on the refractive index only concern
$A^\mu_\parallel$, that is, the so-called ``transverse magnetic''
polarization in which the oscillating magnetic field $b$ is transverse to
the plane of incidence, therefore perpendicular to  $B$.
This is a typical situation for electron spin resonance
(a linearly polarized electromagnetic wave
can be decomposed into two opposite circular polarized waves and
 only one can trigger the resonance depending of the electron spin  $+\frac12$ or
$-\frac12$)
\footnote{The possibility of anisotropic electron spin resonance was
already evoked in the pioneering work \cite{Semenoff} in a different setup
in which the angle that varies is the one between $B$ and the surface of
graphene.}.

This phenomenon takes place when  the angular speed $\omega=2\pi \nu$
of the photon  ($\nu$ is its frequency, which lies, for the visible
spectrum, in the interval $[4.3\,10^{14}\,Hz, 7.9\,10^{14}\,Hz]$)
 matches the Larmor  speed of precession of
the magnetic moment of the electron $\frac{eB}{m^\ast}$, therefore if
the electromagnetic wave ``sees'' an electron with effective mass 
$m^\ast = \frac{eB}{2\pi \nu}= \frac{\hbar eB}{q_0}$.
Such a phenomenological formula, in which $m^\ast \propto eB$, is more
reminiscent of magnetic catalysis in 2+1 dimension (see for example
\cite{Shovkovy}) than of the one explored for example in  \cite{Gorbar} in
``reduced'' $QED_{4,3}$ in which $m^\ast$ is expected to be proportional
to $\sqrt{\hbar eB v_F^2}(/c^2)$. 
Now, even in the absence of $B$, chiral symmetry breaking can also occur,
for $\alpha > \alpha_c$, through a  modification of the Coulomb potential
by polarization effects \cite{Gamayun}.
In any case, the conjectural sequence, that of course needs to be put on
firmer grounds, is that the electrons of the 
virtual $e^+ e^-$ pairs acquire a small mass and resonate by
the action of the two orthogonal magnetic fields.
In this picture, the light beam plays the dual role of the trigger (via the
oscillating $b$) and the probe (via the refractive index) of the resonance.

For $\nu = 6\,10^{14}\,Hz$ and $B=20\,T$ one gets $m^\ast \approx
\frac{m_e}{1000}$, much smaller than the cyclotron mass
 $m_c=\frac{\sqrt{\hbar eB}}{v_F\sqrt{2}}\approx .014\,m_e$ evaluated at
$v_F=\frac{c}{300}$. This comforts the choices that we made at the
start, to consider graphene-born electrons  at $m \approx 0$,
and to write their propagator with
$c \vec p.\vec\gamma$ instead of $v_F \vec p. \vec\gamma$: 
the average time $t_g \sim \frac{a^2 m^\ast}{\hbar}$ that they spend
 inside the medium gets still much smaller than when evaluated
with the cyclotron mass $m_c$ as in subsection \ref{subsub:geneprop}.

\section{The case $\boldsymbol{B=0}$}\label{section:noB}

Studying this limiting case shows that, in the absence of $B$, 
the optical properties of graphene are essentially controlled by the
sole transmittance $V$. In the (narrow) domains where
the approximations of the calculations can presumably
be trusted, no large effect seems to occur, which can be interpreted
as the absence of any resonant phenomenon.
Paradoxically, the perturbative series looks more
difficult to handle than in the presence of $B$.

\subsection{The tensor $\boldsymbol{T^{\mu\nu}_{\xcancel B}(\hat q)}$}

Like for $B\not=0$, the electrons created inside graphene are constrained
to have a vanishing momentum along ``$z$'' and a vanishing mass.
After the traces of Dirac matrices have been done, unlike in the presence
of $B$, the integration over the transverse degrees of freedom cannot be
factorized and done separately. One has to introduce a Feynman parameter
$x$ to
combine the denominators. All calculations can be done exactly (no
expansion is performed), and give
\begin{equation}
\begin{split}
& iT^{11}_{\xcancel B}(\hat q) = i\frac{e^2}{8}\left(
\sqrt{\hat q_E^2}-\frac{q_1^2}{\sqrt{\hat q_E^2}}\right),\quad
iT^{22}_{\xcancel B}(\hat q) = i\frac{e^2}{8}\left(
\sqrt{\hat q_E^2}-\frac{q_2^2}{\sqrt{\hat q_E^2}}\right),\cr
& iT^{33}_{\xcancel B}(\hat q) = i\frac{e^2}{4} \sqrt{\hat q_E^2},\quad
iT^{00}_{\xcancel B}(\hat q) = -i\frac{e^2}{8} \frac{q_1^2+q_2^2}{\sqrt{\hat
q_E^2}},\cr
& iT^{12}_{\xcancel B}(\hat q) = -i\frac{e^2}{8} \frac{q_1q_2}{\sqrt{\hat
q_E^2}},\quad
iT^{01}_{\xcancel B}(\hat q) = -\frac{e^2}{8} \frac{q_0^E q_1}{\sqrt{\hat
q_E^2}},\quad
iT^{02}_{\xcancel B}(\hat q) = -\frac{e^2}{8} \frac{q_0^E q_2}{\sqrt{\hat
q_E^2}},\cr
& T^{03}_{\xcancel B}(\hat q) =T^{13}_{\xcancel
B}(\hat q)=T^{23}_{\xcancel B}(\hat q)=0.
\end{split}
\label{eq:PinoB}
\end{equation}
in which $q_0=iq_0^E$ and $(\hat q^E)^2=(q_0^E)^2 + q_1^2 + q_2^2$.
We recall that, in our setup, $q_2=0$, therefore $n_y=\frac{q_2}{q_0}=0$.
This gives $\sqrt{\hat q_E^2} =   q_0\sqrt{n^2s^2_\theta-1}$.

$\star$\ $T^{ij}_{\xcancel B}$ is proportional to $\pi \alpha$
while, in the presence of $B$, it was proportional to $\alpha$. The extra
$\pi$ comes from $\int_0^1 dx\;\sqrt{x(1-x)}= \frac{\pi}{8}$.

$\star$\ One  checks on (\ref{eq:PinoB}) that, like in the presence of
$B$, $T^{\mu\nu}$ is not transverse. The same remarks apply here, in
particular that $T^{\mu\nu}$ is only an intermediate step in the
calculation of the vacuum polarization $\Pi^{\mu\nu}$.
$q_0 T^{00}_{\xcancel B}+q_1 T^{10}_{\xcancel B}+q_2 T^{20}_{\xcancel B}
+q_3 T^{30}_{\xcancel B}= 0$,
$q_0 T^{01}_{\xcancel B}+q_1 T^{11}_{\xcancel B}+q_2 T^{21}_{\xcancel B}
+q_3 T^{31}_{\xcancel B}=0$,
$q_0 T^{02}_{\xcancel B}+q_1 T^{12}_{\xcancel B}+q_2 T^{22}_{\xcancel B}
+q_3 T^{32}_{\xcancel B}=0$. The last condition
$q_0 T^{03}_{\xcancel B} +q_1 T^{13}_{\xcancel B}+q_2 T^{23}_{\xcancel
B} +q_3 T^{33}_{\xcancel B}=0$
reduces to $q_3 T^{33}_{\xcancel B}=0$, which is not satisfied unless
$q_3=0$.

Eqs.~(\ref{eq:PinoB}) also write
\begin{equation}
\begin{split}
T^{11}_{\xcancel B} &= - \frac{\pi\,\alpha}{2} q_0
\frac{1}{\sqrt{n^2 s^2_\theta-1}},\quad
T^{22}_{\xcancel B} = + \frac{\pi\,\alpha}{2} q_0
\sqrt{n^2s^2_\theta-1},\cr
T^{33}_{\xcancel B} &= + \pi\,\alpha\, q_0
\sqrt{n^2s^2_\theta-1},\quad
T^{00}_{\xcancel B} = - \frac{\pi\,\alpha}{2} q_0
\frac{n^2s^2_\theta}{\sqrt{n^2s^2_\theta-1}},\cr
T^{12}_{\xcancel B} &=0,\quad T^{i3}_{\xcancel B}=0,\cr
T^{01}_{\xcancel B} &= + \frac{\pi \alpha}{2} q_1
\frac{1}{\sqrt{n^2 s^2_\theta-1}},\quad
T^{02}_{\xcancel B} =0.
\end{split}
\label{eq:PinoB2}
\end{equation}

\subsection{The light-cone equations and the refractive index}

The plane of incidence, defined by
$\vec q$ and the direction perpendicular to the graphene surface, is the
same as in the presence of $B$. Hence, we keep the distinction
between the two polarizations
$\epsilon_\parallel$ (transverse magnetic) and $\epsilon_\perp$ (transverse
electric).
The light-cone equations (\ref{eq:lcgeneral}) together with (\ref{eq:PinoB2})
  yield
\begin{equation}
\begin{split}
& \star\ for\ A^\mu_\perp: (1-n^2)\left[ 1+
\frac{\alpha}{2\eta}\sqrt{n^2s^2_\theta-1}
\;\frac{V(u,\rho,n,\theta,\eta)}{\pi}
\right]=0,\cr
&  \star\ for\ A^\mu_\parallel:(1-n^2)\left[1+ \frac{\alpha}{\eta}
\Big(-\frac{c^2_\theta}{2\sqrt{n^2s^2_\theta-1}} +
s^2_\theta\sqrt{n^2s^2_\theta-1}\Big)
\;\frac{V(u,\rho,n,\theta,\eta)}{\pi}
\right]=0.
\end{split}
\label{eq:lcnoB}
\end{equation}
in which $V$ is the same transmittance function as before, given by
(\ref{eq:UVdef}) and (\ref{eq:transmit}).

\subsection{Solutions for $\boldsymbol{A^\mu_\parallel}$
 with $\boldsymbol{n\in{\mathbb R}}$}

Like in the presence of $B$, no non-trivial ``reasonable''
 solution exists for the transverse polarization.
The  difference is, however, that this absence of non-trivial solution
for $A^\mu_\perp$ is not due here to $|T^{33}| \gg |T^{11}|, |T^{22}|$ and
cannot be related {\em a priori} to any dimensional reduction.
We therefore focus hereafter on $A^\mu_\parallel$.

\subsubsection{No solution $\boldsymbol{n< \frac{1}{s_\theta}}$}

In this case the two poles of $V$ given in (\ref{eq:poles}) are real and one
again defines $V$ as a Cauchy integral, which yields (\ref{eq:Vrealpoles}).
In addition to numerical calculations, a simple argument, which uses the
very weak dependence on $u$, shows that no solution exists.
Up to corrections in odd powers of $(\sigma_1 u)$ and $(\sigma_2 u)$, which
vanish at $u=0$, $V$ is a purely real function. The light-cone equation
(\ref{eq:lcnoB}) for $A^\mu_\parallel$ is therefore of the form
$(1-n^2)\left[1\pm i\frac{\alpha}{\eta}*(real\ number) \right]=0$, which
has no non-trivial solution.

\subsubsection{Solutions $\boldsymbol{n> \frac{1}{s_\theta}}$}

\begin{figure}[h]
\begin{center}
\includegraphics[width=6 cm, height=5 cm]{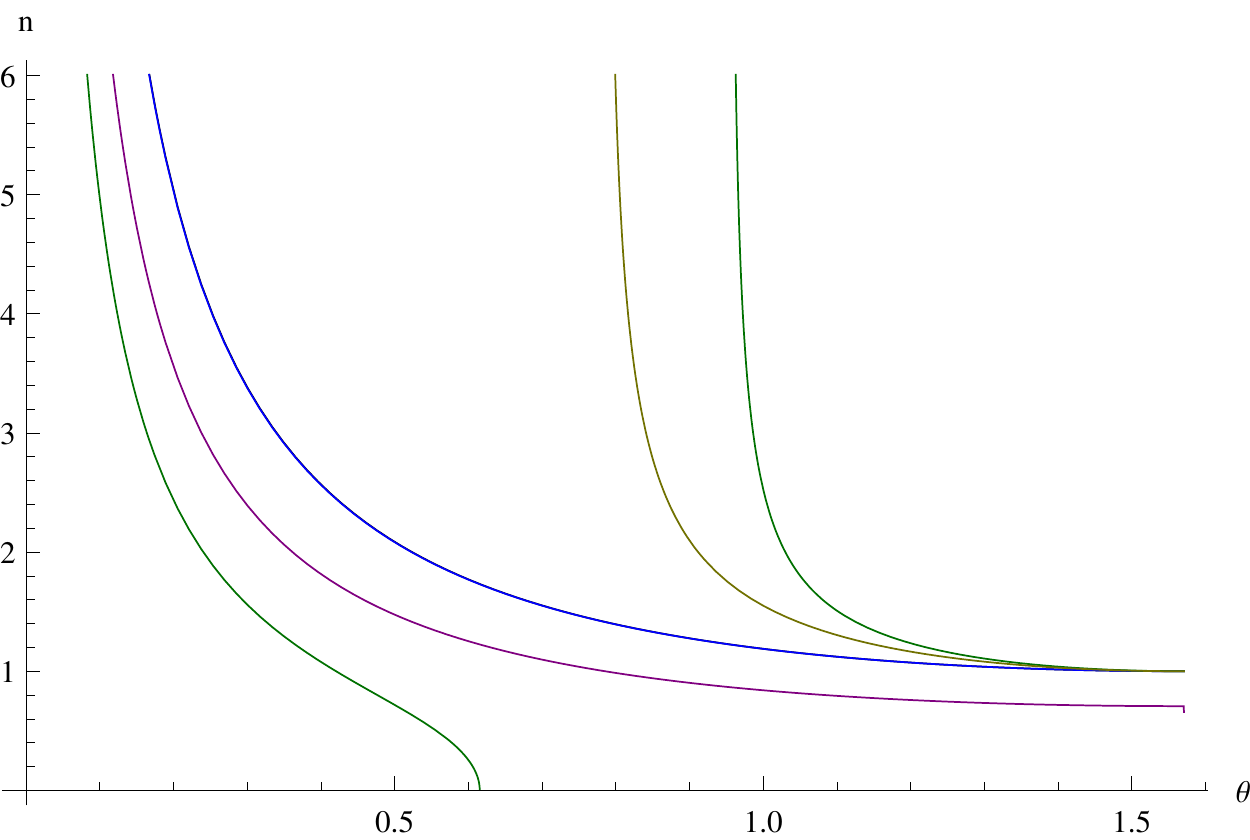}
\end{center}
\caption{The real solution of the light-cone equation
(\ref{eq:lcnoB}) for $A^\mu_\parallel$
 as a function of $\theta$ when no external $B$ is
present.  We vary $\alpha =
1/137\;(blue)$, $1\;(purple)$, $1.5\;(green)$, $2\;(yellow)$. The  black
($\simeq$ blue) curve is $1/\sin\theta$.}
\label{fig:nnoBreal}
\end{figure}

We approximate, at $\eta \ll 1$, according to
(\ref{eq:Vexp}), $V \approx  -\frac{\eta\pi}{\sqrt{n^2s^2_\theta -1}}$.
The corresponding light-cone equation writes
\begin{equation}
1+\alpha \left(\frac{c_\theta^2}{2(n^2s_\theta^2-1)}- s^2_\theta
\right)=0,
\label{eq:lcnoBpar}
\end{equation}
the solution of which is
\begin{equation}
\begin{split}
n^2=\frac{1}{s^2_\theta}\left(1-\frac{\alpha c^2_\theta}{2(1-\alpha
s^2_\theta)}\right).
\end{split}
\label{eq:nnoB}
\end{equation}
It is plotted on Figure~\ref{fig:nnoBreal}.
It only depends on $\alpha$ and we plot it for $\alpha=\frac{1}{137}$ (blue),
$\alpha=1$ (purple) and $\alpha=1.5$ (green), $\alpha=2$ (yellow) together with
$n=\frac{1}{s_\theta}$ (black), the latter being in practice
indistinguishable from $\alpha=\frac{1}{137}$.

It is conspicuous that one cannot trust the results 
when $\theta \to 0$ because they diverge. Furthermore,
all curves below $\frac{1}{s_\theta}$ are to be rejected since we have
shown that no such solution can exist. Last, one notes the presence of a pole at
$s_\theta^2 = \frac{1}{\alpha}$ for $\alpha >1$.

All these restrictions make the approximation of considering $n\in {\mathbb R}$
obviously very hazardous.
This is why we shall perform in subsection \ref{subsec:compnoB}
a detailed study with $n\in {\mathbb C}$.

\subsection{Solutions for $\boldsymbol{A^\mu_\parallel}$
with $\boldsymbol{n\in{\mathbb C}}$}
\label{subsec:compnoB}

\subsubsection{There is no
 solution with  $\boldsymbol{n_1<\frac{1}{s_\theta}}$}
\label{subsub:limtetabig}

When supposing  $n\in {\mathbb R}$, we have seen that the solution with $n<
\frac{1}{s_\theta}$ was unstable, in particular above $\theta_{max}$ such
that $(\sin\theta_{max})^2 =\frac{1}{\alpha}$ were it did not exist
anymore.

Careful investigations for $n\in {\mathbb C}$ show that, like in the
presence of $B$, no solution with $n_1 <\frac{1}{s_\theta}$ exists
\footnote{For $n_1<\frac{1}{s_\theta}$
the expansion of the transmittance $V$ at small $\eta$ and $n_2$ is no
longer given by  (\ref{eq:Vexpcomp1}) but  writes
\begin{equation}
\begin{split}
\star\ \frac{1}{\pi}\Re(V) &= \frac{u n_1 c_\theta }{\sqrt{1-n_1^2
s^2_\theta}}\eta^2 + \frac12 (1+u^2)\eta^2 -\frac{n_1 s_\theta^2}{(1-n_1^2
s^2_\theta)^{\frac32}}\eta n_2 + \ldots\cr
\star\ \frac{1}{\pi}\Im(V) &=
-\frac{\eta}{\sqrt{1-n_1^2s^2_\theta}}+\frac{uc_\theta(2n_1^2
s^2_\theta-1)}{(1-n_1^2 s^2_\theta)^{\frac32}}\eta^2 n_2 +\ldots
\end{split}
\label{eq:Vcomp2}
\end{equation}
that we plug into the light-cone equation (\ref{eq:lcnoB}) for
$A^\mu_\parallel$.}.

\subsubsection{The solution with  $\boldsymbol{n_1>\frac{1}{s_\theta}}$}
\label{subsub:limtetanul2}

In the presence of an external $B$, we have seen that the solution with a
quasi-real index suddenly disappears below an angle $\theta_{min}\approx
\frac{1}{\Upsilon}$. In the present case with no external $B$, there is no
$\theta_{min}$ but the index becomes ``more and more complex'' (that is the
ratio of its imaginary and real parts increase) when $\theta$ becomes
smaller and smaller.

To demonstrate this, we study the light-cone equation (\ref{eq:lcnoB}) for
$A^\mu_\parallel$ with $n=n_1 + in_2,\ n_1,n_2 \in{\mathbb R}$. For
practical reasons, we shall limit ourselves to the expansion of $V$ at
small $\eta$ and $n_2$, valid when the two poles of $V$ lie in different 1/2
planes, given in (\ref{eq:Vexpcomp1}). 

The results are displayed in Figure~\ref{fig:nnoBcomp} below, for $\alpha=1$ (blue),
$\alpha=1.5$ (purple) and $\alpha=2$ (green). The values of $n_1$ are
plotted on the left and the ones of $n_2$ on the right. The value of the
other parameters are $u=.5,\eta=\frac{5}{1000}$.
For $\alpha=\frac{1}{137}$, $n_1$ is indistinguishable from
$\frac{1}{s_\theta}$.

\begin{figure}[h]
\begin{center}
\includegraphics[width=6 cm, height=5 cm]{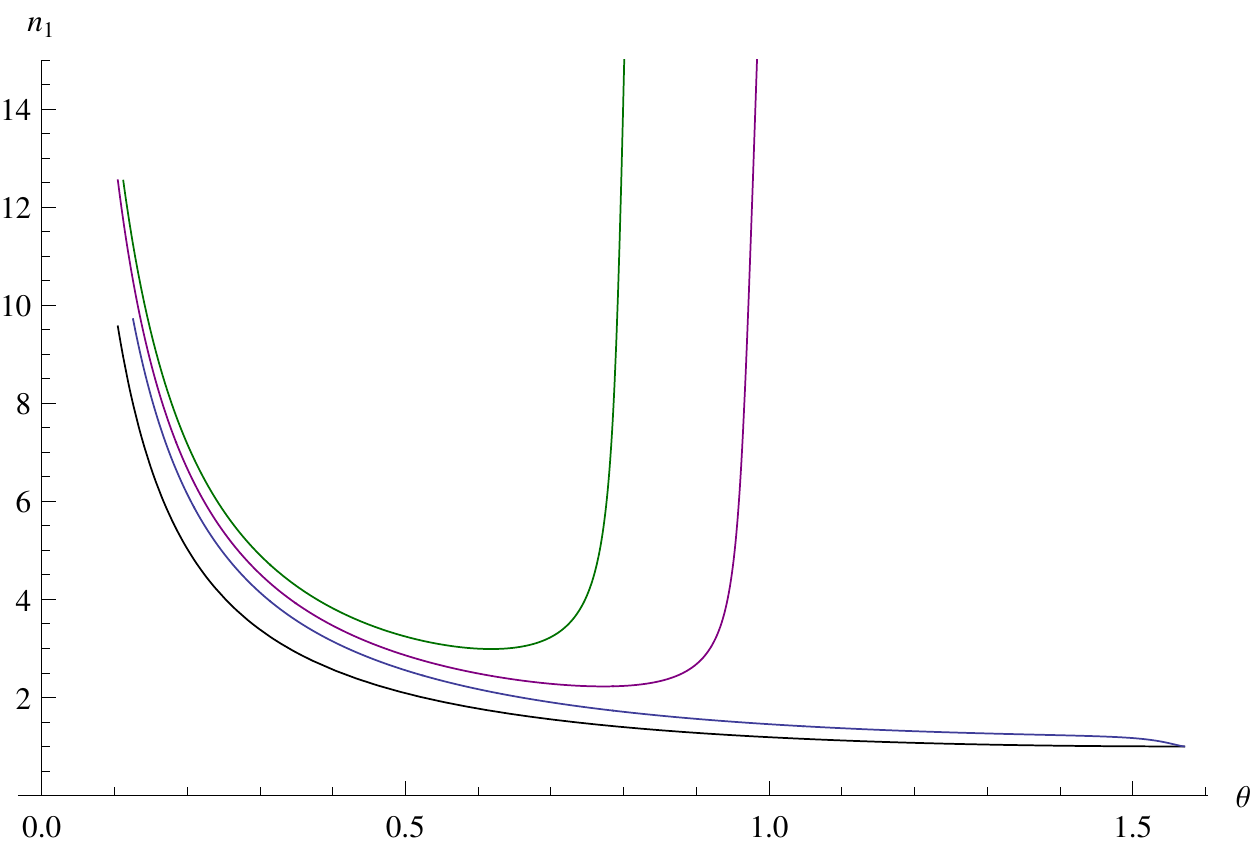}
\hskip 2cm
\includegraphics[width=6 cm, height=5 cm]{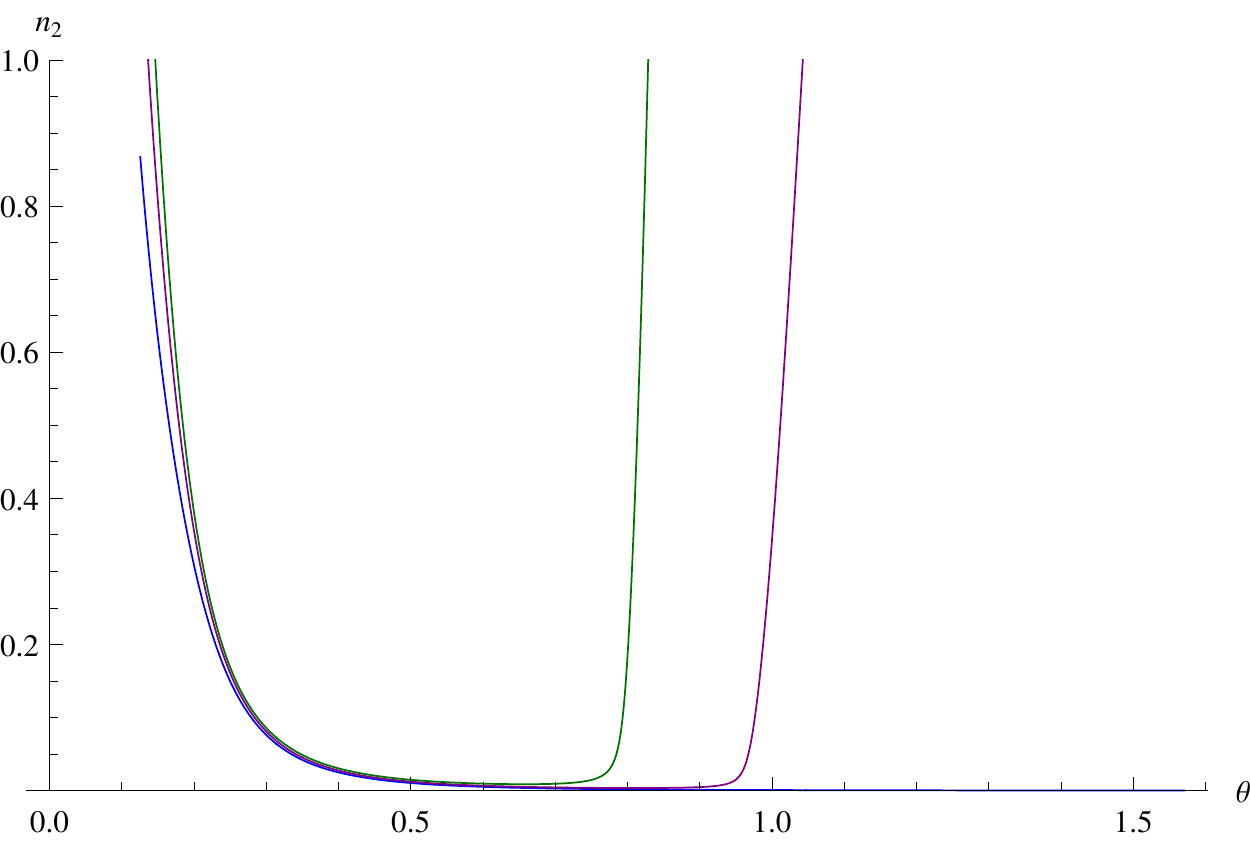}
\end{center}
\caption{The real part $n_1$ (left) and imaginary part $n_2$ (right) of the
solution $n$ of the light-cone equation (\ref{eq:lcnoB}) for
$A^\mu_\parallel$
in the absence of external $B$. The blue curves correspond to $\alpha=1$,
the purple curves to $\alpha=1.5$ and the green curves to $\alpha=2$. The
black curve on the left is $n_1=\frac{1}{s_\theta}$.}
\label{fig:nnoBcomp}
\end{figure}

$\bullet$\ As $\theta$ gets smaller and smaller, the index
becomes complex with larger and larger values of both its components.
It is of course bounded as before to $|n| < \frac{1}{\eta}$ by quantum
considerations. The brutal transition at
 $\theta_{min} \simeq \frac{1}{\Upsilon}$  is replaced by a smooth
transition (which could be anticipated since, in the absence of $B$,
 the parameter $\Upsilon$ does not exist).

$\bullet$\ A divergence occurs at large
$\theta$ for $\alpha > 1$,
 obviously reminiscent of the one that occurred in the
approximation $n\in {\mathbb R}$ at $\theta = \theta_{max},
(\sin\theta_{max})^2 = \frac{1}{\alpha}$ for the solution $n<
\frac{1}{s_\theta}$ (we had noticed that this condition could no
longer be satisfied since, for $s^2_\theta > \frac{1}{\alpha}$, $n$ could
only be larger than $\frac{1}{s_\theta}$).

Three explanations come to the mind concerning this singularity.
 The first is that,
for large values of $n_2$, the expansion (\ref{eq:Vexpcomp1})
 that we used for $V$ is no longer valid; however, using the exact expression for
the transmittance leads to the same conclusion. The second,
 and also very likely one,
is  that the perturbative series becomes very hazardous for
$\alpha >1$ \cite{BarnesSarma}
(2-loop corrections become larger than 1-loop etc ); using
a 2-loop calculation of the vacuum polarization without external $B$
 seems feasible but also goes beyond the scope of this work.
The third is that this divergence is the sign that some
 physical phenomenon occurs, like total reflexion,  for $\theta > \theta_{max}$,
which can only be settled by experiment. It is also known \cite{Gamayun} that chiral
symmetry breaking can occur for $\alpha > \alpha_c$.

$\bullet$\ These calculations show in which domain the approximation $n \in
{\mathbb R}$ is reliable since it requires $n_2 \ll 1$: for example $n_2 <
.1$ needs $.3 \leq \theta \leq \theta_{max}$, which leaves (except for
$\alpha \leq 1$ in which case $\theta_{max} \geq \frac{\pi}{2}$) only a
small domain for $\theta$.

\paragraph{$\bullet$\ A very weak dependence on $\boldsymbol \alpha$ for
$\boldsymbol{\theta < \theta_{max}}$} 

{\ }

In the absence of external $B$ and away from the ``wall'' at large $\theta$,
 the index is seen to depend very little on
$\alpha$.
The dependence of $n$ on $\theta$ is practically only due to the
transmittance function $V$ and to the confinement of electrons inside
graphene. Notice in particular that, when $\alpha=\frac{1}{137} \ll 1$, the
curve is indistinguishable from that of $\frac{1}{s_\theta}$.

The fairly large dependence on $\alpha$  that we uncovered
in the presence of $B$ is therefore triggered by $B$ itself.

\paragraph{$\bullet$\ The dependence on the energy of the photon}

\begin{figure}[h]
\begin{center}
\includegraphics[width=6 cm, height=5 cm]{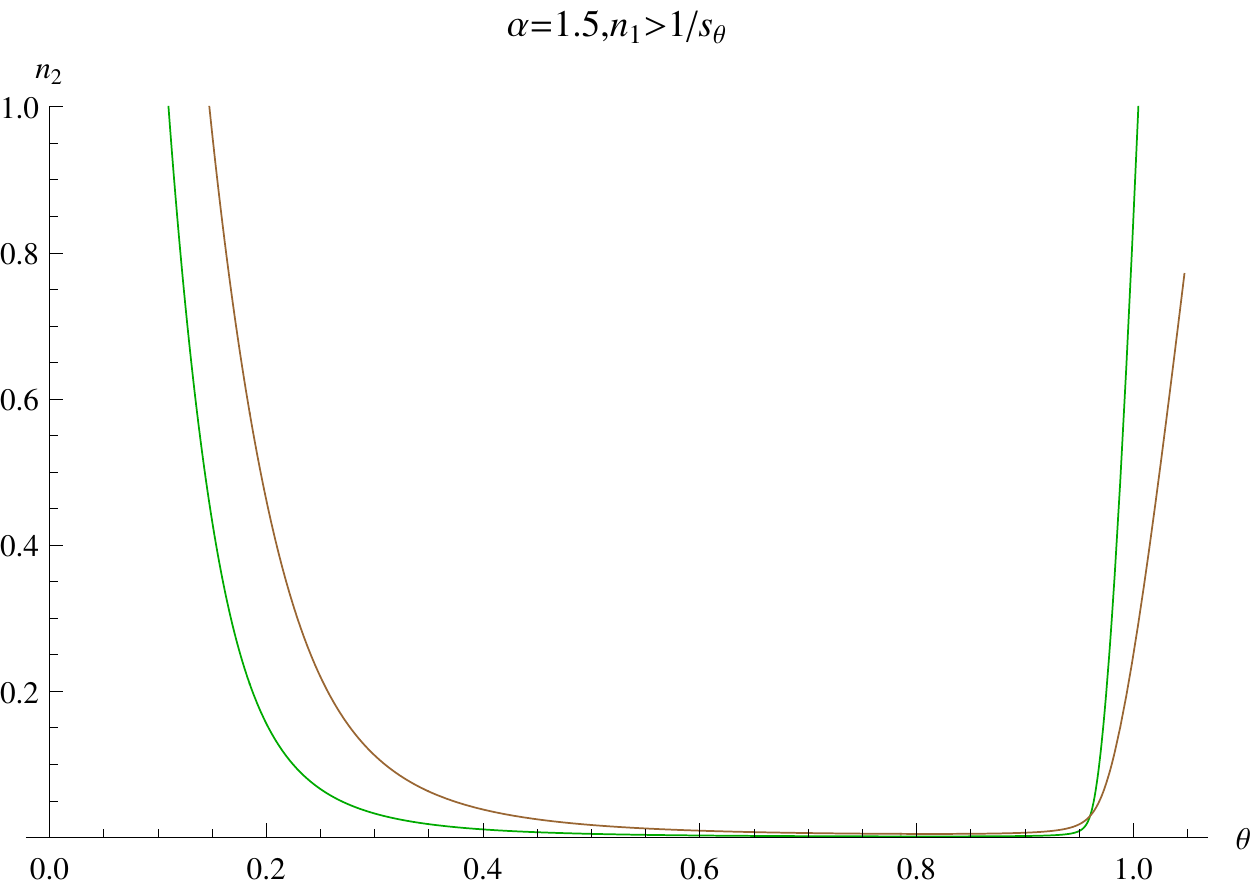}
\end{center}
\caption{$n_2$ as a function of $\theta$ for $\eta=\frac{2}{1000}$
(green) and $\eta=\frac{7}{1000}$ (brown), in the case $\alpha=1.5$.}
\label{fig:n2noB}
\end{figure}

{\ }
The dependence on $\eta$ only occurs in the imaginary part $n_2$ of $n$.
This is shown in Figure~\ref{fig:n2noB}, in which we vary $\eta$ in the visible spectrum,
$\eta \in [\frac{2}{1000}, \frac{7}{1000}]$ at $\alpha=1.5$ (unlike in
Figure~\ref{fig:nnoBcomp}, $\theta$ has not been extended above $\theta_{max}$).

\subsection{The limit of very small $\boldsymbol \theta$ ; absorption of
visible light and  experimental opacity}

\subsubsection{At small $\boldsymbol \theta$}

Since absorption of visible light by graphene
 at close to normal incidence has been measured \cite{Nair},
let us show that our simple model
gives predictions that are compatible with these measurements.
To that purpose, we calculated numerically the index $n$ at
the lowest value of $\theta$ at which  the
2 poles of $V$  lie in different 1/2 planes. We did not make any expansion
for $V$ (the price to pay is of course that no analytical expression is
available)  and obtained
\begin{equation}
\begin{split}
\ast\ & for\ \alpha=1\  and\ \theta = \frac{\pi}{105.9} : n= 41.20 +
.7\times i,\cr
\ast\ & for\ \alpha=2\ and\ \theta = \frac{\pi}{89} : n = 40 + 1 \times
i.
\end{split}
\end{equation}
The two corresponding angles are small enough to be considered close to
normal incidence.

The real part of the index is seen to grow to large values, but it is not
our concern here since the opacity is determined by $n_2$.
The transmission coefficient along $z$ (therefore for $c_\theta \approx 1$)
 is given by
\begin{equation}
T = e^{-8\pi\eta n_2 c_\theta} \approx 1- 8\pi \eta n_2,
\end{equation}
while experimental measurements \cite{Nair} are compatible with 
\begin{equation}
T \approx 1 - \pi\,\alpha_{vac},\ \alpha_{vac}=\frac{1}{137}.
\end{equation}
This requires
\begin{equation}
n_2 \approx \frac{\alpha_{vac}}{8\eta} \in [.57,.31]\ for\ \eta \in
\left[\frac{1.6}{1000}, \frac{2.9}{1000}\right],
\end{equation}
in which the values of $\eta$ correspond to the ones evaluated for visible
light in subsection \ref{subsec:magnitude}.

We get therefore  a reasonable order of magnitude for $n_2$. The factor
$\sim 2$ of discrepancy
between our prediction and the experimental value can be thought as an
estimate of higher order corrections to the vacuum polarization.

\subsubsection{At $\boldsymbol{\theta=0}$}

Like at $B\not=0$ we come back to the light-cone equation (\ref{eq:lc0}).
The transmittance $V$ has the same expression at small $\eta$ given in
(\ref{eq:Vtetanull}) and, using 
$T^{22}_{\xcancel{B}}=-T^{11}_{\xcancel{B}}= \pm i
\frac{\pi\alpha}{2}q_0$ obtained from (\ref{eq:PinoB2}),
 one gets finally
\begin{equation}
q_0(1-n^2)\left[(\beta_1^2+\beta_2^2)q_0 \pm i\frac{\alpha}{4}\eta^2
p_3^m(\beta_2^2-\beta_1^2)\right]=0
\end{equation}
which has $n=1$ for only solution. So, like at $B\not=0$, the index goes to
its trivial value at exactly normal incidence.

Like for $B\not=0$, we are at a loss to give a reliable description
of  the transition between $\theta$ small and $\theta=0$:
our model and the approximations that we made certainly fail at some point
since continuity looks very hard to achieve in this narrow domain.

\subsection{Comparison with the case  $\boldsymbol{B\not=0}$}

Like in the presence of $B$, no non-trivial solution exists for the
transverse polarization of the electromagnetic wave. Though the dimensional
reduction that occurs in the presence of $B$ can no longer be invoked,
this makes, in practice, the solution for $A^\mu_\parallel$ only depend
 on $\Pi^{11}$ and $\Pi^{33}$ (the latter being no longer equal to $-\Pi^{00}$). 
 
When $B\not=0$, we suggested that  the large
modifications to the propagation of photons inside
graphene are due to the magnetic resonance of the spins of electrons,
by the combined action of the
static $B$ and of the oscillating $b$ perpendicular to $B$.
When  $B=0$, no such enhancement is then expected to occur,
which is confirmed by our results. They  only display a weak
dependence on $\alpha$.

Notice that, paradoxically, the case  $B=0$ looks more tedious to
handle. The behavior of the perturbative series at ``fixed
order'' seems indeed to become rapidly uncontrollable when $\alpha$ grows.
This phenomenon has already been noticed \cite{BarnesSarma}, and
techniques going beyond standard perturbation theory (Random Phase
Approximation, Dyson-Schwinger equations \ldots)  are then probably  needed.

In subsection \ref{subsec:MSM} we shall give other arguments why, in
connexion with the massless Schwinger model, 1-loop calculations in the
presence of a large external $B$ maybe more reliable.

\section{Conclusion and prospects}\label{section:conclusion}

We would like to summarize  not only the salient properties and
 achievements  of our description of graphene in external magnetic field,
 but also its odds and weirds, and its limitations.

\subsection{Outlook}

We have shown that, in the presence of a constant uniform external magnetic field,
the refractive index of graphene  is very sensitive to 1-loop quantum
corrections.  The effects, which only concern the ``transverse magnetic''
polarization of photons, are large for optical wavelengths and for
magnetic fields even below 20 Tesla. They only depend (at least for the real
part of the refractive index), on the ratio
$c\frac{\sqrt{2\hbar eB}}{q_0}$. In particular, refractive effects grow
like $\sqrt{eB}$ (as compared with a growth $\propto eB$ in the vacuum for
supercritical magnetic fields demonstrated in \cite{Shabad}), but new
quantum effects are expected at $B \geq B^m$ which will probably modify the
behavior of $n$.

At the opposite,  in the absence of external $B$,  quantum effects stay
small and the optical properties of graphene are mainly controlled by the
transmittance function which incorporates the geometry of the sample and
the confinement along $z$.

The behavior of $n$ as $\theta$ becomes small  has been
found to be different whether $B\not=0$ or $B=0$.
When $B\not= 0$ a brutal transition at
$\theta_{min} \approx \frac{1}{\Upsilon}$ occurs below which the quasi-real
solution valid above this threshold disappears, presumably (but this is
still to be proved rigorously) in favor of a complex solution with large
values of $n_1$ and $n_2$ (see subsection \ref{subsec:bigsol}).
At $B=0$ the transition is smooth:
$n$ becomes gradually complex with larger and larger values of its real and
imaginary components. Unfortunately, the domains of reliability of our
calculations do not overlap such that the transition $B\to 0$ cannot be
achieved smoothly from the case $B\not=0$ which is only reliable at $B$ ``large''.
Efforts are therefore  needed to perform
calculations valid in a wider range of $B$, which allows in particular
 a continuous transition to $B=0$.

Our description of graphene differs from what is usually done and
it may be useful to summarize it.
It has been considered, in position space, as $3+1$ dimensional,
with a very small thickness $2a$.
Electrons at the Dirac points  have been described
as  massless Dirac-like particles with a vanishing ``classical''
momentum along $z$. An
important feature of our calculation is 
confining the $\gamma\, e^+e^-$ vertices inside the very narrow graphene strip
thanks to a calculation in position space of the photon propagator.
This confinement in the direction of $B$  goes along
with quantum fluctuations of the corresponding electronic momenta 
and for the momentum of the photon, which play important roles.
This makes our approach depart  not only from a description of
electrons by a QFT in 2+1 dimensions, but also from a too restrictive
brane-like model in which electrons live in 2+1 dimensions
 while gauge fields live in 3+1. In this respect, the sole calculation of
the genuine vacuum polarization $\Pi_{\mu\nu}$,
 would it be in ``reduced $QED_{4,3}$''
\cite{KotikovTeber}, skips the transmittance $U$ and may not
fully account for the  optical properties of graphene.
This looks specially true at $B=0$, where the index $n$ is mainly controlled
by $U$.
However, the situation could improve at very large $B$ because, as can be
seen on Figure~\ref{fig:nBreal}, for $\alpha \geq 1$ and
 inside the zone of confidence,
$n$ only displays a weak  dependence  on $\theta$ and seems 
rather weakly constrained by the ``leading'' $\frac{1}{s_\theta}$
behavior coming from  $U$. Since $U$, unlike 
$T^{\mu\nu}$, is independent of both $B$ and  $\alpha$,
their relative influence  should decrease as they themselves increase:
$n$ might then mostly depend on (seemingly resonant) effects
controlled by  $T^{\mu\nu}$.

We furthermore  used $c$ and not $v_F$  inside the electron propagators
because, at the idealized limit of a graphene strip with infinite
horizontal extension $L \to \infty$ that we are considering,
 the Coulomb energy of 
an electron inside the medium is expected to vanish and the quantum
fluctuations of their momentum along
$B$ make them mostly propagate outside graphene.
When $c$ is  ``decreased'', which corresponds to electrons more and more
``confined into graphene'', we have found that the effects of $B$ on
the refractive index increase. 

Because of the approximations that we have made, and that we list below, we
cannot pretend to have devised a fully realistic quantum model.
We have indeed:\newline
*\ truncated the perturbative series at 1-loop; there are hopes, however
(see subsection \ref{subsec:MSM}) that, for a strong external $B$,
 this is a reasonable approximation;
 \newline
*\ truncated  the expansion of the electron propagator for large $B$
 at next-to-leading order;\newline
*\ approximated an incomplete $\beta$ function $F(x)= (-2)^{(-1+x)} \beta(-2,1-x,0)
\approx \frac{1}{1-x}$, which in particular forget about poles at $p_0^2=
2neB$ except for $n=1$; this is however  safe for electrons
 with energy less than
$13\sqrt{B(T)}\; (eV)$, which is always achieved when they are created from
photons in the visible spectrum;\newline
* chosen the Feynman gauge for the external photons;\newline
*\ studied light-cone equations only through their expansions at large
$\Upsilon=c\frac{\sqrt{2\hbar eB}}{q_0}$ and small $\eta=aq_0 /\hbar c$.
The small values of $\eta$ that occur in the visible spectrum guarantee
that virtual created from photons  have energies
small enough to stay in the linear (Dirac) part of the spectrum.

In our favor, that we have gone beyond the limit $B\to \infty$ in the
electron propagator is very fortunate because 
the effects induced on the refractive index are due to sub-leading terms.

We have worked in domains of wavelengths and magnetic fields in which
our specific expansions and approximations are under control.
 Magnetic fields smaller or equal to
$20\,T$ are fairly common practice today, and at $20\,T$
the degeneracy of the Landau level at the Dirac point is not yet lifted.
\newline
The large effects that we have
obtained appear less surprising when they are realized to  occur
at suitable conditions for electron spin resonance.

Some additional remarks are due concerning the dimensionality of the problem
(see also subsection \ref{subsec:MSM}).
When $B\to \infty$, the Larmor radius of electrons vanishes like
$\frac{1}{\sqrt{eB}}$ such that, the degrees of freedom orthogonal to
$B$ shrinking to $0$, the physics becomes $1+1$ dimensional (see for
example \cite{ShabadUsov}).
This is generally associated, in standard QED,
 with the projector $(1-i\gamma^1\gamma^2)$
that controls, at this limit, the electron propagator. 
We have seen that, for graphene at large $B <\infty$,
the situation is more subtle but dimensional reduction still arises through
 the integration over the electronic transverse degrees of freedom.
The resulting
two extra powers of $eB$ counter-balance  the inverse powers occurring in the
non-leading terms of the electron propagator, always dropped at $B=\infty$,
and promote $\Pi^{00}$ and $\Pi^{33}$ as the two dominant components
 of the vacuum polarization. The transverse motion of electrons 
therefore plays for graphene an important role.

The direction ``3'' parallel to the external $B$ definitely
also  plays an essential role, not only
by the prominence of $\Pi^{33}$ in the light-cone equation for
$A^\mu_\parallel$, but also, by its ``squeezing'' to $2a$ and,
through the confinement of the $\gamma\,e^+ e^-$ vertices, by
the large quantum fluctuations of the electron momentum
that control  the leading $1/s_\theta$ behavior of the refractive
index.\newline
The three directions of motion of the virtual electrons therefore
collaborate to produce the effects on $n$ that we calculated.

A tantalizing question concerns of course the magnitude of higher order
corrections. If 1-loop corrections to $n$ are large, how
can we trust the result, unless  all higher orders are proved to be
much smaller? At present we have no answer to this.
That $\alpha \simeq 2$
inside graphene is already a  bad ingredient for a reliable
 perturbative treatment
\footnote{In the case of the hydrogen atom at $B\to \infty$
 it was shown in \cite{Godunov}
that 2-loop effects are negligible. It is also instructive to look at
\cite{BarnesSarma} which shows that, in the framework of the Random Phase
Approximation, graphene, despite
a large value of $\alpha$, behaves at 2-loops like a weakly coupled system.
However, in this last study, no external magnetic field is present.}
and,
furthermore, the corrections to $n$ do not look like a standard
series in powers of $\alpha$. Comparisons can be made for example with the
results obtained in the case of non-confined massive electrons with
the effective Euler-Heisenberg Lagrangian.
The equations (2.17) (2.18) of \cite{DittrichGies1} show quantum corrections
to $n$ proportional to $\alpha\left(\frac{e B}{m_e^2}\right)^2$.
 In the study of the
hydrogen atom \cite{MachetVysotsky}\cite{GodunovMachetVysotsky}, typical
corrections are proportional to $\alpha \frac{eB}{m_e^2}$. In the present
study, electrons are massless, and dimensionless factors are built
with $q_0$ in place of $m_e$. Quantum corrections to the leading
$\frac{1}{s_\theta}$ behavior of the index come out proportional to
$\left(\frac{\alpha}{\pi}\right)^2\frac{eB}{q_0^2}$ (see
(\ref{eq:solpar})).
While, at $B=0$ the situation looks very delicate to handle, the case
of a large external $B$ looks more promising. We shall make a few remarks
that motivate this optimism in subsection \ref{subsec:MSM} in connection with
the dimensional reduction $D \to D-2$ already evoked and with the massless
Schwinger model.

A delicate and unresolved issue is the transition to $\theta=0$, at
$B\not=0$ as well as at $B=0$. 
The only solution to the light-cone equation at $\theta=0$ is
the trivial  $n=1$ while  at small $\theta \not=0$ the  non-trivial
solution that we exhibited  behaves like $\frac{1}{s_\theta}$.
Nature may choose everywhere the solution $n=1$,
but, then, one should understand why the 
one on which we focused gets rejected. A possibility is also that the two
solutions coexist  down to $\theta=\theta_{\min}$, below which light
can only propagate with $n=1$.

What happens to an external light beam intersecting a graphene strip is
also unclear. That
we stand very far from geometrical optics (see section \ref{section:intro}),
and that the Snell-Descartes laws of refraction for plane waves 
 cannot be satisfied (we have indeed seen that $n\sin\theta >1$ inside graphene),
are  signals that the situation is not standard.

Last, hints exist that we  only studied in this work
most simple aspects of the optical behavior of graphene in an external magnetic
field. We just make in subsection \ref{subsec:bigsol} below  a few remarks
concerning the possible existence of other solutions to the light-cone
equations. This may however prove a Pandora box that we do not intend to
open here.

\subsection{Are there other solutions with a large absorption ?}
\label{subsec:bigsol}

We have seen that, as $\theta$ decreases, a transition occurs at $\theta
\sim \frac{1}{\Upsilon} = \frac{q_0}{c\sqrt{2\hbar eB}}$: the quasi-real solution
that we have exhibited for larger angles ``disappears''.

\begin{figure}[h]
\begin{center}
\includegraphics[width=6 cm, height=5 cm]{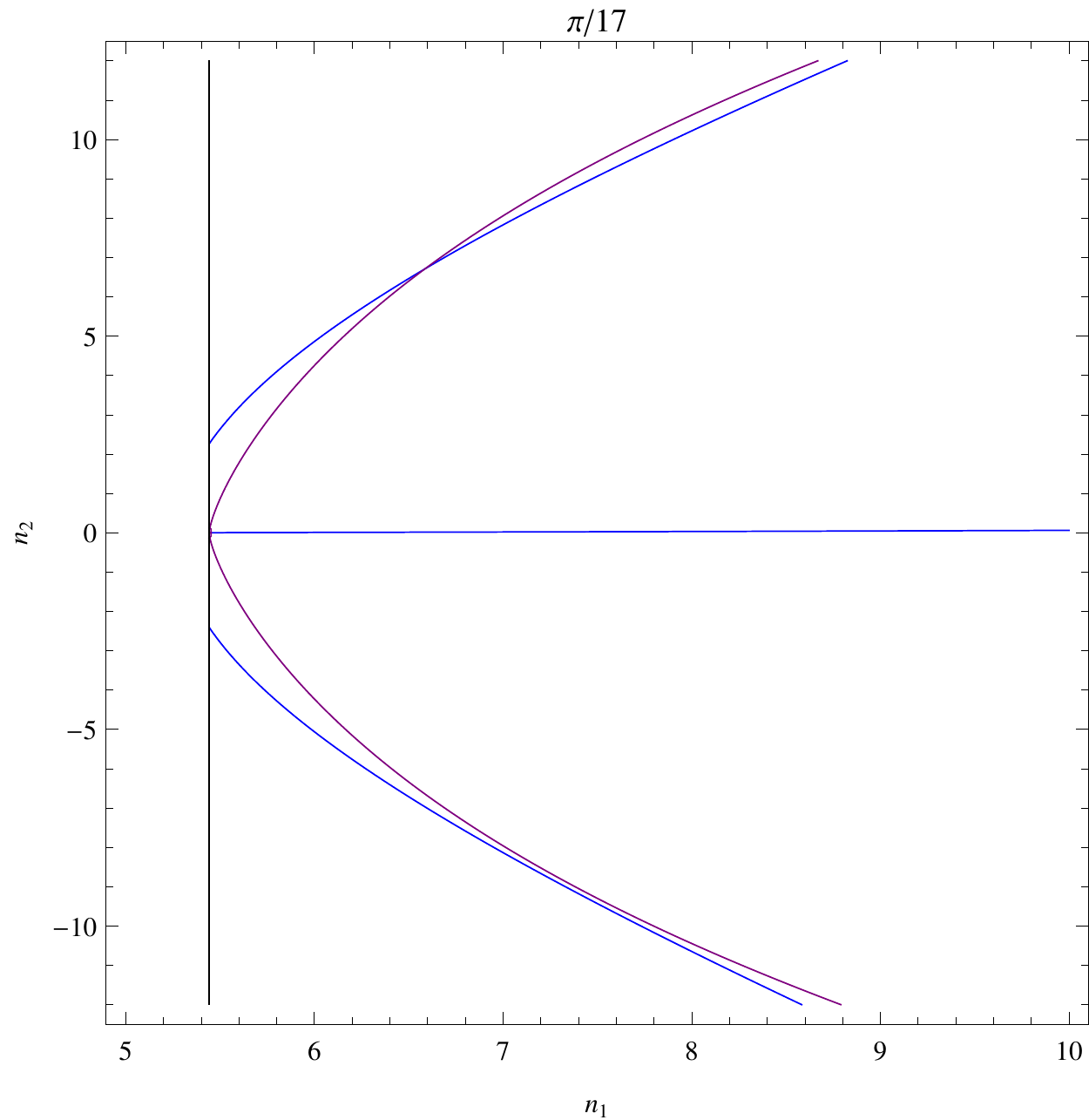}
\end{center}
\caption{Solutions $(n_1,n_2)$ of the real part (purple) and imaginary part (blue)
of the light-cone equation (\ref{eq:lccomp}) for $A^\mu_\parallel$ at
$\theta= \frac{\pi}{17}$ in the presence of $B$.
 The black vertical line on the left corresponds to
$n_1=\frac{1}{s_\theta}$.}
\label{fig:othersol}
\end{figure}

If one considers, below the threshold, at the same $\theta=\frac{\pi}{17}$
the same Figure~\ref{fig:wall1} drawn on a much larger domain for $n_1$ and $n_2$, one gets
Figure~\ref{fig:othersol}. One solution (at least) occurs for the light-cone equation
(\ref{eq:lccomp}),
which corresponds to $n_1 \approx 6.5, n_2 \approx 7$.

This suggests that below $\theta_{min}$, the system goes to a large index
with a large absorption. This type of solution is incompatible with the
 approximations that we have made to find them $|n_2| \ll n_1$ etc, such that
drawing a definitive conclusion requires using more elaborate numerical
methods.

Such solutions with large index/absorption may exist even for $\theta>\theta_{min}$,
that is jointly with the quasi-real solutions that we have focused on in this work.

\subsection{Physics in strong external $\boldsymbol B$ and
the  Schwinger model}
 \label{subsec:MSM}

The electromagnetic coupling $\alpha$ being as large as $2.3$ in suspended
 graphene, the fate of any
perturbative attempt looks {\em a priori} very gloomy, and
the prospects are indeed  dark in the absence of external $B$.

However, in presence of a large external $B$, the situation could be much
better.
If we forget about graphene,  4-dimensional ``standard''  Quantum Electrodynamics
of massive electrons in strong external $B$
shares many properties with that of QED in 2 dimensions without $B$
(see for example \cite{Vysotsky}\cite{MachetVysotsky} and references
therein).
The limit of massless fermions is specially attractive since
the massless Schwinger model has remarkable
characteristics: radiative corrections to the photon propagator
stop at 1-loop (it is anomalous and keeps non-vanishing and constant
while higher orders are both convergent and
proportional to the (vanishing) electron mass), and
the vacuum polarization tensor
is not transverse (while preserving gauge invariance). The latter writes, up to
1-loop, $\Pi^{\mu\nu}(q)=
(g^{\mu\nu} q^2 -q^\mu q^\nu)\left(1-\frac{e^2}{\pi\, q^2}\right)+ q^\mu
q^\nu$, and corresponds to a photon with mass $m_\gamma =
\frac{e}{\sqrt{\pi}}$ \cite{MSM}.

When dealing with graphene, we are precisely,
on the Dirac cones,   concerned with massless
electrons. Furthermore, as we have seen, 
 the dimensional reduction $D \to D-2$ still operates
\footnote{and it does not depends of using $c$ or $v_F$ inside the electron
propagator.}.

However, the calculation that we have done differs from  standard QED
since, in particular, the log divergent integral typical of
the massless Schwinger model has been split in $\mu=\int dp_3$
that factored out and was regularized with the
cutoff $p_3^m$, $\times$ convergent integrals $B,C,D$ (\ref{eq:Bdef})
 over $p_0$.
We have seen furthermore that the leading terms $\Pi^{33}=-\Pi^{00}$
 of the vacuum polarization
tensor are generated by non-leading contributions in the expansion at
large $B$ of the electronic propagators.

This is why more detailed investigations are needed to establish
whether a correspondence exists between graphene in strong external $B$
and the massless Schwinger model.
Among the goals is of course providing reliable arguments
that our 1-loop results stay reliable beyond this approximation,
and that the non-transversality of the vacuum polarization tensor
reveals the presence of a massive photon, which can then be expected to screen
the Coulomb potential. This would provide a sensible access to
electron-electron interactions inside graphene in the presence of an
external $B$.

This will be the object of a subsequent work.

\subsection{Other open issues}

In the course of this study,  we noticed several intriguing features:
for $B\not=0$, the non-trivial solution of the light-cone equation
seems to undergo a brutal transition at $\theta=\theta_{min}$,
new quantum effects can be expected at $B > B^m$, and, even at $B=0$,
something dramatic happens at $\alpha \geq 1$ since $n$ diverges for
 $\theta\geq \theta_{max}$ such that $\sin^2_{\theta_{max}}=\frac{1}{\alpha}$.
These can be only artifacts of the approximations that we
made, adding to the poor reliability of a fixed order perturbative expansion
for a strongly coupled system; but some  could also be signals of phenomena like
total reflection of light,
chiral symmetry breaking, spontaneous pair production,
screening of the Coulomb interaction (massive photon)  \ldots ,
or of yet unsuspected phenomena or phase transitions.
Graphene could then  also prove a
privileged test-ground for the interplay between Quantum Field Theory,
 quantum optics and nanophysics. Experimental testing and  guidance is
of course strongly wished for.


\vskip .5cm

{\em \underline{Acknowledgments:} it is a pleasure to thank M.~Vysotsky
for his continuous interest and  encouragements. We are also very indebted to
B.~Dou{\c c}ot and J.N.~Fuchs for their comments and advice.
Erroneous statements are of course our sole responsibility.}



\appendix

\section{Demonstration of eq.~(\ref{eq:genform})}\label{section:genform}

We start from (\ref{eq:start}), in which, now, the fermion propagator $G$
depends on $B$. The notations are always $v=(v_0,v_1,v_2,v_3)=(\hat
v,v_3), \hat v=(v_0,v_1,v_2)$.
\begin{equation}
\begin{split}
\Delta^{\rho\sigma}(x,y)&=
e^2\int d^3\hat u \int_{-a}^{+a}du_3 \int d^3\hat v \int_{-a}^{+a}dv_3\cr
& \int \frac{d^4q}{(2\pi)^4}\; e^{iq(u-x)}\Delta^{\rho\mu}(q)\,
\gamma_\mu \int \frac{d^4p}{(2\pi)^4}\; e^{ip(u-v)}G(\hat p,B)\,
\gamma_\nu \int \frac{d^4r}{(2\pi)^4}\; e^{ir(v-u)}G(\hat r,B)
\int \frac{d^4s}{(2\pi)^4}\; e^{is(y-v)}\Delta^{\sigma\nu}(s)\cr
& =e^2\int d^3\hat u \int_{-a}^{+a}du_3 \int d^3\hat v \int_{-a}^{+a}dv_3\;
\int \frac{d^3q}{(2\pi)^3}\frac{dq_3}{2\pi}\; e^{i\hat q(\hat u-\hat
x)}e^{iq_3(u_3-x_3)}\Delta^{\rho\mu}(q)\cr
& \hskip 1.5cm
\gamma_\mu \int \frac{d^3\hat p}{(2\pi)^3}\, \frac{dp_3}{2\pi}\;
e^{i\hat p(\hat u-\hat v)} e^{ip_3(u_3-v_3)}G(\hat p,B)\,
\gamma_\nu \int \frac{d^3\hat r}{(2\pi)^3}\, \frac{dr_3}{2\pi}\; e^{i\hat
r(\hat v-\hat u)}
e^{ir_3(v_3-u_3)}G(\hat r,B)\cr
& \hskip 1.5cm
 \int \frac{d^3\hat s}{(2\pi)^3}\,\frac{ds_3}{2\pi}\; e^{i\hat s(\hat y
-\hat v )}e^{is_3(y_3-v_3)}\Delta^{\sigma\nu}(s)\cr
& =e^2\underbrace{\int d^3\hat u\; e^{i\hat u(\hat q+\hat p-\hat
r)}}_{(2\pi)^3\delta(\hat p + \hat q - \hat r)}
\int d^3\hat v\; e^{i\hat v(-\hat p+\hat r-\hat s)}
\int_{-a}^{+a}du_3 \int_{-a}^{+a}dv_3\;
\int \frac{d^3\hat q}{(2\pi)^3}\,\frac{dq_3}{2\pi}\; e^{i\hat q(-\hat
x)}e^{iq_3(u_3-x_3)}\Delta^{\rho\mu}(q)\cr
&\hskip 1.5cm
\gamma_\mu \int \frac{d^3\hat p}{(2\pi)^3}\, \frac{dp_3}{2\pi}\;
e^{ip_3(u_3-v_3)}G(\hat p,B)\,
\gamma_\nu \int \frac{d^3\hat r}{(2\pi)^3}\, \frac{dr_3}{2\pi}\;
e^{ir_3(v_3-u_3)}G(\hat r,B)\cr
 & \hskip 1.5cm
\int \frac{d^3\hat s}{(2\pi)^3}\, \frac{ds_3}{2\pi}\; e^{i\hat s(\hat
y)}e^{is_3(y_3-v_3)}\Delta^{\sigma\nu}(s)\cr
& =e^2\underbrace{\int d^3\hat v\; e^{i\hat v(\hat q-\hat
s)}}_{(2\pi)^3\delta(\hat q-\hat s)}
\int_{-a}^{+a}du_3 \int_{-a}^{+a}dv_3\;
\int \frac{d^3\hat q}{(2\pi)^3}\,\frac{dq_3}{2\pi}\; e^{i\hat q(-\hat
x)}e^{iq_3(u_3-x_3)}\Delta^{\rho\mu}(q)\cr
&\hskip 1.5cm
\gamma_\mu \int \frac{d^3\hat p}{(2\pi)^3}\, \frac{dp_3}{2\pi}\;
e^{ip_3(u_3-v_3)}G(\hat p,B)\, \gamma_\nu \int  \frac{dr_3}{2\pi}\;
e^{ir_3(v_3-u_3)}G(\hat p+\hat q,B)\cr
& \hskip 1.5cm
\int \frac{d^3\hat s}{(2\pi)^3}\,\frac{ds_3}{2\pi}\; e^{i\hat s(\hat
y)}e^{is_3(y_3-v_3)}\Delta^{\sigma\nu}(s)\cr
& =e^2\int_{-a}^{+a}du_3 \int_{-a}^{+a}dv_3\;
\int \frac{d^3\hat q}{(2\pi)^3}\, \frac{dq_3}{2\pi}\; e^{i\hat q(-\hat
x)}e^{iq_3(u_3-x_3)}\Delta^{\rho\mu}(\hat q,q_3)\cr
&\hskip 1.5cm
\gamma_\mu \int \frac{d^3\hat p}{(2\pi)^3}\, \frac{dp_3}{2\pi}\;
e^{ip_3(u_3-v_3)}G(\hat p,B)\,
\gamma_\nu \int  \frac{dr_3}{2\pi}\;
e^{ir_3(v_3-u_3)}G(\hat p+\hat q,B)
 \int \frac{ds_3}{2\pi}\; e^{i\hat q(\hat y)}e^{is_3(y_3-v_3)}
\Delta^{\sigma\nu}(\hat q, s_3)\cr
& =e^2\int \frac{dr_3}{2\pi} \int \frac{ds_3}{2\pi}\int_{-a}^{+a}du_3\;
e^{iu_3(q_3+p_3-r_3)}
\int_{-a}^{+a}dv_3\; e^{iv_3(-p_3+r_3-s_3)}\cr
& \hskip 1cm
\int \frac{d^3\hat q}{(2\pi)^3}\, \frac{dq_3}{2\pi}\; e^{i\hat q(-\hat
x)}e^{iq_3(-x_3)}\Delta^{\rho\mu}(\hat q,q_3)\,
\gamma_\mu \int \frac{d^3\hat p}{(2\pi)^3}\, \frac{dp_3}{2\pi}\;
G(\hat p,B)\, \gamma_\nu\, G(\hat p+\hat q,B)
 e^{i\hat q(\hat y)}e^{is_3(y_3)}\Delta^{\sigma\nu}(\hat q, s_3)\cr
& =\int \frac{dp_3}{2\pi} \int \frac{dq_3}{2\pi}\int \frac{dr_3}{2\pi}
\int \frac{ds_3}{2\pi}\int_{-a}^{+a}du_3\; e^{iu_3(q_3+p_3-r_3)}
\int_{-a}^{+a}dv_3\; e^{iv_3(-p_3+r_3-s_3)} \cr
&\hskip 1.5cm
\int \frac{d^3\hat q}{(2\pi)^3}\;
e^{i\hat q(\hat y -\hat x)}
e^{iq_3(-x_3)} e^{is_3(y_3)}\Delta^{\rho\mu}(\hat q,q_3)
\Delta^{\sigma\nu}(\hat q, s_3)\
\underbrace{e^2 \int \frac{d^3\hat p}{(2\pi)^3}\;
\gamma_\mu\, G(\hat p,B)\, \gamma_\nu\, G(\hat p+\hat q,B)}_{iT_{\mu\nu}
(\hat q,B)},
\end{split}
\end{equation}
which is eq.~(\ref{eq:genform}).

\section{Approximate Coulomb energy of a graphene electron}
\label{section:coulomb}

Let us consider, to simplify, a circular graphene strip of radius $L$. We
need to evaluate the Coulomb energy of an electron inside graphene, 
due to the rest of the medium with global charge $+1$.
We consider that it is  uniformly spread such that the charge
density per unit surface is $\frac{e}{\pi L^2}$ and, inside the two circles of
radii $r$ and $r + dr$ lies a charge $\frac{e}{\pi L^2} 2\pi r dr$. It
contributes to the Coulomb energy of the electron by $-\frac{e}{r}
\frac{e}{\pi L^2} 2\pi r dr = -\frac{2 e^2}{L^2}dr$. The total
electrostatic energy of this electron is accordingly $-\int_0^L dr\;
\frac{2e^2}{L^2}= -\frac{2e^2}{L}$, which vanishes when $L \to \infty$.

The value given by the naive formula $-\frac{e^2}{4\pi
\epsilon_0\, a }$ is $\simeq 8.2\,eV$. It corresponds to a charge $+1$
localized at distance $a$ from the electron.

The ionization energy of an electron inside graphene
 is given in \cite{Ghadiry}, at the limit when the number of carbon atoms
goes to infinity, by $E_t \simeq .1\frac{2\pi \hbar
v_F}{3\,a_c \sqrt{3/4}} \approx 1.13\,eV$, in which $a_c=1.4\,10^{-10}\,m$
is the interatomic spacing. This value is smaller than all photon energies
that we are considering (see subsection \ref{subsec:magnitude}).

\newpage

\begin{em}

\end{em}

\end{document}